\documentclass[12pt]{article}

\usepackage{multirow}
\usepackage{calc}
\usepackage{amsmath,amssymb,amsthm,amscd}
\numberwithin{equation}{section}
\usepackage[bf]{caption}
\usepackage{longtable}
\usepackage{array}
\usepackage{enumerate}
\usepackage{float}
\usepackage{subfig}
\usepackage{mathtools}
\usepackage{multicol}
\usepackage{multirow}
\usepackage{epsfig}
\usepackage{graphicx}
\usepackage{pict2e}
\usepackage{color}
\usepackage{authblk}
\usepackage{cite}
\usepackage[all]{xy}
\usepackage[width=1.09\textwidth]{caption}
\usepackage{bbm}
\usepackage{hyperref}
\usepackage[makeroom]{cancel}

\usepackage[utf8]{inputenc}
\usepackage[T1]{fontenc}
\usepackage[english]{babel}
\usepackage{xspace}
\usepackage{pifont}
\usepackage{ytableau}

\newcommand*{\Scale}[2][4]{\scalebox{#1}{$#2$}}%

\newcommand{\U}[1]{\text{U(#1)}\xspace}
\newcommand{\SU}[1]{\text{SU(#1)}\xspace}
\newcommand{\SO}[1]{\text{SO(#1)}\xspace}
\newcommand{\Sp}[1]{\text{Sp(#1)}\xspace}

\newcommand{\x}{\textsl{x}}

\def\cS{{\mathcal S}}

\def\cW{{\mathcal W}}

\newcommand{\one}[0]{\ensuremath{\mathbf{1} }\xspace}
\newcommand{\two}[0]{\ensuremath{\mathbf{2} }\xspace}

\newcommand{\three}[0]{\ensuremath{\mathbf{3} }\xspace}

\newcommand{\beq}{\begin{equation}}
\newcommand{\eeq}{\end{equation}}
\newcommand{\bea}{\begin{eqnarray}}
\newcommand{\eea}{\end{eqnarray}}

\newcommand{\todo}[1]{}
\renewcommand{\todo}[1]{{\color{red} TODO: {#1}}}

\begin{document}

\baselineskip=15pt

\begin{titlepage}
\begin{flushright}
\end{flushright}

\begin{center}

\vspace*{ 0.0cm}
{\Large {\bf Weak coupling limit of F-theory models with MSSM spectrum and massless \U1's
 }
}
\\[0pt]
\vspace{-0.1cm}
\bigskip
\bigskip {
{{\bf  Dami\'an Kaloni Mayorga Pe\~na}{\it$^{\,\clubsuit}$}},
{{\bf  Roberto Valandro}{\it$^{\,\spadesuit,\,\blacklozenge,\,\clubsuit}$}}
\bigskip }\\[3pt]
\vspace{0.cm}
{\it \small 
${}^{\clubsuit}$ The Abdus Salam International Centre for Theoretical Physics,\\ Strada Costiera 11, 34151, Trieste,
Italy\\[8pt]
${}^{\spadesuit}$ Dipartimento di Fisica, Universit\`a di Trieste,\\ Strada Costiera 11, 34151 Trieste, Italy\\[8pt]
${}^{\blacklozenge}$ INFN, Sezione di Trieste, Via Valerio 2, 34127 Trieste, Italy\\[0.5cm]}

{\small dmayorga@ictp.it,
roberto.valandr@ts.infn.it}
\\[0.4cm]
\end{center}

\begin{center} {\bf Abstract } \end{center}
\vspace{0.4Em}

We consider the Sen limit of several global F-theory compactifications, some of which exhibit an MSSM-like spectrum. We show that these indeed have a consistent limit where they can be viewed as resulting from an intersecting brane configuration in type IIB. We discuss the match of the fluxes and the chiral spectrum in detail. 
We find that some D5-tadpole canceling gauge fluxes do not lift to harmonic vertical four-form fluxes in F-theory. We discuss the connection between splitting of curves at weak coupling and remnant discrete symmetries.

\begin{flushright}
\end{flushright}
\end{titlepage}
\clearpage
\setcounter{footnote}{0}
\setcounter{tocdepth}{2}
\tableofcontents
\clearpage

\section{Introduction}
\label{sec:intro}
F-theory \cite{Vafa:1996xn,Morrison:1996na,Morrison:1996pp} is the most suitable setup to describe type IIB backgrounds with 7-branes. These objects backreact on the closed string background by making the type IIB (complexified) string coupling $\tau$ (also called the axio-dilaton) vary over 
spacetime. The underlying idea of F-theory is to identify $\tau$  with the complex structure of an auxiliary two-torus. When $\tau$ depends on the coordinates of the type IIB  internal manifold $B_3$, the corresponding supersymmetric background in F-theory will be an elliptic fibration $Y_4$ over the base space $B_3$ (if the fibration admits a section, otherwise it will just be a 
genus-one fibration \cite{Braun:2014oya,Morrison:2014era}). 

The power of F-theory resides not only in its unifying language, able to capture the backreaction of the 7-branes and to explore regions of the moduli space where the string coupling is of order one, but also in model building. In fact, an \SU5 GUT model can be constructed with order one top Yukawa coupling \cite{Donagi:2008ca,Beasley:2008dc,Beasley:2008kw,Donagi:2008kj}, generated at a codimension-three locus in the base of the elliptic fibration where the \SU5 singularity enhances to $E_6$. 
This is a great advantage with respect to perturbative type~IIB compactifications on smooth Calabi-Yau (CY) threefolds, where the coupling is forbidden perturbatively.
This fact 
boosted numerous efforts to pursue $\SU{5}$ GUT model building in F-theory. As a result nowadays we have global examples \cite{Marsano:2009ym,Blumenhagen:2009yv,Grimm:2009yu,Marsano:2009gv,Marsano:2009wr,Krause:2011xj,Mayrhofer:2012zy,Borchmann:2013hta,Braun:2013nqa,Braun:2013yti,Cvetic:2013uta,Garcia-Etxebarria:2014qua,Krippendorf:2014xba,Krippendorf:2015kta,Lawrie:2015hia} 
exhibiting three generations of matter and realistic Yukawa textures.

In spite of all these positive features of F-theory, it is in general important to be able to connect the F-theory description to the type~IIB one. In fact, perturbative type~IIB string theory techniques are very powerful and allow to address questions that in F-theory are not still completely understood (e.g. the low energy effective couplings and moduli stabilization). 
Having a range of parameters where both descriptions are available is essential to approach problems that are understood on one side but not on the other.
This is possible when the string coupling is small almost everywhere in the ten-dimensional space-time. The axio-dilaton $\tau$ depends on the complex structure moduli of the F-theory fourfold $Y_4$. 
To reach a weak coupling regime therefore, one needs to take a proper limit in the complex structure moduli space. 
This limit is called the {\it weak coupling limit}, discussed first by Sen \cite{Sen:1997gv} and recently refined in \cite{Clingher:2012rg}. 
In this limit, all the 7-branes are D7-branes or orientifold O7-planes and the CY threefold compactification manifold is easily defined, allowing a direct match with the perturbative type~IIB configuration. 

In the last ten years, several globally consistent semi-realistic F-theory models were constructed, with techniques refined through the years. 
It is then of great interest to apply this limit to these models: On one side, some open issues in F-theory models can be better addressed in type IIB, leading to intuition
on how to 
approach them in the F-theory language. On the other side, 
new type IIB phenomena can be discovered starting from the known F-theory models. 
A recent example of this
can be found in \cite{Collinucci:2016hgh}, where it was shown that the F-theory $E_6$ point Yukawa coupling is possible also in the perturbative type IIB string theory: It is generated by a D1-instanton in the corresponding perturbative type~IIB setup if the CY threefold has a conifold singularity. 

For some of the global models that are present in litterature, expecially those supporting an \SU5 GUT spectrum, the weak coupling limit has been studied \cite{Collinucci:2008pf,Collinucci:2008zs,Collinucci:2009uh,Blumenhagen:2009up,Collinucci:2010gz,Braun:2011zm,Krause:2012yh,Collinucci:2012as,Esole:2012tf,Braun:2014nva,Braun:2014pva,Collinucci:2014taa,Martucci:2015dxa,Martucci:2015oaa,Collinucci:2016hgh,Greiner:2017ery}, leading to the fruitful exchange described above. 
More recently also globally consistent MSSM-like models \cite{Choi:2010su,Choi:2010nf,Choi:2013hua,Grassi:2014zxa,Cvetic:2015txa,Halverson:2015jua}  or $\U1$ extensions of it \cite{Lin:2014qga,Lin:2016vus} have been constructed in F-theory. Similarly, alternative unification schemes such as Pati-Salam grand unification and Trinification have been contemplated \cite{Klevers:2014bqa,Cvetic:2015txa}, in addition to previous constructions based on $\SO{10}$ GUTs \cite{Chen:2010ts,Antoniadis:2012yk,Tatar:2012tm}.
It is then a natural question whether it is possible or not to have analogous versions of them in the perturbative type~IIB picture and if so, whether these are competitive with the F-theory regime from a phenomenological point of view. 

Following these motivations, in this work we study the weak coupling limit of MSSM-like models constructed recently within the context of F-theory. We will concentrate only on models where the type IIB CY threefold is smooth. For the \SU5 models, this requirement discarded the top Yukawa coupling also in the F-theory compactification. In contrast to this, in the considered MSSM-like models the would-be conifold points do not correspond to any of the Yukawa points. Therefore, excluding these points does not prevent from any interesting phenomenological features. 

We start our analysis by taking the $\SU3\times\SU2\times\U1$ MSSM-like model constructed in \cite{Klevers:2014bqa,Cvetic:2015txa}. Here we apply the methods developed in \cite{Lin:2016vus} to compute the possible vertical $G_4$-fluxes 
keeping the base of the elliptic
fibration generic. We then apply the weak coupling limit to the elliptically fibered CY fourfold and find the corresponding type IIB configuration. With this at hand, we are able to work out all the possible gauge fluxes that satisfy the D5-tadpole cancellation condition in type IIB. An analogous procedure has been 
implemented in \cite{Krause:2012yh} for 
$\SU5\times \U1_X$ models; where the corresponding type IIB setup was found to be made up of a \U5 D7-brane stack (plus its orientifold image) and a \U1 D7-brane (plus its orientifold image). The
diagonal $\U1$ gauge bosons of the two stacks  are massive due to the so called geometric St\"uckelberg mechanism \cite{Blumenhagen:2008zz,Grimm:2011tb,Krause:2012yh}; however, a linear combination of them  remains massless and maps to the massless $\U1_X$ in F-theory. 
The authors were able to match all the vertical harmonic $G_4$ fluxes with the type IIB D5-tadpole canceling fluxes, including the only flux along a massive \U1 that does not induce a D5-tadpole. On the other hand, the massive \U1 fluxes are believed to be described generically in F-theory  by non-harmonic four-forms \cite{Grimm:2011tb}. In \cite{Krause:2012yh} the question was then raised whether the F-theory D5-tadpole condition found in \cite{Grimm:2011tb} was able to cancel the non-harmonic part of all such fluxes, as in their example this actually happened. 
After our analysis we are able to answer to this question. We have in fact found a massive \U1 flux in type IIB that is {\it not} described by a harmonic vertical flux in F-theory.

We study also the more refined MSSM-like models of \cite{Lin:2014qga,Lin:2016vus} where an extra massless \U1 is added and a 
richer structure of matter curves is realized. 
As a preparation for this analysis we 
 study a simpler $\U1\times\U1$ model \cite{Borchmann:2013jwa,Cvetic:2013nia,Cvetic:2013uta,Borchmann:2013hta,Cvetic:2013jta}. In both cases we again math the 7-brane configurations, the fluxes and the chiral spectrum. Again we find a massive \U1 flux in type IIB that is not described by a harmonic vertical $G_4$ flux.

Finally we explore the weak coupling limit of some other interesting models: these are toric hypersurface fibrations with fibers in the toric ambient spaces $\mathbb{P}_{F_3}$ and $\mathbb{P}_{F_1}$. The first is a model exhibiting a single $\U1$ symmetry with a particular charge spectrum, since in addition to singlets with charge one and two, it also contains a massless singlet with charge three \cite{Klevers:2014bqa}. 
We discuss the weak coupling limit of this model and find the D7-brane setup which realizes the charge three singlet (the 3 comes merely from the massless linear combination of the standard massive \U1's in type IIB). Interestingly, we notice that in type IIB it is not possible to Higgs these models to produce a 3-index states as it is instead possible in F-theory  \cite{Klevers:2016jsz}.

The latter model, which is based on $\mathbb{P}_{F_1}$ exhibits a $\mathbb{Z}_3$ discrete gauge symmetry \cite{Klevers:2014bqa,Cvetic:2015moa}, is also shown to have a weak coupling limit where the discrete symmetry stems as a discrete remnant of a global $\U{1}$ symmetry. By using the weak coupling limit we are able to derive the number of chiral states that in F-theory is very hard to compute.

In our analysis we also encounter something peculiar: some matter curves that in type IIB are distinguished by a massive \U1 symmetry, join into one curve in F-theory. This is a manifestation that  the corresponding global \U1 symmetry is non a true symmetry of the full setup. In fact, this symmetry is known to be broken to a discrete subgroup (possibly trivial) by non-perturbative effects also in the type IIB context 
\cite{Blumenhagen:2006xt,Ibanez:2006da,Florea:2006si,Blumenhagen:2009qh,BerasaluceGonzalez:2011wy,BerasaluceGonzalez:2012zn}. 
Our claim is that the two distinguished curves one finds at weak coupling
have matter with the same charge under the surviving discrete subgroup. We check this in the simple model mentioned above, where the  discrete $\mathbb{Z}_3$ can be detected directly in F-theory. 

This paper is organized as follows: In Section \ref{sec:Senlimit} we introduce the Sen limit, presenting a simple exemplifying model with one massless \U1. In Section \ref{sec:ToricMSSM} we consider the weak coupling limit of the MSSM-like model of \cite{Cvetic:2015txa}, we discuss the matter content and the possible vertical $G_4$-fluxes and we work out the corresponding perturbative type IIB setup; we finally apply the results to 
a model with a specific base space. 
In Section \ref{sec:F5} we discuss the weak coupling limit of a two (massless) $\U{1}$ model which is constructed by considering an elliptic fibration with the fiber cut as a hypersurface in the 2D toric ambient space $\mathbb{P}_{F_5}$. This constitutes a preamble to Section \ref{sec:F5Top} where we consider the $\U{1}$ extended MSSM-like model of \cite{Lin:2014qga,Lin:2016vus}. There, by a careful match of geometric properties as well as the flux directions we show that these models also exhibit a weak coupling limit. 
In Section \ref{sec:F3F1}  we explore the Sen limit of some other interesting models with charge three states and discrete symmetries. 
Finally in Section \ref{sec:conclusion} we present our conclusions.


%
\section{F-theory models in the perturbative type IIB limit}
\label{sec:Senlimit}

Supersymmetric F-theory compactifications to four dimensions require a Calabi-Yau fourfold that is elliptically fibered over a base manifold $B_3$.
When the elliptic fibration has a section, the fourfold can be described by a Weierstrass model:
\begin{equation} \label{WM}
y^2=x^3 + f\,x\, z^4 +g\,z^6  \:.
\end{equation}
The fiber coordinates $x,y,z$  are embedded into
$\mathbb{P}^2_{123}$. Let us call $F$ the line bundle associated with the hyperplane section of that space. $x,y,z$ are taken to be sections of $(\bar{K}_B\otimes F)^{\otimes 2}$,  $(\bar{K}_B\otimes F)^{\otimes 3}$, $F$ respectively, where $K_B$ is the canonical bundle of the base $B_3$. 
It follows that $f$ and $g$ are sections of $\bar{K}_B^{\otimes 4}$ and $\bar{K}_B^{\otimes 6}$, respectively. For later convenience, they can be rewritten in terms of $b_2$, $b_4$ and $b_6$ where $b_i$ is a section of $\bar{K}_B^{\otimes i}$:
\begin{equation}\label{fgWCL}
 f=-\frac{b_2^2}{3}+2b_4 \qquad\qquad
g=\frac{2}{27}b_2^3-\frac23 b_2b_4+b_6 
\end{equation}
The discriminant locus, where the 7-branes are located, is given by
\begin{equation}
\Delta=4f^3+27g^2=4\,b_2^2\left(b_2 b_6-b_4^2\right)- 36\, b_2b_4b_6+ 32\, b_4^3+ 27\, b_6^2 \, 
\end{equation}
and the $j$-function is
\begin{equation}
j(\tau) = \frac{4\left(24f\right)^3}{\Delta} \:.
\end{equation}

The Sen's weak coupling limit \cite{Sen:1997gv} is a limit in the complex structure moduli space that makes the axio-dilaton become constant almost everywhere in the Type IIB space-time. Let us scale the $b_i$'s with a parameter $\epsilon$ in the following way:
\begin{equation}\label{WClimbi}
 b_2 \rightarrow \epsilon^0 \, b_2\, , \qquad b_4 \rightarrow \epsilon^1\, b_4\, , \qquad b_6 \rightarrow \epsilon^2 \, b_6 \:.
\end{equation}
When $\epsilon \rightarrow 0$, the $j$-function of the elliptic fiber grows as $\epsilon^{-2}$ (away from the vicinity of $b_2=0$); correspondingly, the string coupling becomes small almost everywhere over the base space $B_3$.
In fact, for small $\epsilon$, the discriminant becomes
\begin{equation}
 \Delta \rightarrow -4\,\epsilon^2 b_2^2 \Delta_E + \mathcal{O}(\epsilon^3) \qquad\mbox{where}\qquad \Delta_E\equiv b_4^2-b_2 b_6\:,
\end{equation}
i.e. the discriminant locus factorizes into two components:
\begin{equation}
  b_2=0 \qquad \mbox{and}\qquad \Delta_E=0 \:.
\end{equation}
By looking at the monodromy of the elliptic fiber around such loci, one discovers \cite{Sen:1997gv} that the first one is an O7-plane and the second one gives the location of perturbative D7-branes. Since the O7-plane is the fixed point locus of the orientifold involution, the perturbative type IIB double cover CY three-fold (covering twice the base $B_3$) is 
\begin{equation}\label{WCCY3eq}
\xi^2 = b_2 \:,
\end{equation}
with $\xi$ a section of $\bar{K}_B$ and where the orientifold involution is $(-1)^{F_L}\Omega_p \sigma$, with $\sigma:\, \xi\mapsto -\xi$.

If we introduce the coordinate $\textsl{x} =x-\frac{1}{3}b_2z^2$ and rewrite the Weierstrass equation by using the parametrization \eqref{fgWCL} for $f$ and $g$ we find\footnote{When we rescale the $b_i$'s as in \eqref{WClimbi}, this equation describes a family of Calabi-Yau fourfolds over the $\epsilon$-plane. At $\epsilon=0$, the elliptic fiber degenerates over all points of $B_3$. What is worse $b_4$ and $b_6$ become zero, i.e.
the information on the location of the D7-brane locus, is lost completely. In  \cite{Clingher:2012rg,Donagi:2012ts} it has been studied how to deal with such a degeneration.} 
\begin{equation}\label{W5sing}
 y^2 = \textsl{x}^3 + b_2 \,\textsl{x}^2\, z^2+ 2b_4\, \textsl{x}\,z^4 + b_6 \,z^6 \:.
\end{equation}
In this form, the sections $b_i$'s defining the perturbative O7 and D7 data appear in a simple way.
We will use this form of the elliptic fibration in the rest of the paper.

Keeping the $b_i$'s form generic one has a smooth Calabi-Yau fourfold. At weak coupling one finds only one invariant D7-brane described by the equation $b_4^2-\xi^2b_6$ and supporting no massless gauge boson. Due to the form of the equation this has been called in literature a Whitney brane. To obtain a more interesting 7-brane setup, one needs to specialize the form of the $b_i$'s. In F-theory one obtains then singularities of the elliptic fibrations; at weak couplings the Whitney brane splits into stack of D7-branes supporting Abelian or non-Abelian gauge groups and charged matter. We will see a relevant example in the next section.

\subsection{Example: one massless \U1}\label{Sec:MPexample}

Massless \U1's gauge bosons are obtained in F-theory if the elliptic fibration has an extra (possibly rational) section. 
In \cite{Morrison:2012ei}, the generic form of an elliptic fibration with one extra section has been worked out. The corresponding Weierstrass model in the notation of \cite{Morrison:2012ei} is
\begin{equation}\label{EqMPW}
y^2 = x^3+ \left( c_1c_3-b^2c_0 -\frac{c_2^2}{3}\right) x + c_0c_3^2 - \frac13 c_1c_2c_3+\frac{2}{27}c_2^3-\frac23 b^2c_0c_2 +\frac{b^2c_1^2}{4}\:,
\end{equation}
where we set $z=1$ (the interesting physics happens in this patch).
Written in terms of the variable $\textsl{x} =x-\frac{1}{3}b_2z^2$, the defining equation takes the form
\begin{equation}\label{EqMP}
  y^2 = \textsl{x}^3 +c_2\textsl{x}^2 + \left( c_1c_3-b^2c_0 \right) \textsl{x} + c_0c_3^2-b^2c_0c_2+\frac{b^2c_1^2}{4}\:,
\end{equation}
where $b$, $c_0$, $c_1$, $c_2$ and $c_3$ are generic sections of the line bundles $\mathcal{B}$, $\bar{K}^{\otimes 4}\otimes \bar{\mathcal{B}}^{\otimes 2}$, $\bar{K}^{\otimes 3}\otimes \bar{\mathcal{B}}$, $\bar{K}^{\otimes 2}$ and $\bar{K}\otimes \mathcal{B}$ respectively  (with $\mathcal{B}$ an arbitrary line bundle on $B_3$).

This fourfold has two conifold-like singularities along two curves on the base $B_3$. They are both resolvable and hence signals the presences of a massless $U(1)$ gauge boson in the low dimensional effective theory \cite{Grimm:2011tb}. The extra-divisor giving the massless $U(1)$ gauge bosons (from the redution of $C_3$) is the new rational section. Matter fields live on these curves that are charged under the $U(1)$ gauge group. The fields living on one curve have double the charge of the fields living on the other curves. Setting charge equal to 1 for the latter, the former are charge 2 fields \cite{Morrison:2012ei}.

Let us see now the weak coupling limit. The $b_i$'s take now the particular form
\begin{eqnarray}
b_2 &=& c_2 \:, \\
b_4 &=& \frac12 \left( c_1c_3-b^2c_0 \right) \:, \\
b_6 &=& c_0c_3^2-b^2c_0c_2+\frac{b^2c_1^2}{4} \:.
\end{eqnarray}
We need to rescale the the sections $b$ and $c_i$'s such that the $b_i$'s scale as \eqref{WClimbi}. We want to do this in the most generic way, i.e. without generating extra matter and extra gauge group factor with respect to the F-theory setup under consideration. A proper choice is\footnote{An equivalent choice is
$
 \qquad b \rightarrow \epsilon^1 \, b\, , \qquad  c_0 \rightarrow \epsilon^0 \, c_0\, , \qquad c_1 \rightarrow \epsilon^0 \, c_1\,, \qquad c_2 \rightarrow \epsilon^0 \, c_2\, ,  \qquad c_3 \rightarrow \epsilon^1\, c_3\:.
$
}
\begin{equation}\label{WClimbci}
 b \rightarrow \epsilon^0 \, b\, , \qquad  c_0 \rightarrow \epsilon^2 \, c_0\, , \qquad c_1 \rightarrow \epsilon^1 \, c_1\,, \qquad c_2 \rightarrow \epsilon^0 \, c_2\, ,  \qquad c_3 \rightarrow \epsilon^0\, c_3\:.
\end{equation}
We notice that at weak coupling the $b_4$ loses one term and factorizes as $b_4=\frac{c_1c_3}{2}$. After the limit, the D7-brane configuration we ended up with is:
\begin{equation}
 \Delta_E = 0 \qquad\mbox{with}\qquad \Delta_E =\left( c_3^2 - c_2 b^2 \right)\, \left( \frac{c_1^2}{4}-c_2c_0 \right)\:.
\end{equation}
The O7-plane is at the locus $c_2=0$. On the CY $\xi^2=c_2$, the D7-brane locus becomes
\begin{equation}
\Delta_E = \left( c_3 - \xi b \right)\, \left( c_3 + \xi b \right)\,  \left( \frac{c_1^2}{4}-\xi^2c_0 \right)
\end{equation}
We recognizes a system of one \U1 brane and its orientifold image, plus a Whitney brane. The two $\U1$ branes are in the same homology class and hence the \U1 gauge bosons is massless (if no gauge flux is switched on). The matter occurs at the D7-brane intersections. We have a charge $1$ state at the intersection of the \U1 brane with the Whitney brane and one charge $2$ state at the intersection of the $\U1$ brane with its image. We obtain the same spectrum as at strong coupling, i.e. we are describing the same physical configuration at weak and strong coupling. This is an important requirement to fulfill in order to claim to have a weakly coupled description of the F-theory setup. For several cases, a weak coupling limit is possible (in the sense that the string coupling is small everywhere)
but at the price of generating extra gauge groups (see \cite{Esole:2012tf}).

\section{An MSSM-like F-theory model}
\label{sec:ToricMSSM}

In this Section we  study a phenomenologically more interesting case, i.e. an elliptic fibration supporting the Standard Model spectrum. 

\subsection{F-theory description}
\subsubsection{Geometric setup}

\begin{figure}[H]
\centering
\begin{minipage}{.56\textwidth}
  \centering
  \includegraphics[scale=.4]{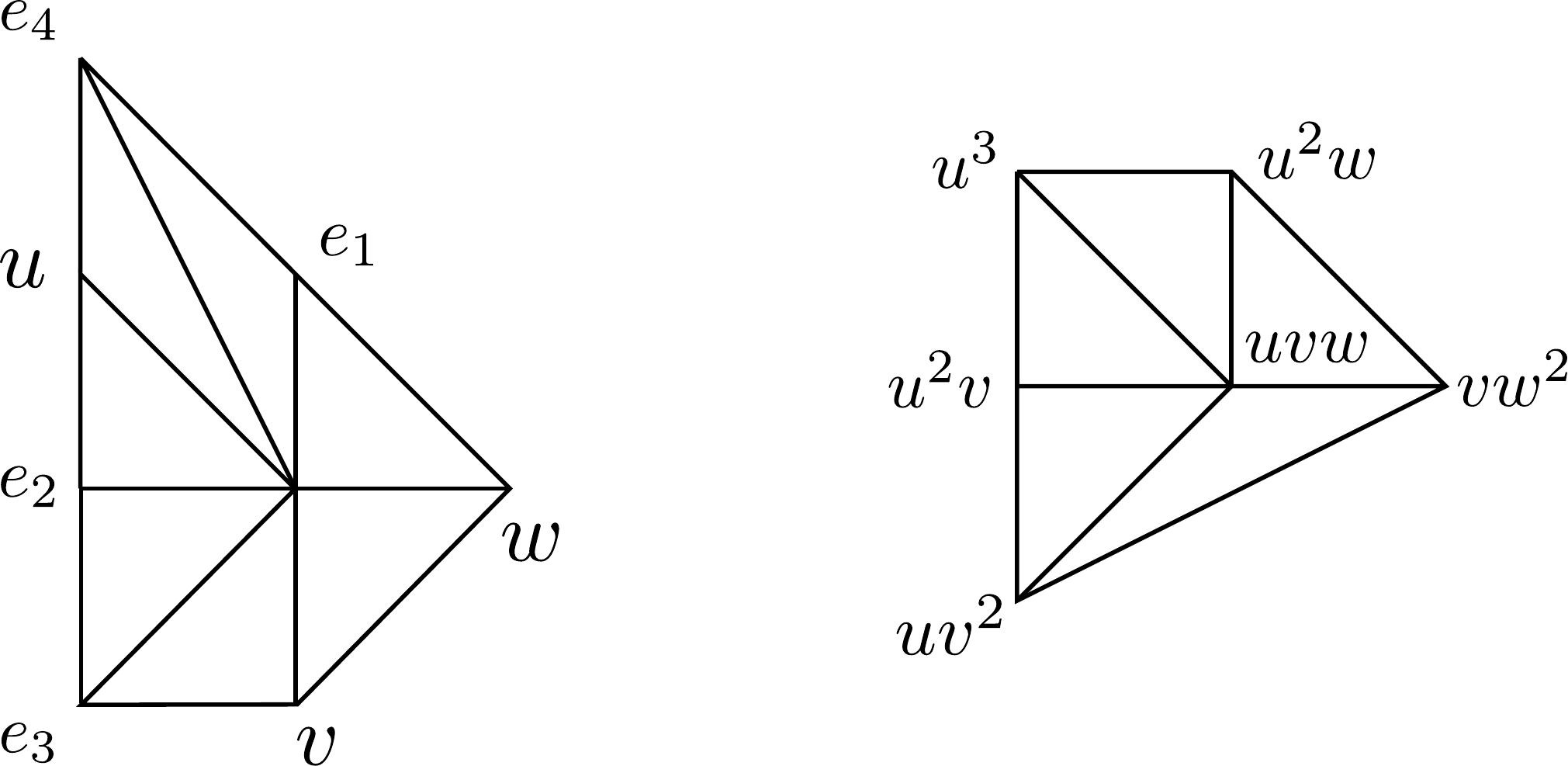}
\end{minipage}%
\begin{minipage}{.44\textwidth}
{\footnotesize
  \begin{tabular}{|c|c|}\hline
Section & Line Bundle\\ \hline
$u$ & $\mathcal{O}(H-E_1-E_2-E_4+\cS_9+K_B)$ \\ \hline
$v$ & $\mathcal{O}(H-E_2-E_3+\cS_9-\cS_7)$\\ \hline
$w$ & $\mathcal{O}(H-E_1)$\\ \hline
$e_1$ & $\mathcal{O}(E_1-E_4)$\\ \hline
$e_2$ & $\mathcal{O}(E_2-E_3)$\\ \hline
$e_3$ & $\mathcal{O}(E_3)$\\ \hline
$e_4$ & $\mathcal{O}(E_4)$\\ \hline
\end{tabular}}
\end{minipage}
\caption{\label{fig:poly11_toric} The polytope $F_{11}$ and its dual. The  table contains the divisor classes of the coordinates in $\mathbb{P}_{F_{11}}$.}
\end{figure}

In this  model 
the elliptic fiber is described as an hypersurface in the 2D toric ambient space $\mathbb{P}_{F_{11}}$. This is associated to the polytope $F_{11}$ depicted in Figure \ref{fig:poly11_toric}. In the associated table, we read the coordinates and the associated  line bundles (whose they are sections). These last ones are written as $\mathcal{O}(\mathcal{D})$, with $\mathcal{D}$ a divisor of the fourfold, written as a linear combination of the base divisors\footnote{We will often use the same symbol for the  divisor $D$ in $B_3$ and the vertical divisor in $Y_4$ given by the elliptic fibration over $D$. We will call $D$ also its Poincar\'e dual two-form on $B_3$ and the corresponding pullback $\hat{\pi}^\ast (D)$ that is  Poincar\'e dual to the vertical divisor ($\hat{\pi}:Y_4\rightarrow B_3$ is the projection from the elliptic fibration to the base manifold).} $K_B$ (the canonical class of $B_3$), $\cS_7$ and $\cS_9$ and the divisors $H,E_1,E_2,E_3,E_4$. 
The defining equation is $p_{F_{11}}=0$, with
\begin{equation}\label{eq:pF11}
p_{F_{11}}=s_1 e_1^2 e_2^2 e_3 e_4^4 u^3 + s_2 e_1 e_2^2 e_3^2 e_4^2 u^2 v + s_3 e_2^2 e_3^2 u v^2 + s_5 e_1^2 e_2 e_4^3 u^2 w 
+s_6 e_1 e_2 e_3 e_4 u v w + s_9 e_1 v w^2 \,,
\end{equation}
where $s_i$ are sections of suitable line bundles, chosen  such that $p_{F_{11}}=0$ is a Calabi-Yau manifold. One can write the corresponding classes in terms of the anticanonical class of the base, and two extra divisor classes $\cS_7$ and $\cS_9$: 
\beq
\label{eq:sis}
\text{
\renewcommand{\arraystretch}{1.2}
\begin{tabular}{|c|c|c|c|c|c|}
\hline
$s_1$ & $s_2$ & $s_3$ & $s_5$ & $s_6$ & $s_9$  \\ \hline
$3\bar{K}_B-\cS_7-\cS_9$ & $2\bar{K}_B-\cS_9$ & $\bar{K}_B+\cS_7-\cS_9$ & $2\bar{K}_B-\cS_7$ & $\bar{K}_B$ & $\cS_9$ \\
\hline
\end{tabular}
}\:.
\eeq

By means of Nigell's algorithm one can work out the Weierstrass form \eqref{WM} of the CY \eqref{eq:pF11},
with the following expressions for $f$, $g$ and the discriminant $\Delta$
\begin{eqnarray}
\label{SMfexpr}f&=& -\frac13 \left(\frac{s_6^2}{4} - s_2s_9  \right)^2+2\left( - \frac14 s_3s_9(s_5s_6-2s_1s_9)  \right)
\,, \\
\label{SMgexpr} g&=& \frac{2}{27} \left(\frac{s_6^2}{4} - s_2s_9  \right)^3 -\frac23  \left(\frac{s_6^2}{4} - s_2s_9  \right)  \left(  - \frac14 s_3s_9(s_5s_6-2s_1s_9)  \right)
+\frac14 s_3^2s_5^2s_9^5
   \,,
\end{eqnarray}
\begin{align}
\begin{split}
  \Delta=\frac{1}{16} s_3^2 s_9^3 [ & s_3 s_5^3 s_6^3 - s_2 s_5^2 s_6^4 + s_1 s_5 s_6^5 + 
   27 s_3^2 s_5^4 s_9 - 36 s_2 s_3 s_5^3 s_6 s_9 + 8 s_2^2 s_5^2 s_6^2 s_9\\ 
  & +30 s_1 s_3 s_5^2 s_6^2 s_9 - 8 s_1 s_2 s_5 s_6^3 s_9 - s_1^2 s_6^4 s_9 - 
   16 s_2^3 s_5^2 s_9^2 + 72 s_1 s_2 s_3 s_5^2 s_9^2   \\
   &\quad  +16 s_1 s_2^2 s_5 s_6 s_9^2 - 96 s_1^2 s_3 s_5 s_6 s_9^2 + 
   8 s_1^2 s_2 s_6^2 s_9^2 - 16 s_1^2 s_2^2 s_9^3 + 64 s_1^3 s_3 s_9^3]\,. 
\end{split}
\end{align}
Notice that  \eqref{eq:pF11} is the resolved version of the given Weierstrass model. We will call both spaces $Y_4$ in the following, which one we mean will be clear from the context.

Following  Kodaira's classification, the vanishing order of the above quantities confirm that the fiber degenerates to an $I_2$-fiber over the locus $\{s_3=0\}$ and to an $I_3$-fiber over the locus $\{s_9=0\}$. The matter content for this model has been computed in refs \cite{Klevers:2014bqa,Cvetic:2015txa} and it is summarized in Table \ref{tab:poly11_matter}. By looking at the Tate form of the present fourfold, one moreover realizes that the $I_2$-fiber is associated to a $\Sp1\cong \SU2$ gauge group, while the $A_2$ singularity is `split' and the gauge group is $\SU3$ \cite{Bershadsky:1996nh}.\footnote{One can see this by shifting the $x$ coordinate to $\textsl{x}$ and realizing that the coefficient of $\textsl{x}^2$ becomes a square on top of $s_9=0$.}

\begin{table}[h]
\begin{center}
\renewcommand{\arraystretch}{1.2}
\begin{tabular}{|c|@{}c@{}|}\hline
Representation & Locus \\ \hline
$(\three,\two)_{1/6}$ & $\{s_3\,,\,s_9 \}$ \\ \hline
$(\one,\two)_{-1/2}$ & $ \{s_3\,,\,s_2s_5^2 + s_1(s_1 s_9-s_5s_6)\}$ \\ \hline
$(\overline{\three},\one)_{-2/3}$ & $\{ s_5\,,\,s_9 \}$ \\ \hline
$(\overline{\three}, \one)_{1/3}$ & $  \{ s_9\,,\,s_3 s_5^2 + s_6(s_1 s_6-s_2 s_5)\}$ \\ \hline
$(\one,\one)_{1}$ & $\{ s_1\,,\,s_5\}$ \\ \hline
\end{tabular}
\caption{\label{tab:poly11_matter} Matter representations of $\SU3 \times \SU2 \times \text{U}(1)$ appearing in the $X_{F_{11}}$-model, together with their associated codimension-two loci. The charge under the $\U1_Y$ generator is indicated by a subscript.}
\end{center}
\end{table}

By comparing \eqref{SMfexpr} and \eqref{SMgexpr} with \eqref{fgWCL}, one extracts the expressions for the sections $b_i$'s:
\begin{eqnarray}\label{SMbi}
b_2 &=& \frac{s_6^2}{4} - s_2s_9  \:, \\
b_4 &=&- \frac14 s_3s_9(s_5s_6-2s_1s_9) \:, \\
b_6 &=& \frac14 s_3^2s_5^2s_9^2 \:.
\end{eqnarray}
The defining equation for our CY fourfold is then
\begin{equation}\label{CT4SM}
y^2 = \x^3 + \left(\frac{s_6^2}{4} - s_2s_9 \right)\,\x^2\,z^2 - \frac12 s_3s_9(s_5s_6-2s_1s_9)\,\x\,z^4  +\frac14 s_3^2s_5^2s_9^2\,z^6 \:.
\end{equation}

Over the codimension-one loci in $B_3$ where the non-Abelian gauge group live, the singular point in the fiber is given by:
\begin{align}
 \{s_3=0\}\,:&\quad [\x:y:z]=\left[0:0:1\right]\,,\\
 \{s_9=0\}\,:&\quad [\x:y:z]=\left[-\frac{s_6^2}{12}:0:1\right]\,.
\end{align}
The Weierstrass model includes naturally the zero section
\begin{align}
 S_0\,:\quad [\x:y:z]=[1:1:0]\,.
 \end{align}
In order to devise the location of the extra section one can rewrite \eqref{CT4SM} in the factorized form (in the patch $z=1$)
\begin{equation}\label{eq:WSF11}
\left(y-\frac12 s_3 s_5 s_9 \right)\left(y+\frac12 s_3 s_5 s_9 \right) = \x\left( \x^2 + \left(\frac{s_6^2}{4} - s_2s_9 \right)\,\x - \frac12 s_3s_9(s_5s_6-2s_1s_9)  \right)\:,
\end{equation}
from which one can obtain the fiber coordinates of the extra section:\footnote{One can actually read two extra sections, the second being at $[\x:y:z]=\left[ 0:-\frac{s_3 s_5 s_9}{2}:1\right]$. This second one however is not independent from the given ones.}
\begin{align}
 S_1\,:\quad [\x:y:z]=\left[ 0:\frac{s_3 s_5 s_9}{2}:1\right]\,.
 \label{eq:S1}
\end{align}
From these two inequivalent sections one obtains the Shioda map for the generator of a geometrically massless $\U1$ symmetry:
 \begin{align} \label{eq:ShiodaF11}
\sigma_1=S_1-S_0 +  K_B + \frac{1}{2} D^{\SU2}	+ \frac{1}{3}\big(D^{\SU3}_{1}+2D^{\SU3}_{2}\big) \, .
\end{align}
The exceptional divisors $D^{\SU2}$ and $D^{\SU3}_{1},D^{\SU3}_{2}$ are given by the following linear combinations of the divisors in Fig.\ref{fig:poly11_toric}:
 \begin{align} 
D^{\SU2}=[e_1]\,\quad	D^{\SU3}_{1}=[e_2]\,,\quad D^{\SU3}_{2}=[u] \,.
\end{align}

\subsubsection{Fluxes and chiral matter}
\label{sec:FluxesF11}
In order to obtain the chiral indices associated to the different matter curves present in this model, we have to construct the $G_4$ fluxes that lie inside the primary vertical cohomology $H^{(2,2)}_V(Y_4)$ \cite{Greene:1993vm}, where $Y_4$ is the resolved foufold defined by the equation \eqref{eq:pF11}.\footnote{If one is interested in the full massless spectrum, including  vector-like matter, more refined techniques must be used \cite{Bies:2014sra,Bies:2017fam}.} 
The relevance of the vertical fluxes for the chiral spectrum was first notice in \cite{Grimm:2010ez,Braun:2011zm,Marsano:2011hv,Krause:2011xj} where explicit examples were constructed and the generated chiral spectrum computed.
For the chosen setup, the vertical fluxes have been explictly constructed for a particular base choice, $B_3=\mathbb{P}^3$, in \cite{Cvetic:2015txa}. In this work we follow instead the methods of \cite{Krause:2012yh,Lin:2015qsa,Lin:2016vus}, which enables us to address the issue of fluxes in a base independent way. 

The vertical cohomology $H^{(2,2)}_V(Y_4)$ is constructed as a quotient ring at grade two. Its elements are linear combinations of products $D_A \wedge D_B$, with $\{D_A\}$ ($A=0,...,h^{(1,1)}(Y_4)-1$) a basis for $H^{(1,1)}(Y_4)$. The vertical flux can then  be written as
\beq
G_4=c_{AB} D_A\wedge D_B\,,
\eeq
with some coefficients $c_{AB}$. In the following we will often omit the $\wedge$ symbol. 

In all cases of our interest the Calabi-Yau fourfold is described as a a toric hypersuface, where all the two-forms of the Calabi-Yau $Y_4$ are pullbacks of two-forms in its corresponding ambient space $X_5$. Of particular importance are the quartic intersections in $Y_4$, which in our case can be related to the quintic intersections in the ambient space. The quintic intersections can be computed as a polynomial ring at grade five, modulo the Stanley-Reisner ideal (SR) and modulo linear relations (LIN) that can be read off from the toric diagram of the fiber ambient space. After imposing a few (known) explicit fiber intersections one can readily express any quintic intersection in terms of cubic intersections in the base $B_3$
\beq
D_{A_1} D_{A_2} D_{A_3} D_{A_4} D_{A_5}=c_{\alpha_1\alpha_2\alpha_3}^{A_1A_2A_3A_4A_5} D_{\alpha_1} D_{\alpha_2} D_{\alpha_3}\,,
\eeq
with $D_\alpha$ being base divisors. 

The computation of the quartic intersections in the Calabi-Yau fourfold $Y_4$ relies on the quintic intersections in the ambient space $X_5$. Taking any product of divisors at degree four $D_{A_1} D_{A_2} D_{A_3} D_{A_4}$ together with the class of the hypersurface $[p_T]$ gives a quintic  intersection in the ambient space which corresponds to the quartic intersection $D_{A_1} D_{A_2} D_{A_3} D_{A_4}$ in the fourfold,~i.e. 
\beq
H^{(4,4)}(Y_4)\cong \frac{\mathbb{Q}[D_A]^4 \wedge [P_{F_{11}}]}{{\rm SRI\,+\,LIN}}\subset H^{(5,5)}(X_5)\,.
\eeq
With the quartic intersection numbers at hand we can discuss the physical requirements that have to be imposed on the $G_4$ flux. These are called transversality constraints and correspond to demanding that certain Chern-Simons coefficients vanish \cite{Dasgupta:1999ss,Marsano:2011hv,Grimm:2011fx,Grimm:2011sk,Cvetic:2012xn,Cvetic:2013uta}:
\begin{align}\label{eq:cs1}
\Theta_{0\alpha} &=\frac{1}{2}\int_{Y_4} G_4 \wedge S_0 \wedge D_\alpha = 0\,,\\\label{eq:cs2}
\Theta_{\alpha\beta} &=\frac{1}{2}\int_{Y_4} G_4 \wedge D_\alpha \wedge D_\beta = 0\,.
\end{align}
Here $D_\alpha$ denote the vertical divisors. The previous conditions are necessary in order to ensure that the resulting four dimensional theory is Lorentz invariant \cite{Dasgupta:1999ss}. Additionally we have to ensure that all non-Abelian gauge symmetries remain unbroken, which is guaranteed by the following condition 
\begin{align}\label{eq:cs3}
 \Theta_{m\alpha} &=\frac{1}{2}\int_{X} G_4 \wedge E_m \wedge D_\alpha = 0\,,
\end{align}
with $E_m$ being the exceptional divisors associated to the non-Abelian factors. 

The vertical flux takes the form 
\beq
G_4=c_I G_4^{(I)}\,,
\eeq
with $\{G_4^{(I)}\}$ being a set independent solutions to Eqs. \eqref{eq:cs1}-\eqref{eq:cs3}. 
For the case we are concerned with, the basis of divisors reads 
\beq 
D_A=\{S_0,S_1,D_\alpha,D^{\SU{2}},D^{\SU{3}_1},D^{\SU{3}_2}\}
\eeq
with  $D_\alpha$ the vertical divisors associated with the elements of a basis for $H^{(1,1)}(B)$, i.e. $\alpha=1,\ldots,h^{(1,1)}(B)$, and $D^{\SU{2}},D^{\SU{3}_1},D^{\SU{3}_2}$ the exceptional divisors in the resolved fourfold. Out of all vertical divisors $D_\alpha$ we distinguish among the subspace $\mathfrak{f}=\langle \bar{K}_B,\cS_7,\cS_9\rangle$ whose generators determine the fibration structure. For the purposes of our entire discussion it is irrelevant whether or not they are linearly dependent. One can always express a base divisor as a linear combination of the elements of $\mathfrak{f}$ and of some remaining independent divisors, that we denote as $D_\alpha^\prime$, $\alpha=1,\ldots,h^{(1,1)}(B)-{\rm rk}(\mathfrak{f})$. 

A simple Mathematica code can be used to compute the quintic intersections in the ambient space upon reduction of quintic monomials in a Groebner basis for the Stanley-Reisner ideal. As said above, the quartic intersections on the fourfold $Y_4$ can be easily computed by intersecting four divisors with the clas of $Y_4$ in the ambient space. After imposing the transversality constraints and getting rid of redundant flux pieces, we find the following expression for the $G_4$ flux over a generic base $B_3$:\footnote{For the case of base $B_3=\mathbb{P}^3$, with $\bar{K}_B =4 H$, $\cS_7=n_7 H$, $\cS_9=n_9 H$ and $H$ being the hyperplane class in $\mathbb{P}^3$, we can show that the flux expression correctly reduces to the one found in \cite{Cvetic:2015txa}.}
\begin{align}\label{SMgenericG4}
 G_4=\mathcal{F}\wedge \sigma_1+\Lambda \left(6\bar{K}_B^2 + \bar{K}_B  S_0 + S_0^2 - 5 \bar{K}_B  \cS_7 + \cS_7^2 - 2 \bar{K}_B  \cS_9 + \cS_7 \cS_9 \right)\,,
\end{align}
where  $\sigma(S_1)$ is the Shioda map of the section $S_1$ given in \eqref{eq:ShiodaF11}, and $\mathcal{F}=\gamma_\alpha D_\alpha$ (with $\alpha=1,\ldots,h^{(1,1)}(B)$) is a vertical divisor. The coefficients $\gamma_\alpha$ and $\Lambda$ are subject to the flux quantization condition and must also be in agreement with the cancellation of the D3-tadpole \cite{Witten:1996md}. 

We can finally use the flux to compute the chiral indices for the matter representations. These are given by integrating the flux $G_4$ on the corresponding matter surfaces \cite{Donagi:2008ca,Hayashi:2008ba}. 
These matter surfaces can be described as algebraic four-cycles in $X_5$ (the pushforward of the surfaces on $Y_4$ via the embedding map). For a given representation $\mathbf{R}$ there will be a six form $[\gamma_\mathbf{R}]$ Poincar\'e dual to the corresponding four-cycle, such that the chirality can be computed as:
\beq
\chi(\mathbf{R})=\int_{\gamma_\mathbf{R}}G_4=\int_{X_5}G_4\wedge [\gamma_\mathbf{R}]\,.
\eeq
The chiralities for the charged states are summarized in Table~\ref{tab:chiralities0}. By the procedure outlined above, they can be written as cubic intersections in the base $B_3$.
\begin{table}[h]
\begin{center}
\renewcommand{\arraystretch}{1.4}
{\footnotesize
\begin{tabular}{|c|c|c|}\hline
Representation & $G_4=\mathcal{F}\wedge \sigma_1$ &  $ \begin{array}{c}G_4=\Lambda(6\bar{K}_B^2 + S_0^2+\cS_7 (\cS_7 + \cS_9)\\ \quad\quad\quad+ \bar{K}_B (S_0 - 5 \cS_7 - 2 \cS_9))\end{array}$ \\ \hline

$(\three,\two)_{1/6}$ & $\frac16 \mathcal{F} \bar{K}_B (\bar{K}_B + \cS_7 - \cS_9)$  & 0\\ \hline

$(\one,\two)_{-1/2}$ & $\frac12 \mathcal{F}( \bar{K}_B + \cS_7 - \cS_9) ((-6  \bar{K}_B + 2 \cS_7 + \cS_9) $ & $\begin{array}{c}2 \Lambda ( \bar{K}_B+ \cS_7 - \cS_9)  \\
\times(2  \bar{K}_B- \cS_7) (3 \bar{K}_B- \cS_7 - \cS_9) \end{array} $ \\ \hline
      
$(\overline{\three},\one)_{-2/3}$  & $-\frac23 \mathcal{F}\cS_9 (2 \bar{K}_B - \cS_7) $ &
   $\Lambda \cS_9(2 \bar{K}_B  - \cS_7) (3\bar{K}_B  - \cS_7 - \cS_9) $ \\ \hline
   
$(\overline{\three}, \one)_{1/3}$ & $\frac13 \mathcal{F} \cS_9 (5 \bar{K}_B - \cS_7 - \cS_9)$   &   $-\Lambda \cS_9(2 \bar{K}_B  - \cS_7) (3\bar{
K}_B  - \cS_7 - \cS_9)$\\ \hline

$(\one,\one)_{1}$ & $\mathcal{F}(2 \bar{K}_B  - \cS_7) (3\bar{K}_B  - \cS_7 - \cS_9) $  & $\begin{array}{c}\Lambda(2  \bar{K}_B - \cS_7) (3 \bar{K}_B- \cS_7 - \cS_9) \\ \times(-4 \bar{K}_B+ 2 \cS_7 + \cS_9) \end{array} $ \\ \hline

\end{tabular}
}
\caption{\label{tab:chiralities0} Chiralities for the charged matter in the $F_{11}$ model over a generic base. It is understood that the triple intersection numbers are to be computed over the base $B_3$.}
\end{center}
\end{table}

We finish this section with an observation.
In the absence of the $\SU{2}$ singularity the gauge symmetry reduces to $\SU{3}\times\U1$. One can show that in this case the $\Lambda$ flux simplifies to 
\beq
G_4^\Lambda =12 \bar{K}_B^2 + \bar{K}_B  S_0 + S_0^2 - 10\bar{K}_B \cS_9 + 2 \cS_9^2\,,
\eeq
 and it becomes equivalent to a flux of the form $(-6\bar{K}_B+3\cS_9)\wedge\sigma_1$. This is in agreement with the observations of \cite{Krause:2012yh}, stating that for ${\rm SU}(n)\times\U1$ models only fluxes of the form $\mathcal{F}\wedge \sigma$ are allowed for $n<5$.

\subsection{The weak coupling limit}
\label{sec:WCMSSM}

In order to take the weak coupling limit we follow the same procedure we used in the example in Section \ref{Sec:MPexample}. We need to find a scaling of the sections $s_i$'s such that the $b_i$'s in \eqref{SMbi} scale as~\eqref{WClimbi}. A proper choice is\footnote{One may also choose to scale only $s_3$; however this choice is equivalent to \eqref{SMwcscalings}, as $s_3$ appears always in product with either $s_1$ or $s_5$. }
\begin{equation}\label{SMwcscalings}
 s_1 \rightarrow \epsilon^1 s_1, \qquad s_3 \rightarrow \epsilon^0 s_3, \qquad s_5 \rightarrow \epsilon^1 s_5, \qquad s_6 \rightarrow \epsilon^0 s_6, \qquad 
  s_9 \rightarrow \epsilon^0 s_9\:.
\end{equation}

The double cover CY threefold  $X_3$ is described by the following equation
\begin{equation}\label{SMCYeq}
 \xi^2 = \frac{s_6^2}{4}-s_2 s_9 \:,
\end{equation}
where we used \eqref{WCCY3eq} and \eqref{SMbi}.
In order to prevent a conifold singularity along the locus $\xi=s_6=s_9=s_2=0$ we will consider bases $B_3$ for which this locus is empty. Dealing with smooth CYs allows us to compute the relevant quantities through standard techniques; this makes the match with the F-theory result easier to check. However, one may deal with such singular CY threefolds by using non-commutative resolution techniques, as explained in \cite{Collinucci:2016hgh}. From equation \eqref{SMCYeq}, one sees that the divisor $\{s_9=0\}$ (and $\{s_2=0\}$ as well) splits into two components:
\begin{equation}\label{WWteqs}
 W\,\equiv\,\,\,\, \left\{ s_9=0, \,\,\, \xi-\frac12 s_6 = 0 \right\} \qquad\mbox{and}\qquad
\tilde W \,\equiv\,\,\,\, \left\{ s_9=0, \,\,\, \xi+\frac12 s_6 = 0 \right\} \:.
\end{equation}
 These two divisors are in different homology classes in $X_3$ and are mapped to each other by the orientifold involution $\xi\mapsto -\xi$.

\subsubsection{D7-brane setup}

One can now look at the D7-brane locus $\Delta_E=0$ at weak coupling:
\begin{align}\label{eq:DeltaE}
 \Delta_E=\frac{s_3^2 s_9^3}{4} \left(s_1 s_5 s_6 - s_2 s_5^2 - s_1^2 s_9\right)\:.
\end{align}

Intersecting the D7-brane locus $\Delta_E=0$ with the CY equation \eqref{SMCYeq} one obtains the following D7-branes:
\begin{itemize}
 \item \textbf{\U3 stack:} There are three D7-branes wrapping the divisor $W$ and their images wrapping the divisor $\tilde W$.  Since the two divisors are in different homology classes, a geometric St\"uckelberg mechanism occurs making the diagonal $\U1$ massive \cite{Ibanez:1998qp,Poppitz:1998dj,Aldazabal:2000dg}. 
\item \textbf{\SU2 stack:} There are two D7-branes wrapping the invariant irreducible divisor $U\equiv \{s_3=0\}$. The two branes are image to each other and they support an $\Sp1\cong \SU2$ massless gauge boson.  The diagonal $\U1$ vector multiplet 
is projected out of the spectrum by the orientifold, even though there is still the possibility to have a gauge flux associated with this $\U1$ that is the pull-back of a CY odd form to the invariant divisor $U$.
\item \textbf{\U1 stack:} The remaining locus in $\Delta_E$ is 
\begin{align}\label{eq:U10}
\Delta_E^{\rm rem}\equiv s_2 s_5^2-s_6 s_1 s_5 + s_9 s_1^2=0\:.
\end{align}
When we intersect this locus with the CY equation \eqref{SMCYeq} it factorizes into two divisors~\cite{Krause:2012yh}. These can be described algebraically as non-complete intersections by the following equations
\begin{eqnarray}
\label{Veqn}  V &\equiv &  \left\{s_1 s_9 + s_5\left( \xi-\frac{s_6}{2}\right)=0\,,\,\,\,  s_5 s_2- s_1 \left(\xi + \frac{s_6}{2}\right) =0\,, \,\,\, {\rm Eq.}\,\, \eqref{SMCYeq} \right\}\,,\\ \vspace*{1mm}
\label{Vteqn} \tilde V &\equiv &  \left\{s_1 s_9 - s_5 \left( \xi + \frac{s_6}{2}\right)=0\,,\,\,\,  s_5 s_2+ s_1\left( \xi - \frac{s_6}{2}  \right)=0\,, \,\,\, {\rm Eq.}\,\, \eqref{SMCYeq} \right\} \,.
\end{eqnarray}
Again these loci are in different homology classes and hence the associated $\U1$ is massive. We will see that however a linear combination of the two massive $\U1$s (the second being the one supported on the $\U3$ locus) is in fact massless \cite{Krause:2012yh}.
\end{itemize}
\begin{figure}[h!]
  \centering
  \setlength{\unitlength}{0.1\textwidth}
  \begin{picture}(4,3.5)
    \put(0,0){\includegraphics[scale=.5]{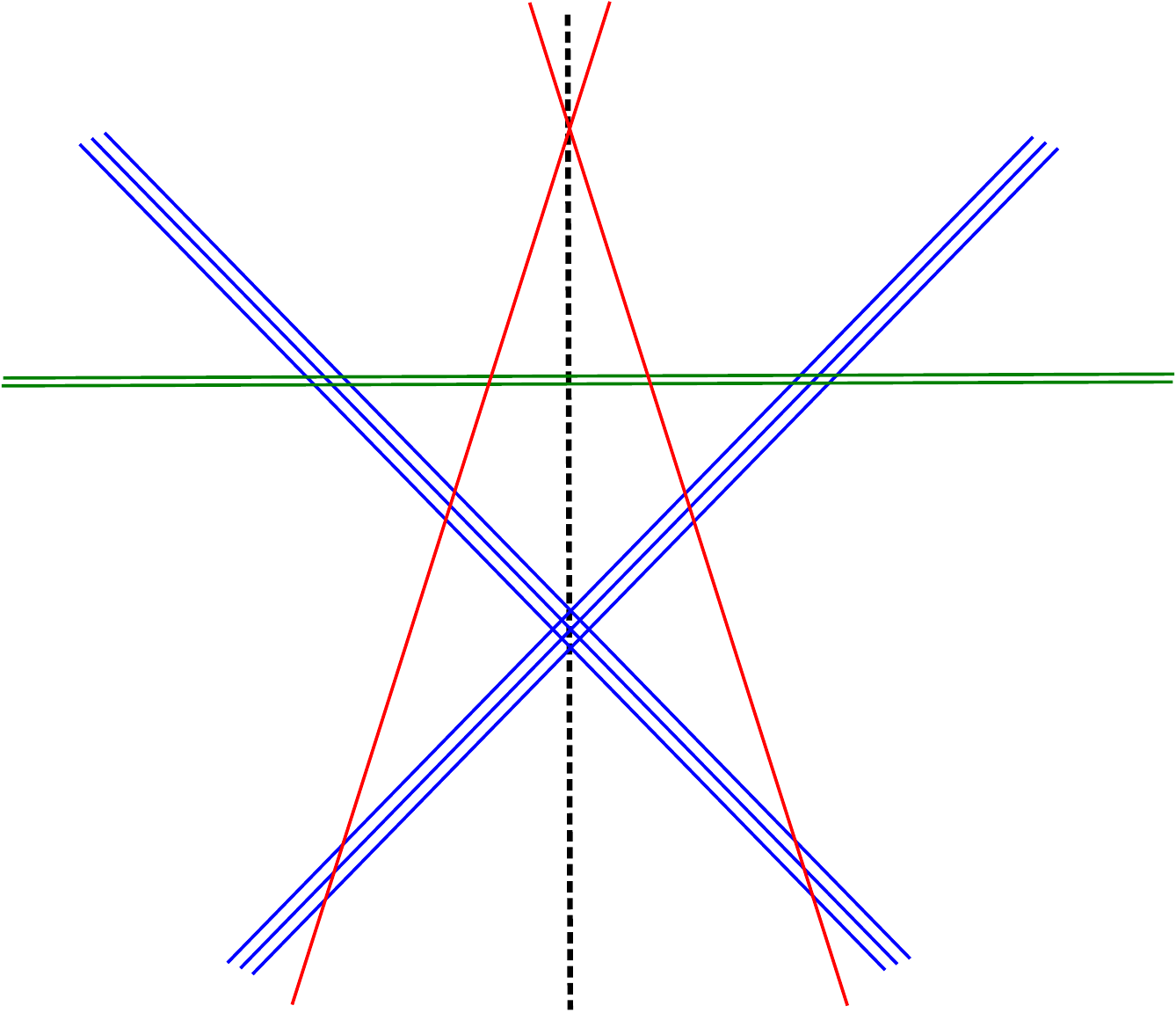}}
    \put(1.7,3.5){\small{\color{red} $V$}}
    \put(2.07,3.5){\small{\color{red} $\tilde V$}}
    \put(0.25,3.05){\small{\color{blue} $W$}}
    \put(3.52,3.05){\small{\color{blue} $\tilde W$}}
   \put(-0.15,2.1){\small{\color{green} $U$}}
   \put(2,0.1){\small $O7$}
  \end{picture}
\caption{\label{fig:weak} Schematics of the brane stacks in the weakly coupled type IIB limit.}
\end{figure}
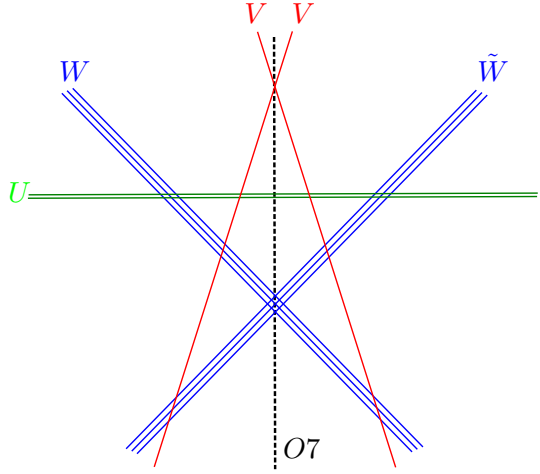

At this point it is convenient to remark some homological relations among the four-cycles we have just specified. First of all, it is possible to relate divisor classes of the base $B_3$ to divisor classes of the double cover the Calabi-Yau threefold $X_3$. The pullback to $X_3$ of the anticanonical divisor of $B_3$ coincides with the class of the orientifold plane 
\beq
\pi^*(\bar{K}_B)=[\xi]=D_{O7}\,,
\label{eq:Dict0}
\eeq
where $\pi:\,X_3\rightarrow B_3$ is the two-to-one projection.
Similarly, one can show that 
\beq
\pi^*([s_9])=W+\tilde W\,\quad {\rm and }\quad \pi^*([s_3])=U\,. 
\label{eq:Dict}
\eeq

As stated above, we will focus only on models with non-singular $X_3$ at weak coupling. Hence we require the absence of the conifold singularity\footnote{In $\SU5$ F-theory models, the absence of this point is related to the absence/smallness of the top Yukawa $\mathbf{10} \cdot\mathbf{10}\cdot \mathbf{5}$ \cite{Donagi:2009ra,Collinucci:2016hgh}. For the $\SU3 \times \SU2 \times \U1$ model the conifold singularity happens away from any of the Yukawa points $\{s_3=s_5=s_9=0\}$ and $\{s_3=s_9=s_1 s_6-s_2 s_5=0\}$, corresponding to the top and bottom quark Yukawas, respectively.} at
\begin{align}
 s_9=s_6=s_2=0\,.
\end{align}
In other words, we have the vanishing of the following triple intersection \cite{Krause:2012yh}
\begin{equation}
 \int_{X_3}D_{O7}(2D_{O7}-W_+)W_+=\int_{X_3}D_{O7}(2D_{O7}-(W+\tilde{W}))(W+\tilde{W})=0\,,
  \label{eq:that}
\end{equation}
where we have defined $W_\pm\equiv W\pm \tilde{W}$ as orientifold even and odd combinations. One can additionally prove the following relations \cite{Krause:2012yh}:
\begin{align}
 D_{O7}W=D_{O7}\tilde{W}=W\tilde{W}\quad\Rightarrow\qquad W_+^2-W^2_-=2D_{O7}W_+\,,\quad D_{O7}W_-=0\,.
 \label{eq:this}
\end{align}
Note that \eqref{eq:this} contains more information and automatically implies Eq. \eqref{eq:that}. 

One can also find the class of $V$ and $\tilde V$: Consider the locus $s_9\Delta_E^{\rm rem}=0$. It is in the class $2(4D_{O7}-U-(W+\tilde W))$. If we intersect it with the CY equation \eqref{SMCYeq}, this locus factorizes as
\begin{equation}
s_9\Delta_E^{\rm rem} = \left(s_1 s_9 + s_5\left( \xi-\frac{s_6}{2}\right)\right)\cdot\left(s_1 s_9 - s_5 \left( \xi + \frac{s_6}{2}\right)\right)\:.
\end{equation}
Both components are in the homology class $4D_{O7}-U-(W+\tilde W)$. By using the defining equations for $W,\tilde{W}$, i.e. \eqref{WWteqs}, and for $V,\tilde{V}$, i.e.  \eqref{Veqn} and \eqref{Vteqn}, one sees that the first component is in the class $V+W$, while the second is in the class $\tilde{V}+\tilde{W}$. Hence we conclude that 
\begin{align}
V=4D_{O7}-(2W+\tilde{W})-U\,,\quad \tilde{V}=\sigma^*V\,,
\end{align}
which of course is consistent with the D7 tadpole cancellation condition 
\begin{align}
 (V+\tilde{V})+2U+3(W+\tilde{W})=8 D_{O7}\,.
 \end{align}
  
Let us now comment on the \U1 symmetries. As we noticed above,  in the type IIB model we see two Abelian gauge symmetries: one is on the locus $V$ (and its image $\tilde{V}$), while the other is the diagonal \U1 of $\U3$ on the locus $W$ (and its image $\tilde{W}$). Since the homology classes of these divisors are different from the ones of their orientifold images, the corresponding \U1 gauge bosons are massive \cite{Ibanez:1998qp,Poppitz:1998dj,Aldazabal:2000dg}. 
On the other hand, in F-theory we have one massless \U1 gauge boson.  Actually this happens also in the type IIB setting; in fact, one linear combination 
of the massive \U1's is massless. The D7-brane worldvolume coupling that gives mass to the gauge bosons is the following:
\begin{equation}\label{CSgeomCoup}
 \int_{D7} C_6 \wedge \mathsf{F} \:.
\end{equation}
$\mathsf{F} $ is the four-dimensional gauge boson field strength. $C_6$ is the RR six-form potential (dual to $C_2$). It is odd under the orientifold projection, hence it gives zero modes when expanded along odd forms. The relevant zero modes for our discussion is a two-form potential that appears in the expansion of $C_6$ along odd four-forms. In the present example $h^{2,2}_-(X_3)=h^{1,1}_-(X_3)=1$, so that we have only one zero mode: $C_6= c_2 \wedge \omega_4^{(-)}$ with $\omega_4^{(-)}\in H^{2,2}_-(X_3)$. From \eqref{CSgeomCoup} we obtain the following terms in the four-dimensional action:
\begin{equation}
  \sum_{I=W,V} \int_{\mathbb{R}^{3,1}} n^I \mathsf{F}_I \wedge c_2\:.
\end{equation}
Here $\mathsf{F}_I$ is the four-dimensional field strengths of the $\U1$ gauge boson living on the $I$-th D7-brane wrapping the divisor $D_I$. $n^I$ is the coefficient of $D_I-\sigma^\ast D_I$ along the odd generator $W_-$ times the number of branes wrapping such divisor; in the present case, $n^W=3$ and $n^V=-1$.
Upon dualization, the four dimensional two-form $c_2$ becomes an axion scalar that is eaten by one gauge boson, giving it a mass through what has been called the geometric St\"uckelberg mechanism \cite{Ibanez:1998qp,Poppitz:1998dj,Aldazabal:2000dg,Grimm:2010ez,Grimm:2011tb}. Since there is only one axion field, only one linear combination of the two $\U1$ gauge bosons become massive, leaving the orthogonal combination massless. The massless (hypercharge) $\U1$ generator is then
\begin{align}\label{SMU1masslessGen}
 Q_Y=\frac{1}{6}(3Q_V+Q_W)\,,
\end{align}
with $Q_V$ being the charge under the \U1 associated to $V$ and $Q_W$ the charge under the diagonal $\U1\subset \U3$.

The matter fields live at the intersections of of the brane stacks. In the following we list the matter content and the curves where they live. The fields will be labeled by their transformation under the gauge groups: $({\bf R}_{\SU3},{\bf R}_{\SU2})_{(Q_V,Q_W)}^{Q_Y}$, where $({\bf R}_{\SU3},{\bf R}_{\SU2})$ is the representation under the $\SU2\times\SU3$ gauge group, again $(Q_V,Q_W)$ denote the charges under the $\U1_V$ and the $\U1_W\subset \U3$ and $Q_Y$ is their massless combination (even though it is determined by $Q_V$ and $Q_W$, we write it explicitly for later use).
\begin{itemize}
 \item $WU\,$: $(\three,\two)_{(0,1)}^{1/6}$. The associated locus is given by the vanishing of the elements of the following ideal
 \begin{align}\label{eq:32}
  \left\{s_3\,,\,s_9\,,\,\xi-\frac{s_6}{2}\right\}\:.
 \end{align}
\item $\tilde V W\,$: $(\overline{\three},\one)_{(-1,-1)}^{-2/3}$. The associated locus is given by the intersection of the $\SU3$ stack with the that of $\U1$ which is given in Eq. \eqref{Veqn}:
 \begin{align}
  \left\{s_9\,,\,s_5\,,\,\xi-\frac{s_6}{2}\right\}\:.
 \end{align}
 \item $ VW\,$: $(\overline{\three},\one)_{(1,-1)}^{1/3}$. The corresponding locus reads
 \begin{align}
  \left\{s_9\,,\,s_2s_5-s_1s_6\,,\,\xi-\frac{s_6}{2}\right\}\:.
 \end{align}
  \item $W \tilde W\,$: $(\overline{\three},\one)_{(0,2)}^{1/3}$.  The intersection is given by
 \begin{align}
 \{s_9,\, s_6\,, \xi\}\:.
 \end{align}
  \item $U V\,$: $(\one,\two)_{(-1,0)}^{-1/2}$. The intersection is given by
 \begin{align}
 \left\{s_3,\,s_1 s_9 + s_5\left( \xi-\frac{s_6}{2}\right),\,  s_5 s_2- s_1 \left(\xi + \frac{s_6}{2}\right), \,\, -\xi^2+\frac{s_6^2}{4}-s_2s_9\right\}\:.
 \end{align}
  \item $V \tilde V\,$: $(\one,\one)_{(2,0)}^{1}$. The intersection is given by
 \begin{align}\label{eq:11}
\left \{s_1,s_5 \,,\,\, -\xi^2+\frac{s_6^2}{4}-s_2s_9\right\} \:. 
 \end{align}
\end{itemize}
The loci of the complex conjugate representations, which appear at the image of the ones above, are obtained by replacing $\xi\mapsto-\xi$, in Eqs. \eqref{eq:32}-\eqref{eq:11}. 

Note that one has an extra triplet in comparison to the F-theoretic spectrum. This is due to the fact that in the limit $\epsilon\rightarrow 0$, the curve $(\overline{\three}, \one)_{1/3}$ splits into two components. Its $\epsilon$ dependent locus reads (see Table \ref{tab:poly11_matter})
\begin{align}
\{ s_9\,,\epsilon^2\,s_3 s_5^2 +\epsilon s_6(s_1 s_6-s_2 s_5)\}\, .
\end{align}
Taking the leading order in $\epsilon$ we find two irreducible components: The curve $\{s_9,\,(s_1 s_6-s_2 s_5)\}$ corresponding to the triplet $(\overline{\three}, \one)_{(1,-1)}^{1/3}$, and the curve $\{s_9,\,s_6\}$ corresponding to $(\overline{\three}, \one)_{(0,2)}^{1/3}$.

\

Let us comment on the splitting of the curves at weak coupling. Notice first that the F-theory curves really splits into two components when $\epsilon$ vanishes, i.e. at zero string coupling. At this value the mass of the \U1 massive gauge boson becomes zero as well \cite{Grimm:2011tb,Braun:2014nva}. For small $\epsilon$ F-theory is already saying that something is happening:
two curves supporting fields that had different global charges join into one curve. However, in F-theory fields living on the same curve should have the same charges under all the global symmetries. Hence one concludes that the continue (massive) $\U1$ symmetry that is visible in perturbative type IIB string theory must be broken, by some non-perturbative effect, to a discrete subgroup $\Gamma$ for which the massive \U1 charges gets identified ($\Gamma$ may also be trivial). %
We will investigate this  issue in a simple model in Section~\ref{sec:QdP7}. 

One may wonder if the extra symmetry at $\epsilon\rightarrow 0$ would prevent some couplings in type IIB that are allowed in F-theory. This actually does not happen: all triple couplings present in F-theory are also allowed in perturbative type IIB. Let us see how it works for some of them: The coupling $(\bar\three,\one)^{-2/3}(\bar\three,\one)^{1/3}(\bar\three,\one)^{1/3}$ is localized in F-theory at the locus
\begin{equation}
\{\, s_5=0,\,\,s_6=0,\,\, s_9=0 \,\}\:,
\end{equation}
where there is a gauge enhancement to $\SO8$. In type IIB this corresponds to the coupling $(\bar\three,\one)^{-2/3}_{(-1,-1)}(\bar\three,\one)^{1/3}_{(1,-1)}(\bar\three,\one)^{1/3}_{(0,2)}$ where there is the intersection of the $\U1$ stack with the $\U3$ stack and the orientifold plane (that actually gives again the enhancement to $\SO8$).
Another important coupling is the down Yukawa coupling $(\three,\two)^{1/6}(\three,\one)^{1/3}(\one,\two)^{-1/2}$. In F-theory it occurs at the locus
\begin{equation}\label{DownYukPts}
\{\, s_3=0,\,\,s_9=0,\,\, s_2s_5-s_1s_6=0 \,\}\:.
\end{equation}
In type IIB we have two curves supporting the right-handed down quark:  $(\bar\three,\one)^{1/3}_{(1,-1)}$ and $(\bar\three,\one)^{1/3}_{(0,2)}$. We have then in principle two types of Yukawa couplings. However the coupling $(\three,\two)^{1/6}_{(0,1)}(\three,\one)^{1/3}_{(0,-2)}(\one,\two)^{-1/2}_{(-1,0)}$
 is forbidden by the (massive) symmetry (and in fact the corresponding curves do not intersect geometrically). On the other hand, the allowed one, i.e. $(\three,\two)^{1/6}_{(0,1)}(\three,\one)^{1/3}_{(1,1)}(\one,\two)^{-1/2}_{(-1,0)}$, is exactly at the same locus as \eqref{DownYukPts}.

Finally, notice that this structure allows a  mechanism that suppresses the down Yukawa coupling with respect to the top one in perturbative type IIB:
If by a proper choice of fluxes we have no chiral zero modes on the curve $(\bar\three,\one)^{1/3}_{(1,-1)}$,  then we could forbid the down Yukawa perturbatively. In F-theory this should correspond to have the field wave function (determined by the same flux choice) localized all away from the Yukawa points. The weak coupling limit will then split the F-theory curve, keeping all the zero modes of the curve $(\bar\three,\one)^{1/3}_{(0,2)}$. At small $\epsilon$ this hierarchy should still work.

\subsubsection{Fluxes and chiralities}

Having worked out the matter representation we proceed to the computation of the corresponding chiralities. For this purpose we first need to deduce the gauge fluxes allowed by the D5-tadpole cancellation condition: 
\beq\label{D5tadpoleCanc}
0=\sum_{I}n_I\left[D_I F_I+(\sigma^* D_I) (\sigma^* F_I)\right]\,,
\eeq
where the index $I$ runs over all brane stacks, $D_I$ is the divisor where the $I$-th brane stack sits and $n_I$ the number of branes wrapping $D_I$. 
$F_I$ are the gauge flux on the brane wrapping $D_I$. In principle it will be an element of $H^2(D_I)$ (subject to a Freed-Witten quantization condition \cite{Minasian:1997mm,Freed:1999vc}). Since we are interested in the chiral spectrum, we will consider only the subset of two-forms on $D_I$ that are the pull-back of two-forms on the CY three-fold $X_3$, i.e. $\iota^*_{D_I}\omega$ with $\omega\in H^2(X_3)$. We will then omit the $\iota^*_{D_I}$ symbol.

For our concrete example, the condition \eqref{D5tadpoleCanc} reads 
\begin{align}
\begin{split}
 0&=3(W_+F_-^W +W_-F_+^W)+(V_+F_-^V+V_-F_+^V)+2 U F_-^U\\ &=W_-(3F_+^W+F_+^V)+3W_+(F_-^W-F_-^V)+2U(F_-^U-F_-^V)\,,
 \end{split}
\end{align}
after writing both fluxes and divisors in terms of orientifold odd and even components. Note that, since the divisor $U$ is orientifold-even, there is no even gauge flux on it that leaves the $SU(2)$ gauge group unbroken. We can distinguish between purely even, purely odd and mixed fluxes satisfying the D5-tadpole cancellation condition:
\begin{itemize}
\item Allowed even fluxes are (we write only the non-zero fluxes)
\begin{align}
\label{eq:Flux1}
 (F_+^W,F_+^V)=(\frac{2\lambda}{3}D_{O7},0)\,,\quad  (F_+^W,F_+^V)=(\frac{1}{6}F,-\frac{1}{2}F)\,,
\end{align}
where $F$ is an even two-form and $\lambda$ a generic rational coefficient (that should satisfy the proper quantization condition).
\item In the orientifold odd sector we can see that the general solution to the D5-tadpole implies $F_-^W=F_-^V=F_-^U$. As the flux is uniformly tuned over all divisors it can be reabsorbed into the $B$ field. This type of flux does not contribute to the chiral index of any matter representation \cite{Blumenhagen:2008zz}; hence, we do not consider these odd type fluxes any further.  
\item Finally, there are in principle two combinations of even and odd fluxes, i.e.
\begin{align}
\label{eq:Flux2}
 (F_+^W,F_-^W)=\alpha(-\frac{1}{3}W_+,\frac{1}{3}W_-)\,,\quad  (F_+^W,F_-^U)=\beta(-\frac{1}{3}U,-\frac{1}{2}W_-)\,,
\end{align}
with $\alpha$ and $\beta$ generic coefficients. However, one
can show that the $\alpha$- and $\lambda$-fluxes living on the $\U3$ stack are not independent: The first one is the two-form $F^W_\alpha=-\frac{2\alpha}{3}\iota^\ast \tilde{W}$, while the second is $F^W_\lambda=\frac{2\lambda}{3}\iota^\ast D_{O7}$ (where we made the pull-back symbol explicit). 
From the identity \eqref{eq:this}, one can see that $\iota^\ast \tilde{W}=\iota^\ast D_{O7}$.
For this reason, to avoid redundancies, we will set $\alpha=0$. Moreover, notice that if the homology class of $U$ is proportional to the class of $D_{O7}$, then $\iota^\ast_U W_-=0$ (see \eqref{eq:this}) and the $\beta$-flux is also equivalent to the $\lambda$-flux.
\end{itemize}
A generic flux choice will be a combination of the inequivalent fluxes described above and it will then depend on the arbitrary data $F,\lambda,\beta$.

Having all the inequivalent allowed fluxes we proceed with the computation of the chiral indices. For a given matter representation at the intersection of brane stacks $a$ and $b$, its corresponding chiral index is given by\footnote{In the case of symmetric and antisymmetric representations living at the intersections of brane/image brane, the chirality equation picks the form:\beq
\chi(\mathbf{N}_a,\mathbf{N}_{a^\prime})=\int_{X_3} D_a (\tilde{D}_a\pm D_{O7})F_a\,,
\eeq
in which the plus sign gives the chirality of the symmetric and the minus, that of the antisymmetric representation.}
\beq
\chi(\mathbf{N}_a,\overline{\mathbf{N}}_b)=\int_{X_3} D_a D_b(F_a-F_b)\,,
\eeq
where $D_a$ and $D_b$ are the divisors on which the brane stacks sit, and $F_a$ and $F_b$ their corresponding world-volume fluxes. For the case under consideration, the chiralities for the pure even and mixed type fluxes are summarized in Table \ref{tab:SMIIBchir}.

\begin{table}[h]
\begin{center}
\renewcommand{\arraystretch}{1.7}
{\scriptsize
\begin{tabular}{|c|c|c|c|}\hline
{\scriptsize Rep.} & {\scriptsize $(F_+^W,F_+^V)=(\frac{1}{6}F,-\frac{1}{2}F)$} & {\scriptsize $(F_+^W,F_+^V)=(\frac{2\lambda}{3}D_{O7},0)$} & {\scriptsize $(F_+^W,F_-^U)=\beta(-\frac{1}{3}U,\frac{1}{2}W_-)$}\\ \hline
$(\three,\two)_{(0,1)}$ & $\frac{1}{12} F U W_+ $ & $\frac{1}{3} \lambda D_{O7}  W_+ U $ & $\frac{1}{12}\beta U (6 D_{O7} - 2 U - 3 W_+) W_+$ \\ \hline
$(\overline{\three},\one)_{(-1,-1)}$& $\frac{1}{3} F(-3 D_{O7}+U+W_+)W_+ $ & $\frac{1}{3}\lambda D_{O7} W_+( U-D_{O7} )$ & $-\frac16 \beta U (5 D_{O7} - U - 2 W_+) W_+$\\ \hline
$(\overline{\three}, \one)_{(1,-1)}$ & $\frac{1}{6}F(5 D_{O7}-U-2 W_+)W_+ $ & $\frac{1}{3}\lambda D_{O7} W_+ ( D_{O7}-U ) $ & $-\frac16 \beta U (3 D_{O7} - U - W_+) W_+$ \\ \hline
$(\overline{\three}, \one)_{(0,2)}$ & $\frac{1}{6} F D_{O7} W_+$ & $\frac{2}{3}\lambda D_{O7}^2 W_+ $ & $\frac13 \beta D_{O7} U W_+$  \\ \hline
$(\one,\two)_{(-1,0)}$ & $\frac{1}{4}F(-8D_{O7}+2U+3W_+) U $ & $0$ & $\frac14 \beta  U W_+(2 D_{O7} - W_+)$ \\ \hline
$(\one,\one)_{(2,0)}$ &   $\begin{array}{c} \frac{1}{2} F (4D_{O7}-U+2W_+) \\ \times(3D_{O7}-U+W_+) \end{array} $  & $0$ & $0$\\ \hline

\end{tabular}
}
\caption{\label{tab:SMIIBchir} Chiralities in the perturbative limit for the allowed D5-tadpole canceling fluxes. In the central column we used the identity 	\eqref{eq:that}.}
\end{center}
\end{table}

\subsubsection*{Type IIB fluxes vs F-theory $G_4$-flux}

We can now check that the type IIB chiralities match with the F-theory result. Since the states $(\overline{\three}, \one)_{(1,-1)}$ and $(\overline{\three}, \one)_{(0,2)}$ exhibit the same hypercharge, their chiralities have to be added in order to match with the chirality of the $(\overline{\three}, \one)_{1/3}$ on the F-theory side. 

Let us also recall that the triple products of divisors in Tables \ref{tab:chiralities0} and  \ref{tab:SMIIBchir} mean triple intersections in the base $B_3$ and the threefold $X_3$, respectively. Using the relations \eqref{eq:Dict0} and \eqref{eq:Dict} together with the fact that in the double cover Calabi-Yau the intersections are twice as in the base, i.e.
\begin{align}
 \int_{X_3}\pi^*(D_a)\pi^*(D_b)\pi^*(D_c)=2\int_{B}D_aD_bD_c\,,
\end{align}
one can immediately show that the first columns of Tables \ref{tab:chiralities0} and  \ref{tab:SMIIBchir} match after setting 
$$F=\pi^*(\mathcal{F})\:.$$

The remaining F-theory $\Lambda$ flux must be a combination of $F$,  $\lambda$ and $\beta$ fluxes on the type IIB side. 
The proper combination matching all the F-theory  chiralieties of the $\Lambda$ flux is 
\begin{align}
F = -6\Lambda \,D_{O7} + 2\Lambda\, U + 3\Lambda\,W_+\,,\quad\qquad\lambda=0\,,\quad\qquad\beta=\Lambda\,. 
\end{align}
The previous equation holds also in cases when $D_{O7}$, $U$ and $W_+$ are not linearly independent.

Notice one important fact: while the massless \U1 fluxes match nicely, not all the type IIB massive gauge fluxes have a counterpart in F-theory. The $\lambda$-flux that is not represented in F-theory by an {\it harmonic vertical divisor }. As mentioned in the introduction, this answer a question raised in \cite{Krause:2012yh}, whether all D5-tadpole canceling fluxes (massless and massive) were at the end represented by harmonic vertical divisors. The answer is then negative: only part of them will behave in this way.

\subsubsection*{Matching D3-tadpole and FI terms}

If the matching prescription done above is correct, the fluxes should contribute the same D3-tadpole in type IIB and F-theory. Moreover they should generate the same FI-term for the \U1 symmetry. Let us check this.

We first start with the matching of the flux-dependent part of the D3-tadpole. On the F-theory side it is given by 
\begin{equation}\label{eq:gg}
 Q_{D3,{\rm F}}=\frac{1}{2}\int_{\hat{Y}_4}G_4\wedge G_4\,,
\end{equation}
while in type IIB it is given by
\begin{equation}\label{eq:ff}
 Q_{D3,{\rm IIB}}=-\frac{1}{4}\sum_i n_i\left(\int_{D_i} F_i^2+\int_{\tilde{D}_i} (\sigma^{*}F_i)^2\right)\,,
\end{equation}
that for the present case takes the form
\begin{equation}\label{eq:tadF11}
 Q_{D3,{\rm IIB}}=-\frac{1}{4}\int_{X_3}3 W_+ [(F_+^{W})^2 -(F_+^{V})^2] + 8 D_{O7} (F_+^{V})^2 + 2 U [-(F_+^{V})^2 + (F_-^{U})^2]\,.
\end{equation}

Let us consider the $\mathcal{F}$ vs $F$ fluxes. In F-theory the flux $G_4=\mathcal{F}\wedge \sigma_1$ has the following D3-charge:
\begin{equation}
 Q_{D3,{\rm F}}^{\mathcal{F}}=\frac{1}{2}\int_{B}\mathcal{F}^2\hat{\pi}^*(\sigma_1\cdot\sigma_1)=-\frac12 \int_{B}\left(2\bar{K}_B-\frac{2}{3}\cS_9-\frac{1}{2}\cS_3\right)\mathcal{F}^2\,,
\end{equation}
where we have used the Neron-Tate height pairing $b_{11}=-\hat{\pi}^*(\sigma_1\cdot\sigma_1)=\bar{K}_B+\frac{1}{3}\cS_9-\frac{1}{2}\cS_3$
 as computed in \cite{Klevers:2014bqa}, with $\cS_3=\bar{K}_B+\cS_7-\cS_9$ and with $\hat{\pi}$ the projection map from the elliptic fibration to the base $B_3$. On the type IIB side we have the flux $(F_+^W,F_+^V)=(\frac{1}{6}F,-\frac{1}{2}F)$. Substituting it into the expression \eqref{eq:tadF11} we obtain
\begin{align}
 Q_{D3,{\rm IIB}}^F&=
 -\frac{1}{16}\int_{X_3}\left(\frac{1}{3}W_++V_+\right) F^2 = -\frac{1}{4}\int_{X_3}\left(2 D_{O7}-\frac{2}{3}W_+-\frac{1}{2}U\right) F^2\,,\\
 &=-\frac{1}{4}\int_{X_3}\left(2 \,\pi^\ast (\bar{K}_B)-\frac{2}{3}\pi^\ast (\mathcal{S}_9)-\frac{1}{2}\pi^\ast (\mathcal{S}_3)\right) \pi^\ast(\mathcal{F})^2\,,
\end{align}
which matches with the F-theory result (we used the fact that $D_{O7}=\pi^\ast \bar{K}_B$, $W_+=\pi^\ast\mathcal{S}_9$, $U=\pi^\ast\mathcal{S}_3$ and $F=\pi^\ast(\mathcal{F})$).

Next let us consider the contribution of the $\Lambda$-flux in F-theory and its counterpart in type IIB. On the F-theory side the contribution to the D3-charge is
\begin{equation}
 Q_{D3,{\rm F}}^{\Lambda}=-\frac{\Lambda^2}{2}\int_{B}\left(6 \bar{K}_B - 2 \cS_3 - 3 \cS_9\right) \left(4 \bar{K}_B- \cS_3 - 2 \cS_9\right) \left(3 \bar{K}_B - \cS_3 - \cS_9\right)\,.
\end{equation}
On the type IIB side this flux is to be matched in part by an $F$-flux, with $F=\Lambda(-6D_{O7}+3W_++2U)$, together with the $\beta$-flux with $\beta=\Lambda$. Plugging $F_+^W=\Lambda(-D_{O7}+\frac12 W_+)$,  $F_+^V=\Lambda(D_{O7}-\frac32 W_+-U)$ and $F_-^U=\Lambda W_{-}/2$ in Eq. \eqref{eq:tadF11}, one obtains 
\begin{align}
 Q_{D3,{\rm IIB}}^\Lambda&=-\frac{\Lambda^2}{4}\int_{X_3}\left(6 D_{O7}- 2 U - 3 W_+\right) \left(4 D_{O7}- U- 2 W_+\right) \left(3 D_{O7}- U - W_+\right)\,. 
\end{align}
that matches the F-theory result (after substituting $D_{O7}=\pi^\ast \bar{K}_B$, $W_+=\pi^\ast\mathcal{S}_9$, $U=\pi^\ast\mathcal{S}_3$ and $F=\pi^\ast(\mathcal{F})$).

A generic flux in F-theory is $G_4=G_4^{\mathcal{F}}+G_4^\Lambda$. Its contribution to the D3-charge also includes a mixed term of the form $\int G_4^{\cal F}\wedge G_4^\Lambda$.
On the other hand also a generic choice for the type IIB fluxes will have some mixed contribution. By following the same type of computation done above, one can show that the mixed term between $\mathcal{F}$- and $\Lambda$-fluxes also matches its type IIB counterpart.  

\

Finally one has to match the FI terms. In F-theory, the FI term for the massless hypercharge $\U1$ is given by: 
\begin{align}
{\rm FI}_{\rm F}^Y\sim\, &\frac{1}{2\mathcal{V}_B}\int_{\hat{Y}_4}J\wedge \sigma_1 \wedge G_4\nonumber\\
\sim\, & \frac{1}{2\mathcal{V}_B}\int_{B}J\left[-\mathcal{F}\left(2\bar{K}_B-\frac{2}{3}\cS_9-\frac{1}{2}\cS_3\right)+\Lambda (4\bar{K}_B - \cS_3 - 2 \cS_9) (3\bar{K}_B - \cS_3 - \cS_9)\right]
\end{align}
From the type IIB perspective, the FI term is obtained as: 
\begin{align}
{\rm FI}_{\rm IIB}^Y\sim\, &\frac{1}{6\mathcal{V}_X}\int_{X_3} (3V_+{\rm tr}(F_+^V)-W_+{\rm tr}(F_+^W))\nonumber\\
\sim\, & \frac{1}{2\mathcal{V}_X}\int_{X_3}J\left[-F\left(2D_{O7}-\frac{2}{3}W_+-\frac{1}{2}U\right)+\Lambda (4D_{O7} -U - 2 W_+) (3D_{O7} - U - W_+)\right]
\end{align}
which matches the F-theory one.

\subsubsection*{Matching geometrical quantities} 

To conclude this section, we show that also the geometric contribution to the D3-charge on the two sides match.
The number of  D3-branes needed to cancel the D3-tadpole is given in F-theory by 
\beq
N_{D3}=\frac{\chi(Y_4)}{24}-Q_{D3,{\rm F}}\,,
\eeq
while in type IIB we have
\beq
N_{D3}=\frac{\chi(D_{O7})}{6}+\frac{\chi_{D7}}{24}-Q_{D3,{\rm IIB}}\,.
\eeq 
In these formulae, $Q_{D3,{\rm F}}$ and $Q_{D3,{\rm IIB}}$ are the flux contributions given in Eqs. \eqref{eq:gg} and \eqref{eq:ff}. Above we have shown that these two quantities are equal to each other.
Therefore, the Euler number of the Calabi Yau fourfold must coincide with the quantity $4 \chi(D_{O7})+\chi_{D7}$, with 
\begin{align}
\chi_{D7}=\sum_I N_I (\chi(D_I)+\chi(\tilde D_I))\,,
\end{align}
and $\chi(D)$ the Euler characteristic of the divisor $D$, i.e.
\begin{align}
\chi(D)=\int_{D}c_2(D)=\int_{X_3}D(D^2+c_2(X_3))\,. 
\end{align}

We start with the F-theory computation. The Euler characteristic of the Calabi-Yau fourfold can be derived from the Chern class, which is computed by adjunction. In the case under consideration we have 
\beq
c(Y_4)=\frac{c(B_3)c(\mathbb{P}_{F_{11}})}{(1+[p_{F_{11}}])}
\eeq
where $[p_{F_{11}}]=\bar{K}_B+c_1(\mathbb{P}_{F_{11}})$ is the class of the hypersurface described by $p_{F_{11}}=0$. Working out the expansion we get $c_4(Y_4)$. The integration of this form on the Calabi-Yau fourfold reduces to cubic intersections on the base by means of the methods highlighted in Section~\ref{sec:FluxesF11}. We finally obtain
\begin{align}\label{FthgeomD3charge}
\begin{split}
\chi(Y_4)=&3 \int_B  (4 c_2(B) \bar{K}_B + 48 \bar{K}_B^3 - 32 \bar{K}_B^2 \cS_3 + 8 \bar{K}_B \cS_3^2 
\\&\phantom{3 \int_B  (}- 56 \bar{K}_B^2 \cS_9 + 
   25 \bar{K}_B \cS_3 \cS_9 - 3 \cS_3^2 \cS_9 + 22 \bar{K}_B \cS_9^2 - 5 \cS_3 \cS_9^2 - 2 \cS_9^3)\,,
\end{split}
\end{align}
in agreement with the result of \cite{Cvetic:2015txa}. On the type IIB side we have 
\begin{align}\label{IIBgeomD3charge}
\begin{split}
4 \chi(D_{O7})+Q_{D7}=\,& 4 \chi(D_{O7})+3(\chi(W)+\chi(\tilde W))+2\chi(U)+(\chi(V)+\chi(\tilde V))\,,\\
=\,& 
3\int_{X_3} (4 c_2(X_3) D_{O7} + 44 D_{O7}^3 - 32 D_{O7}^2 U + 8 D_{O7} U^2 - 16 D_{O7}^2 W_+
\\
&\phantom{3\int_{X_3} (} + 
   25 D_{O7} U W_+ - 3 U^2 W_+ + 2 D_{O7} W_+^2 - 5 U W_+^2 - 2 W_+^3)\:.
\end{split}
\end{align}
We see that the two contributions \eqref{FthgeomD3charge} and \eqref{IIBgeomD3charge} match after we substitute $c_2(X_3)= c_2(B)+\bar{K}_B^2$. In fact, one has (again by adjunction)
\beq
c(X_3)=\frac{c(B)(1+[\xi])}{1+2[\xi]}=1+(c_2(B)+c_1(B)^2)+(-2 c_1(B)^3 - c_1(B) c_2(B) + c_3(B))\,.
\eeq

\subsection{A concrete example: The base as an orbifold of $Q^{(dP_7)^2}$}
\label{sec:QdP7}

We now consider an explicit example where the generic features described above become concrete. On the type IIB side we take the Calabi-Yau threefold known as $Q^{(dP_7)^2}$ \cite{Blumenhagen:2008zz,Collinucci:2009uh,Blumenhagen:2009up}. It is a hypersurface in the toric ambient space defined in Table~\ref{tab:scalings3} (the last column shows the multidegree of the defining equation) 
\begin{table}[h!]
\begin{center}
\renewcommand{\arraystretch}{1.2}
\begin{tabular}{|ccccccc|c|}
\hline
$x_1$ & $x_2$ & $x_3$ & $x_4$ & $x_5$ & $x_6$ & $x_7$ & ${\rm Deg.\,\, eq}_{X_3}$\\
\hline
1 & 1 & 1 & 1 & 1 & 0 & 0 & 5\\
0 & 0 & 0 & 0 & 1 & 1 & 0 & 2\\
0 & 0 & 0 & 1 & 0 & 0 & 1 & 2\\
\hline
\end{tabular}
\caption{\label{tab:scalings3} Scalings of the coordinates for the $Q^{dP_7}$ space together with the multidegree of the equation cutting the double cover Calabi-Yau threefold $X_3$.}
\end{center}
\end{table}
\noindent
and with the Stanley-Reisner ideal given by
$
\{x_1 x_2 x_3,\,x_5 x_6,\,x_4 x_7 \}
$.

A basis for $H^{1,1}(X_3)$ is given by the divisor classes $D_1=\{x_1=0\}\cap X_3$, $D_6=\{x_6=0\}\cap X_3$ and $D_7=\{x_7=0\}\cap X_3$. Their intersection polynomial in the Calabi-Yau reads
\beq
I=2(D_7^3+D_6^3)+2D_1^2(D_7+D_6)-D_7^2(2D_1+D_6)-D_6^2(2 D_1+D_7)+D_1 D_6 D_7\,.
\eeq

The manifold has been constructed 	such that it contains two Del Pezzo surfaces which get interchanged under the orientifold involution
\begin{equation}\label{ExampleOrInvol}
  \sigma\,:\,\qquad  x_4\leftrightarrow x_5\qquad
   x_6\leftrightarrow x_7\:.
\end{equation}
The D7-brane setup will include a $\U3$ stack wrapping the Del Pezzo surface at $x_6=0$ (and its orientifold image at $x_7=0$) and one $\SU2$ stack wrapping an invariant divisor.
To make the quotient we construct the two to one map
\begin{align}
 \begin{split}
  (x_4,x_5,x_6,x_7)
  \, \mapsto\, (s_2,s_6,s_9)= (2x_4 x_5,2(x_5 x_7+x_4 x_6),2x_6 x_7)\:,
 \end{split}
\end{align}
where we have called $s_9$ the last coordinates to be consistent with the generic case, where the $\U3$ stack was at $s_9=0$. The odd combination is given by $\xi=x_5 x_7-x_4 x_6$, i.e. the involution acts as $\xi\mapsto-\xi$. Notice that the new coordinates must satisfy the equation
$$
\xi^2 = \frac{s_6^2}{4}-s_2s_9 \:.
$$
In this representation, the type IIB CY threefold is given as two equations in a five-dimensional toric variety. One of these equations is exactly what we read in \eqref{SMCYeq}.

The quotient $B_3$ is a hypersurface in the four-dimensional toric variety in Table \ref{tab:scalings4}, with the SR ideal $\{x_1 x_2 x_3,\, s_2 s_9 s_6\}$ (notice that the conifold point $s_2= s_9=s_6=0$ is automatically forbidden by this SR ideal). 

\begin{table}[H]
\begin{center}
\renewcommand{\arraystretch}{1.2}
\begin{tabular}{|cccccc|c|}
\hline
$ x_1$ & $x_2$ & $x_3$ & $s_2$ & $s_6$ & $s_9$ & ${\rm Deg.\,\, eq}_{B_3}$\\
\hline
1 & 1 & 1 & 2 & 1 & 0 & 5\\
0 & 0 & 0 & 1 & 1 & 1 & 2\\
\hline
\end{tabular}
\caption{\label{tab:scalings4} Degrees of the coordinates and of the equation defining the base~$B_3$.}
\end{center}
\end{table}

In the base manifold $B_3$, both divisors $D_6$ and $D_7$ are projected down to the divisor $D_{67}$ whose pull-back is $\pi^\ast(D_{67})=D_6+D_7$. In terms of these, we obtain the following classes on $B_3$
\beq
[s_2]=2D_1+D_{67}\,\quad [s_6]=D_1+D_{67}\,,\quad [s_9]=\cS_9=D_{67}\,,
\eeq
where, by abuse of notation, we  called $D_1$ also the divisor $\{x_1=0\}$ on the base.
The intersection numbers on the base $B_3$ are given by:
\begin{align}
D_1^3=0\,,\quad D_1^2 D_{67}=2\,,\quad D_1 D_{67}^2=D_{67}^3=-1\,,
\end{align}
such that $[s_2]\cdot[s_6]\cdot[s_9]=0$ ensuring the absence of the conifold singularity in $X_3$. 

In order to fully specify the fibration we have to give the class of $\cS_7=n_{1}D_1+n_{2}D_{67}$. On the type IIB side, this is equivalent to fixing the class of the invariant divisor $U$ to be $U=(1+n_1)D_1+n_2 D_{67}$. Effectiveness of the fibration restricts the values for $n_{1}$ and $n_{2}$ to lie in the following range 
\begin{align}
-1\leq n_{1}\leq 2\,,\quad 0\leq n_{2}\leq 2\,,
\end{align}
in addition to that, one must ensure that the relevant base sections $s_i$ do not factorize (in their most generic form allowed by their degrees). In our model (see Table~\ref{tab:scalings4}), 
the sections $s_i$ involved in the fiber equation \eqref{eq:pF11}, whose associated class is $[s_i]=n_{i1} D_1+n_{i2} D_{67}$, do not factorize if
$n_{i1}\geq n_{i2}$ (except 
 for the case $n_{i1}=0$ and $n_{i2}=1$). This leads to a much stronger relation on~$\cS_7$: 
\begin{align}
 n_{2}-1\leq n_1\leq n_{2}\,. 
\end{align}

The four-form flux takes the form \eqref{SMgenericG4}, that in the present example becomes
\begin{eqnarray}
G_4&=&a_1 \sigma_1 D_1+a_2 \sigma_1 D_{67}\\
&+&\Lambda \left(D_{67}^2 (n_{2} - 4 n_{1} + 8 ) (n_{2}-2) + D_1^2 ( n_{1}^2 + n_{1} -9  + \tfrac{15}{2} n_{2}- 3 n_{1} n_{2}) +(D_1 + D_{67}) S_0 + 
   S_0^2\right).\nonumber
\end{eqnarray}

Having a concrete model at hand we can explore the possibility of having a complete family structure (i.e. all the SM representations have the same number of chiral modes).  For the particular matter configuration we are considering the complete family structure is equivalent to the requirement of an anomaly free hypercharge. Written in terms of the Chern-Simons coefficients, the anomaly freedom condition reads
\beq
\int_{Y_4} G_4 \wedge \sigma_1 \wedge D_1=\int_{Y_4} G_4 \wedge \sigma_1 \wedge D_{67}=0\,,
\eeq
which is satisfied whenever 
\begin{align}\label{eq:Onepar}
\begin{split}
a_2=&-a_1\frac{192 - 80 n_1 + 16 n_1^2 - 168 n_2 + 48 n_1 n_2 - 
      12 n_1^2 n_2 + 42 n_2^2 + 6 n_1 n_2^2 - 9 n_2^3}{-252 + 
    232 n_1 - 80 n_1^2 + 12 n_1^3 + 66 n_2 - 18 n_1 n_2 - 
    6 n_1^2 n_2 - 18 n_2^2 + 9 n_1 n_2^2}, \\
\Lambda=&-a_1\frac{124 - 56 n_1 + 12 n_1^2 - 30 n_2 - 6 n_1 n_2 + 
      9 n_2^2}{
   2 (-252 + 232 n_1 - 80 n_1^2 + 12 n_1^3 + 66 n_2 - 18 n_1 n_2 - 
      6 n_1^2 n_2 - 18 n_2^2 + 9 n_1 n_2^2} \:.
\end{split}
\end{align}
Eqs. \eqref{eq:Onepar} have one pole at $(n_1,n_2)=(2,2)$, simply implying that at this stratum the fluxes do not allow for a full family structure and therefore the hypercharge gauge boson always gets a mass by fluxed St\"uckelberg mechanism. One can further work out the number of families as a function of the parameter $a_1$ to show that it is zero also for $(n_1,n_2)=(-1,0)$.

Given that the duality has been shown to work in general, we can approach the generic case in type IIB, where we have some split matter curves and one flux more, and from there obtain the F-theory limiting case in which the $\lambda$-flux is turned off. 

First let us consider the a very special case: Note that when $n_2=n_1+1$, the $\SU2$ divisor $U$ and the orientifold plane are proportional: $U=(1+n_1)D_{O7}$. As we mentioned above, this makes the $\beta$- and the $\lambda$-fluxes equivalent. We can therefore make $\lambda=0$ and work out the conditions for a complete family structure in terms of the $\beta$- and $F$-fluxes, with $F=\gamma_1 D_1+\gamma_2 (D_6+D_7)$. We obtain 
\beq\label{eq:gamma2}
\gamma_2 |_{n_2=n_1+1}=\frac14 (-11 + 5 n_1) \gamma_1\,,\quad \beta  |_{n_2=n_1+1}=-\frac{ (103  - 74 n_1 + 15 n_1^2)\gamma_1}{4 (1 + n_1)}\,.
\eeq
Note that as we have conveniently set $\lambda=0$ to remove redundancies, at the strata $n_2=n_1+1$, the type IIB models already match the F-theory ones. 

Away from the strata $n_2=n_1+1$, the $\beta$- and $\lambda$-fluxes are inequivalent. Demanding complete families implies relations for $\beta$ and $\lambda$ in terms of $\gamma_1$ and $\gamma_2$
\beq\label{eq:lambdaf}
\lambda=-\frac{(-12 - 8 n_1 + 4 n_1^2 - 2 n_2 - 2 n_1 n_2 + 3 n_2^2)\gamma_1+(8 + 8 n_1 - 12 n_2)\gamma_2}{3 (1 + n_1 - n_2)}\,,
\eeq
\beq
\beta=-\frac{(-14 + 2 n_1 + 3 n_2)\gamma_1+(-1 + 3 n_1 - 3 n_2)\gamma_2}{1 + n_1 - n_2}\,.
\eeq
We can use these parameters to compute the number of families, for the cases away from $n_2=n_1+1$,  while for $n_2=1+n_1$ we will take \eqref{eq:gamma2} with $\lambda=0$. Additionally, recall that in type IIB the matter curve $(\overline{\three},\one)_{1/3}$ splits, therefore we can compute the difference among the chiralities as well. Both the number of families and the chirality splitting for the $(\overline{\three},\one)_{1/3}$ curve are shown in Table \ref{tab:chirsIIB}. 
\begin{table}[H]
\centering
\begin{minipage}{.40\textwidth}
\centering
{\small
\renewcommand{\arraystretch}{1.1}
\begin{tabular}{|c|c@{\hspace{0.7\tabcolsep}}c@{\hspace{0.7\tabcolsep}}c@{\hspace{0.7\tabcolsep}}c|}
\hline
$n_2\backslash n_1$ & -1 & 0 & 1 & 2 \\
\hline
 0 & $0$ & $16 \gamma_1 - 2 \gamma_2$  &  & \\
1 &  &$-\tfrac{15}{2} \gamma_1$ & $5 \gamma_1$ & \\
2 & &  & $-6 \gamma_1$ & $0$ \\ \hline
\end{tabular}}
\\
\vspace{0.05\textwidth}
(a)
\end{minipage}%
\begin{minipage}{.60\textwidth}
\centering
{\small
\renewcommand{\arraystretch}{1.1}
\begin{tabular}{|c|c@{\hspace{0.7\tabcolsep}}c@{\hspace{0.7\tabcolsep}}c@{\hspace{0.7\tabcolsep}}c|}
\hline
$n_2\backslash n_1$ & -1 & 0 & 1 & 2 \\
\hline
0 & $0$	& $-50 \gamma_1+10 \gamma_2$ &  &  \\
1 &  &
$-\tfrac{45}{2} \gamma_1$  & $-27 \gamma_1+4 \gamma_2$	&  \\
2 &  & 
	 &	$-6 \gamma_1$ &	$-10 \gamma_1$ \\ \hline
\end{tabular}}
\\
\vspace{0.04\textwidth}
(b)
\end{minipage}
\caption{\label{tab:chirsIIB} (a) The net number of families as a function of the parameters $\gamma_1$ and $\gamma_2$ ($F=\gamma_1 D_1+\gamma_2 D_{67}$) (b) The difference among the chiralities for the split matter curves $\Delta \chi(\overline{\three},\one)_{1/3}=\chi(\overline{\three},\one)_{(1,-1)}^{1/3}-\chi(\overline{\three},\one)_{(0,2)}^{1/3}$.}
\end{table}
The type IIB models which have an F-theory version must have $\lambda=0$. The result matches the type IIB one at the strata $n_2=n_1+1$. Away from those strata,  we see from Eq. \eqref{eq:lambdaf},  that this implies a relation between the coefficients $\gamma_1$ and $\gamma_2$ such that the number of families depend on a single parameter as expected from Eq. \eqref{eq:Onepar}. Written in terms of $\gamma_1$, the chiralities as well as the splitting between the families is given in Table \ref{tab:chirsFth}
\begin{table}[h]
\centering
\begin{minipage}{.50\textwidth}
\centering
{\small
\renewcommand{\arraystretch}{1.1}
\begin{tabular}{|c|cccc|}
\hline
$n_2\backslash n_1$ & -1 & 0 & 1 & 2 \\
\hline
0 &0  & 	$13 \gamma_1$	& & \\
1&&	$-\tfrac{15}{2} \gamma_1$ & $5 \gamma_1$	& \\
2&& &	$-6 \gamma_1$	& $0$\\
 \hline
\end{tabular}}
\\
\vspace{0.05\textwidth}
(a)
\end{minipage}%
\begin{minipage}{.50\textwidth}
\centering
{\small
\renewcommand{\arraystretch}{1.1}
\begin{tabular}{|c|cccc|}
\hline
$n_2\backslash n_1$ & -1 & 0 & 1 & 2 \\
\hline
0 & 0 &	$-35 \gamma_1$ &&\\
1 && $-\tfrac{45}{2} \gamma_1$ &	$-10 \gamma_1$	& \\
2 & & &	$-6 \gamma_1$	& 0 \\ \hline
\end{tabular}}
\\
\vspace{0.05\textwidth}
(b)
\end{minipage}
\caption{\label{tab:chirsFth} After setting $\lambda=0$ we can get the number of families in the F-theory limit (a) The net number of families as a function of the parameter $\gamma_1$. (b) The difference among the chiralities for the split matter curves $\Delta \chi(\overline{\three},\one)_{1/3}=\chi(\overline{\three},\one)_{(1,-1)}^{1/3}-\chi(\overline{\three},\one)_{(0,2)}^{1/3}$. Recall that the chiral index for the field  $(\overline{\three},\one)_{1/3}$ is given as $ \chi(\overline{\three},\one)_{1/3}=-\chi(\overline{\three},\one)_{(1,-1)}^{1/3}-\chi(\overline{\three},\one)_{(0,2)}^{1/3}$. }
\end{table}

It was already stressed that in contrast to $\SU5$ F-theory GUTs, the requirement of a type IIB limit for this F-theoretic MSSM-like model does not compromise the structure of the Yukawa couplings. Therefore, both weak and strong coupling limits are suitable grounds for phenomenology. There are however, three main differences between the F-theory and the type IIB models:
\begin{itemize}
\item[1)] 	There is one more flux direction in type IIB. The so-called $\lambda$-flux in type IIB has to be set to zero in order to match with the harmonic vertical fluxes in $H^{(2,2)}_V(Y_4)\subset H^{(2,2)}(Y_4)$. The $\lambda$-flux, that is still chirality inducing, may be represented by a non-harmonica four-form in F-theory, as suggested by \cite{Grimm:2011tb}.
\item[2)] In F-theory we have five matter curves, one for each chiral field appearing in the MSSM. Since the down-type Higgs and lepton doublets are not distinguished in this model, we expect the Higgses to arise from a vector-like pair living at the $(\one,\two)_{-1/2}$ curve. In contrast to that, in the type IIB model we have six curves as a consequence of the $(\overline{\three},\one)_{1/3}$ curve splitting into $(\overline{\three},\one)^{1/3}_{(1,-1)}$ and $(\overline{\three},\one)^{1/3}_{(0,2)}$. Hence, 
one can distribute the chiralities among these two curves, such that in the end their chiralities add up to the net number of families. 
\item[3)] In type IIB, there are two $\U1$ symmetries, one of these is geometrically massless and coincides with that in F-theory. The other $\U1$ is St\"uckelberg massive and leaves behind a global $\U1$ remnant at the perturbative level. Under this global $\U1$ the up type  $(\three,\one)^{1/6}_{(0,1)}\cdot(\one,\two)^{1/2}_{(1,0)}\cdot(\overline{\three},\one)^{-2/3}_{(-1,-1)}$  and down-type $(\three,\one)^{1/6}_{(0,1)}\cdot(\one,\two)^{-1/2}_{(-1,0)}\cdot(\overline{\three},\one)^{1/3}_{(1,-1)}$ Yukawa couplings are allowed. However, the down-type Yukawa of the form $(\three,\one)^{1/6}_{(0,1)}\cdot(\one,\two)^{-1/2}_{(-1,0)}\cdot(\overline{\three},\one)^{1/3}_{(0,2)}$ is forbidden. Therefore, if all down-type quarks are in the representation $(\overline{\three},\one)^{1/3}_{(0,2)}$, so that the $(\overline{\three},\one)^{1/3}_{(1,-1)}$ curve is depleted of chiral states, the down-type Yukawa coupling must be suppressed in comparison with the up-type one. This type of hierarchy is more difficult to see in F-theory at the level of the codimension-three singularities. It is expected that the hierarchy is manifest once we compute the Yukawa couplings as wavefunction overlaps \cite{Font:2008id,Heckman:2008qa,Marsano:2009ym,Cecotti:2009zf,Marchesano:2009rz,Leontaris:2010zd,Aparicio:2011jx,Palti:2012aa,Font:2013ida,Marchesano:2015dfa,Font:2015slq,Carta:2015eoh} or from analyzing the fiber splittings from codimension two to three \cite{Martucci:2015dxa,Martucci:2015oaa}.
\end{itemize}

Looking back at Table \ref{tab:chirsFth}, we identify an F-theory model where the  hierarchy mentioned in~3)~occurs. Note that for the choice $(n_1,n_2)=(1,2)$ the number of families is $-6\gamma_1$. Therefore, if the flux quantization as well as the D3-tadpole allow it, for $\gamma_1=\pm1/2$ we have a three family model with a perturbative as well as an F-theory description. Note now that the $(\overline{\three},\one)_{1/3}$ matter comes all from the $\chi(\overline{\three},\one)_{(0,2)}^{1/3}$ curve in type IIB. Therefore, in this case the hierarchy of masses for up-type and down-type quarks is manifest. 

The next thing to consider when aiming at realistic models are the flux quantization as well as the D3-tadpole cancellation. 
For this particular example the Euler characteristic has been computed and it is reported in Table \ref{tab:Euler}.
\begin{table}[h]
\begin{center}
\renewcommand{\arraystretch}{1.2}
\begin{tabular}{|c|cccc|}
\hline
$n_2\backslash n_1$ & -1 & 0 & 1 & 2 \\
\hline
0 & $\frac{103}{4}$ & $\frac{75}{4}$ &  &  \\
1 & & $\frac{73}{4}$ & $\frac{33}{2}$ & \\
2 &  & & $\frac{31}{2}$ & $19$ \\ \hline
\end{tabular}
\caption{\label{tab:Euler} The quantity $\chi/24$ entering the D3 brane tadpole.}
\end{center}
\end{table}

\newpage
\section{A $\U1\times \U1$ F-theory model}
\label{sec:F5}
In this Section we consider the weak coupling limit of a $\U1\times \U1$ F-theory model which has been studied in \cite{Borchmann:2013jwa,Cvetic:2013nia,Cvetic:2013uta,Borchmann:2013hta,Lin:2014qga,Lin:2016vus}.
\subsection{F-theory description}
\subsubsection{Geometric setup}
\begin{figure}[H]
\centering
\begin{minipage}{.56\textwidth}
  \centering
  \includegraphics[scale=.4]{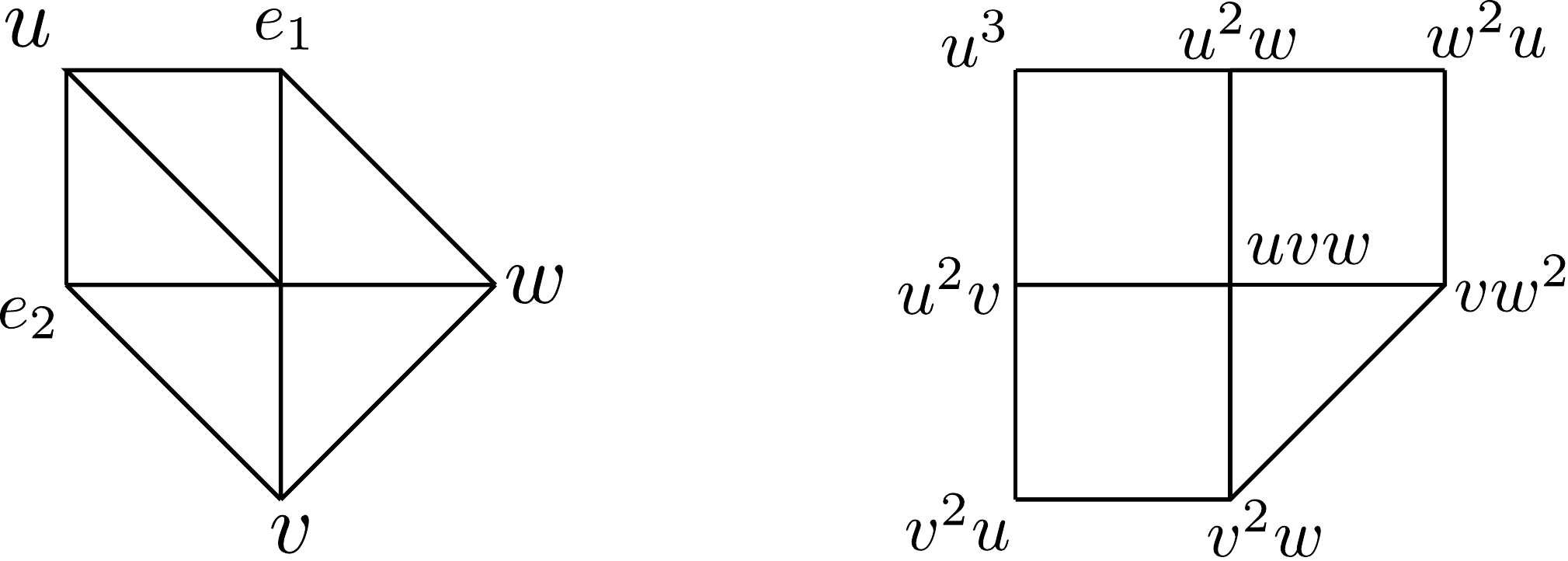}
\end{minipage}%
\begin{minipage}{.44\textwidth}
{\footnotesize
  \begin{tabular}{|c|c|}\hline
Section & Line Bundle\\ \hline
$u$ & $\mathcal{O}(H-E_1-E_2+\cS_9+K_B)$ \\ \hline
$v$ & $\mathcal{O}(H-E_2+\cS_9-\cS_7)$\\ \hline
$w$ & $\mathcal{O}(H-E_1)$\\ \hline
$e_1$ & $\mathcal{O}(E_1)$\\ \hline
$e_2$ & $\mathcal{O}(E_2)$\\ \hline
\end{tabular}}
\end{minipage}
\caption{\label{fig:poly5_toric} The polytope $F_{5}$ and its dual. The table on the right contains the divisor classes of the coordinates in $\mathbb{P}_{F_{5}}$.}
\end{figure}

The fiber is cut as a cubic hypersurface in the toric ambient space $\mathbb{P}_{F_{5}}$ corresponding to a $\mathbb{P}^2$ blown up at two points. The polytope $F_{5}$ as well as its dual are shown in Fig. \ref{fig:poly5_toric}.
The fiber equation is $p_{F_5}=0$, with $p_{F_5}$ given by the following expression:
\beq
\label{eq:pF5}
p_{F_5} = s_1 e^2_2 e^2_1 u^3 + s_2 e^2_2 e_1 u^2 v + s_3 e^2_2 u v^2 + s_5 e_2 e^2_1 u^2 w + s_6 e_2 e_1 u v w 
 + s_7 e_2 v^2 w + s_8 e^2_1 u w^2 + s_9 e_1 v w^2 \, .
\eeq
The sections $s_i$ have the same degrees as in Eq. \eqref{eq:sis}, to which we have to add $s_7$ and $s_8$:
\beq
\label{eq:sis2}
\text{
\renewcommand{\arraystretch}{1.2}
\begin{tabular}{|c|c|}
\hline
$s_7$ & $s_8$   \\ \hline
$\cS_7$ & $\bar{K}_B+\cS_9-\cS_7$  \\
\hline
\end{tabular}\:.
}
\eeq
The Weierstrass equation, written in the form Eq. \eqref{W5sing}, has the following expressions for the coefficients:\footnote{Of course other choices are possible. We will see in the following that this choice is the proper one to define a weak coupling limit that gives the same 7-brane setup in F-theory and in type IIB. } 
\begin{align}
\begin{split}
b_2&=\frac{s_6^2}{4}-s_5 s_7 \:, \\
b_4&=\frac{1}{12}\left(s_3 s_6^2 s_8 + 2 s_3 s_5 s_7 s_8 - 3 s_2 s_6 s_7 s_8 + 6 s_1 s_7^2 s_8 - 
 2 s_3^2 s_8^2 \right.\\
&\phantom{=\frac{1}{12}(}\left.- 3 s_3 s_5 s_6 s_9 + s_2 s_6^2 s_9+ 2 s_2 s_5 s_7 s_9 - 
 3 s_1 s_6 s_7 s_9 + 2 s_2 s_3 s_8 s_9 - 2 s_2^2 s_9^2 + 6 s_1 s_3 s_9^2\right)\:,
 \end{split}
 \end{align}
 \begin{align}
 \begin{split}
 b_6&=\frac{1}{108} \left(3 s_3^2 s_6^2 s_8^2 + 24 s_3^2 s_5 s_7 s_8^2 - 18 s_2 s_3 s_6 s_7 s_8^2 + 
   27 s_2^2 s_7^2 s_8^2 \right. \\
&\phantom{=\frac{1}{108}(} - 72 s_1 s_3 s_7^2 s_8^2 - 8 s_3^3 s_8^3  - 
   18 s_3^2 s_5 s_6 s_8 s_9 + 6 s_2 s_3 s_6^2 s_8 s_9 - 6 s_2 s_3 s_5 s_7 s_8 s_9 \\
&\phantom{=\frac{1}{108}(}  - 
   18 s_2^2 s_6 s_7 s_8 s_9 + 90 s_1 s_3 s_6 s_7 s_8 s_9   - 18 s_1 s_2 s_7^2 s_8 s_9 + 
   12 s_2 s_3^2 s_8^2 s_9 + 27 s_3^2 s_5^2 s_9^2  \\
&\phantom{=\frac{1}{108}(} - 18 s_2 s_3 s_5 s_6 s_9^2  + 
   3 s_2^2 s_6^2 s_9^2 + 24 s_2^2 s_5 s_7 s_9^2  - 54 s_1 s_3 s_5 s_7 s_9^2 - 
   18 s_1 s_2 s_6 s_7 s_9^2\\
&\phantom{=\frac{1}{108}(}  \left.  + 27 s_1^2 s_7^2 s_9^2 + 12 s_2^2 s_3 s_8 s_9^2 - 
   72 s_1 s_3^2 s_8 s_9^2 - 8 s_2^3 s_9^3 + 36 s_1 s_2 s_3 s_9^3\right)\:.
   \end{split}
\end{align}
The fiber exhibits three inequivalent rational points which are related to the three sections of the elliptic fibration. The first is the zero section at 
\begin{align}
 S_0\,:\quad [\x:y:z]=[1:1:0]\,,
 \end{align}
while  the others are 
\begin{align}
 S_1\,:\quad [\x:y:z]=\left[\frac13 (2 s_3 s_8 -s_2 s_9):\frac12 (s_3 s_6 s_8 - s_2 s_7 s_8 - s_3 s_5 s_9 + s_1 s_7 s_9):1\right]
 \label{eq:S1F5}
\end{align}
and
\begin{align}
\begin{split}
 S_2\,:\quad [\x:y:z]=&\,\left[ s_7 \left( s_7s_8^2-s_6s_8s_9+s_5s_9 \right) +\frac13 s_9^2\left( 2s_3s_8-s_2s_9 \right) \right.: \\ 
     & \,\,\,   \frac12\left( 2s_7s_8-s_6s_9  \right) \left( s_7\left( s_7s_8^2-s_6s_8s_9+s_5s_9 \right)  + s_3s_8s_9^2 \right) + 
     \left. \frac12 s_9^4(s_3s_5+s_1s_7) 
\,:\,s_9\,\right]\:.
\end{split}
 \label{eq:S2F5}
\end{align}
We notice that the last one is a rational section.
Having three inequivalent sections, the Mordell-Weil group of the fibrations is two-dimensional. 
The number of massless $\U1$ gauge bosons is then two. 
The corresponding divisors are 
\begin{align}
\label{eq:ShiodaF51}
\sigma_1&= (S_1 - S_0 - \bar{K}_B) \, , \\
\label{eq:ShiodaF52}
\sigma_2 &= (S_2 - S_0 -\bar{K}_B -\cS _9)\, .
\end{align}
Looking at the Weierstrass model, one can confirm that the fibration is not singular at codimension-one. Hence there are no non-Abelian gauge symmetries in this model. Instead, at codimension-two there are six loci along which the fiber degenerates to an $I_2$. Therefore we have six charged\footnote{In literature, these fields are called singlets, to make it clear that they are not charged under a non-Abelian gauge symmetry.} superfields distinguished by their charges under the two $\U1$ symmetries. Their corresponding charges and associated loci are summarized in Table \ref{tab:poly5_matter}.
\begin{table}[t]
\begin{center}
\footnotesize
\renewcommand{\arraystretch}{1.4}
\begin{tabular}{|c|c|}
\hline
Representation &Locus \\ \hline

$\one_{(1,-1)}$ &  $V(I_{(1)}):=\{s_3 = s_7 = 0 \}$ \\ \hline

$\one_{(1,0)}$ & $\begin{array}{c} V(I_{(2)}):=\{s_2 s_7^2 + s_3^2 s_9 - s_3 s_6 s_7=0 \\ s_5 s_3 s_7 - s_3^2 s_8 - s_7^2 s_1 = 0 \}\backslash V(I_{(1)}) \end{array}$ \\ \hline

$\one_{(-1,-2)}$ & $V(I_{(3)}):=\{s_8 = s_9 = 0 \}$ \\ \hline

$\one_{(-1,-1)}$ & $\begin{array}{c} V(I_{(4)}):=\{ s_2 s_8 s_9 - s_3 s_8^2 - s_9^2 s_1=0 \\ s_5 s_9^2 - s_6 s_8 s_9 + s_8^2 s_7 = 0 \} \backslash(V(I_{(3)}) \end{array}$ \\ \hline

$\one_{(0,2)}$ &  $V(I_{(5)}):=\{ s_9 = s_7 = 0 \}$ \\ \hline

$\one_{(0,1)}$ & $\begin{array}{c} V(I_{(6)}):=\{ s_1 s_9^4 s_7^2 + (s_3 s_9^2+s_7 \\ \times (-s_6s_9+s_8s_7)) (s_3s_8s_9^2+s_7 \\ \times (-s_6s_8s_9+s_8^2s_7+s_9^2s_5))= 0 \\ s_2 s_9^3 s_7^2 + s_3^2s_9^4 - s_3s_6s_9^3s_7 \\ -s_7^3(-s_6s_8s_9+s_8^2s_7+s_9^2s_5)= 0 \} \\ \backslash(V(I_{(1)})\cup V(I_{(3)})\cup V(I_{(5)})) \end{array}$ \\ \hline
\end{tabular}
\caption{\label{tab:poly5_matter}Charged singlets under $\U{1}^2$ with the expressions for their corresponding codimension-two loci.}
\end{center}
\end{table}
\subsubsection{Fluxes and chiral matter}
Following a similar method as in Section \ref{sec:FluxesF11} we obtain the independent flux directions in $H^{(2,2)}_V(X)$. In this case the fibral divisors are $S_0$, $S_1$ and $S_2$, while among the vertical divisors we have the special ones  $\{\bar{K}_B,\cS_7,\cS_9\}$, as in the previous case. The most general flux expression consistent with the vanishing of the Chern-Simons terms of the form $\Theta_{0\alpha}$ $\Theta_{\alpha\beta}$ is given by
\begin{align}\label{eq:G4F5}
 G_4=\mathcal{F}_1\wedge \sigma_1+\mathcal{F}_2\wedge \sigma_2+\Lambda \left(S_0^2 + [K_{B}^{-1}](-[K_{B}^{-1}] + S_2) + \cS_9 (-\cS_7 + \cS_9) \right)\,
\end{align}
with $\mathcal{F}_1$ and $\mathcal{F}_2$ being generic vertical divisors. The chiralities for the singlet fields under this flux are reported in Table \ref{tab:poly5_matterbis}. 
\begin{table}[htb]
\begin{center}
\footnotesize
\renewcommand{\arraystretch}{1.4}
\begin{tabular}{|c|@{}c@{}|@{}c@{}|}
\hline
Representation & $G_4=\mathcal{F}_1\wedge \sigma_1+\mathcal{F}_2\wedge \sigma_2$ &  $ \begin{array}{c}G_4=\Lambda(S_0^2 + \bar{K}_B (-\bar{K}_B + S_2) \\ \quad+\cS_9 (-\cS_7 + \cS_9))  \end{array}$ \\ \hline

$\one_{(1,-1)}$ & $(\mathcal{F}_1-\mathcal{F}_2)\cS_7  (\bar{K}_B+\cS_7-\cS_9)$ & $-\Lambda \bar{K}_B  \cS_7 ( \bar{K}_B  + \cS_7 - \cS_9)$  \\ \hline

$\one_{(1,0)}$ & $\begin{array}{c}\mathcal{F}_1( 6\bar{K}_B^2 + \bar{K}_B(4\cS_7-5\cS_9) \\ - 2\cS_7^2 + \cS_7 \cS_9 +\cS_9^2 )\end{array}$  & $-\Lambda \cS_7 \cS_9 (\bar{K}_B - \cS_7 + \cS_9)$  \\\hline

$\one_{(-1,-2)}$ & $-(\mathcal{F}_1+2\mathcal{F}_2)\cS_9 (\bar{K}_B-\cS_7 +\cS_9)$ & $\begin{array}{c}-\Lambda \cS_9(2  \bar{K}_B  +\cS_7 - 2 \cS_9)\\ \quad\times( \bar{K}_B - \cS_7 +\cS_9)\end{array}$ \\ \hline

$\one_{(-1,-1)}$ & $\begin{array}{c}-(\mathcal{F}_1+\mathcal{F}_2)( 6\bar{K}_B^2 + \bar{K}_B(-5\cS_7+4\cS_9) \\ + \cS_7^2 + \cS_7 \cS_9 -2\cS_9^2) \end{array}$ & $\begin{array}{c}\Lambda(-6 \bar{K}_B^3 + 2 (\cS_7 - \cS_9) \cS_9^2 + \bar{K}_B^2 (5 \cS_7 + 2 \cS_9)\\ - 
 \bar{K}_B (\cS_7^2 + 7 \cS_7 \cS_9 - 6 \cS_9^2))\end{array}$   \\\hline

$\one_{(0,2)}$ & $2 \mathcal{F}_2\cS_7 \cS_9$ &  $\Lambda \cS_7 (\bar{K}_B + \cS_7 - \cS_9) \cS_9$ \\ \hline

$\one_{(0,1)}$ & $\begin{array}{c}\mathcal{F}_2( 6\bar{K}_B^2 + \bar{K}_B(4\cS_7+4\cS_9) \\ -2\cS_7^2 -2\cS_9^2) \end{array}$ & $\begin{array}{c}2 \Lambda (3 \bar{K}_B ^2 - \bar{K}_B  \cS_7 + 2 \bar{K}_B  \cS_9 - \cS_9^2)\\ \times (\bar{K}_B  + \cS_7 - \cS_9)\end{array}$  \\ \hline
\end{tabular}
\caption{\label{tab:poly5_matterbis}Charged matter representations under U(1)$^2$ and corresponding codimension-two fibers of $X_{F_5}$.}
\end{center}
\end{table}

\subsection{The weak coupling limit}
In order to take the weak coupling limit we must first specify the $\epsilon$ scalings of the sections $s_i$. A choice that leads to the same setup in type IIB  is 
\beq
 s_8 \rightarrow \epsilon^1 s_8, \quad s_9 \rightarrow \epsilon^1 s_9, \quad s_i \rightarrow \epsilon^0 s_i \quad {\rm (}i\neq 8,9{\rm )}
\eeq
In the limit $\epsilon\rightarrow 0$, the D7-brane locus is $\Delta_E=0$, with
\begin{equation}\label{eq:deltaE2}
\Delta_E=-\frac14  s_7 \cdot \left[ (s_3s_5-s_1s_7)^2-(s_3s_6-s_2s_7)(s_2s_5-s_1s_6)
   \right] \cdot 
   \left[s_7 s_8^2 - s_6 s_8 s_9 + s_5 s_9^2\right]\:.
\end{equation}
Given the above expression for $b_2$, the  double cover Calabi-Yau threefold is given by
\beq\label{eq:XP5}
\xi^2=\frac{s_6^2}{4}-s_5 s_7\,.
\eeq
In order to deal only with smooth CY threefold, we restrict the base space $B_3$ to those spaces for which the conifold point $\xi=s_6=s_5=s_7=0$ is absent.

\subsubsection{D7-brane setup}
To understand how many irreducible D7-branes we have, we need to intersect the three factors in Eq. \eqref{eq:deltaE2} with the Calabi-Yau equation \eqref{eq:XP5}. As we will see shortly, each of the components is going to split in such a way that in the type IIB model we have three $\U1$ gauge symmetries:
\begin{itemize}
 \item  \textbf{$\mathbf{\U1_1}$ stack:} Consider first the locus $\{s_7=0\}$. One can see that in the Calabi-Yau it splits into the following components: 
 \begin{equation}\label{XXteqs}
X\,\equiv\,\,\,\, \left\{ s_7=0, \,\,\, \xi-\frac12 s_6 = 0 \right\} \qquad\mbox{and}\qquad
\tilde X \,\equiv\,\,\,\, \left\{ s_7=0, \,\,\, \xi+\frac12 s_6 = 0 \right\} \:.
\end{equation}
These two divisors are in different homology classes and therefore the $\U1$ symmetry resulting from one brane wrapping $X$ and its image wrapping $\tilde X$ is geometrically massive. 
\item  \textbf{$\mathbf{\U1_2}$ stack:} Let us now consider the second factor in Eq.  \eqref{eq:deltaE2},
\begin{align}\label{eq:disc2}
\Delta_E^{\rm rem,1}=(s_3s_5-s_1s_7)^2-(s_3s_6-s_2s_7)(s_2s_5-s_1s_6)\:.
\end{align}
Once intersected with the Calabi-Yau equation, this locus decomposes into two non-complete intersection four-cycles, one of which is given by the following expression
\begin{eqnarray}\label{eq:Y}
Y\, &\equiv &\left\{ \left(\xi - \frac{s_6}{2}\right) (s_3s_6-s_2s_7) + s_7 (s_3 s_5 - s_1 s_7)\,, \right.\\
&&\quad \left( \xi+\frac{s_6}{2}\right) (s_2s_5 - s_1s_6)  - s_5 (s_3 s_5 - s_1 s_7)\,,\\
 &&\quad \quad \left. \left( \xi-\frac{s_6}{2}\right) (s_3s_5 - s_1s_7) + s_7 (s_2s_5 - s_1s_6) \,,{\rm Eq. \eqref{eq:XP5}} \right\}\,,
 \end{eqnarray}
 the other $\tilde Y=\sigma^* Y$ is obtained from \eqref{eq:Y} upon exchange $\xi\mapsto-\xi$. The divisors $Y$ and $\tilde Y$ are in different homology classes (as we will show below). Hence, the $D7$ branes wrapping such divisors give rise to a massive $\U1$ symmetry. 

\item  \textbf{$\mathbf{\U1_3}$ stack:} The remaining factor in Eq. \eqref{eq:deltaE2} reads 
\begin{align}\label{eq:disc3} 
\Delta_E^{\rm rem,2}=&s_7 s_8^2 - s_6 s_8 s_9 + s_5 s_9^2\:.
\end{align}
When intersected with the Calabi-Yau equation it splits, giving rise to two divisors in different homology classes and defined by the following set of non-transversely intersecting polynomials
\begin{align}
Z\,\equiv\,\,\,\, &\left\{ s_7 s_8 - s_9 \left(  \xi + \frac{s_6}{2} \right), s_5 s_9 + s_8 \left( \xi-\frac{s_6}{2}\right),{\rm Eq \eqref{eq:XP5}}\right\}\,,\\
\tilde Z\,\equiv\,\,\,\, &\left\{ s_7 s_8 + s_9 \left(  \xi - \frac{s_6}{2} \right), s_5 s_9 - s_8 \left( \xi+\frac{s_6}{2}\right),{\rm Eq \eqref{eq:XP5}}\right\}\,.
\end{align}
Again, the associated $\U1$ is massive. 
\end{itemize}

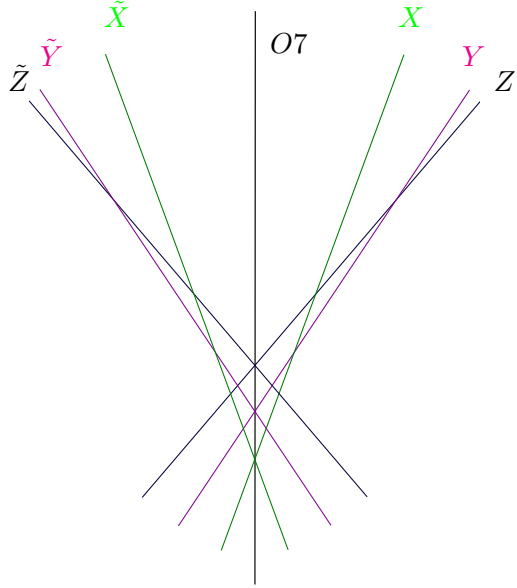
\begin{figure}[htb]
  \centering
  \setlength{\unitlength}{0.1\textwidth}
  \begin{picture}(4,5)
    \put(0,0){\includegraphics[scale=.5]{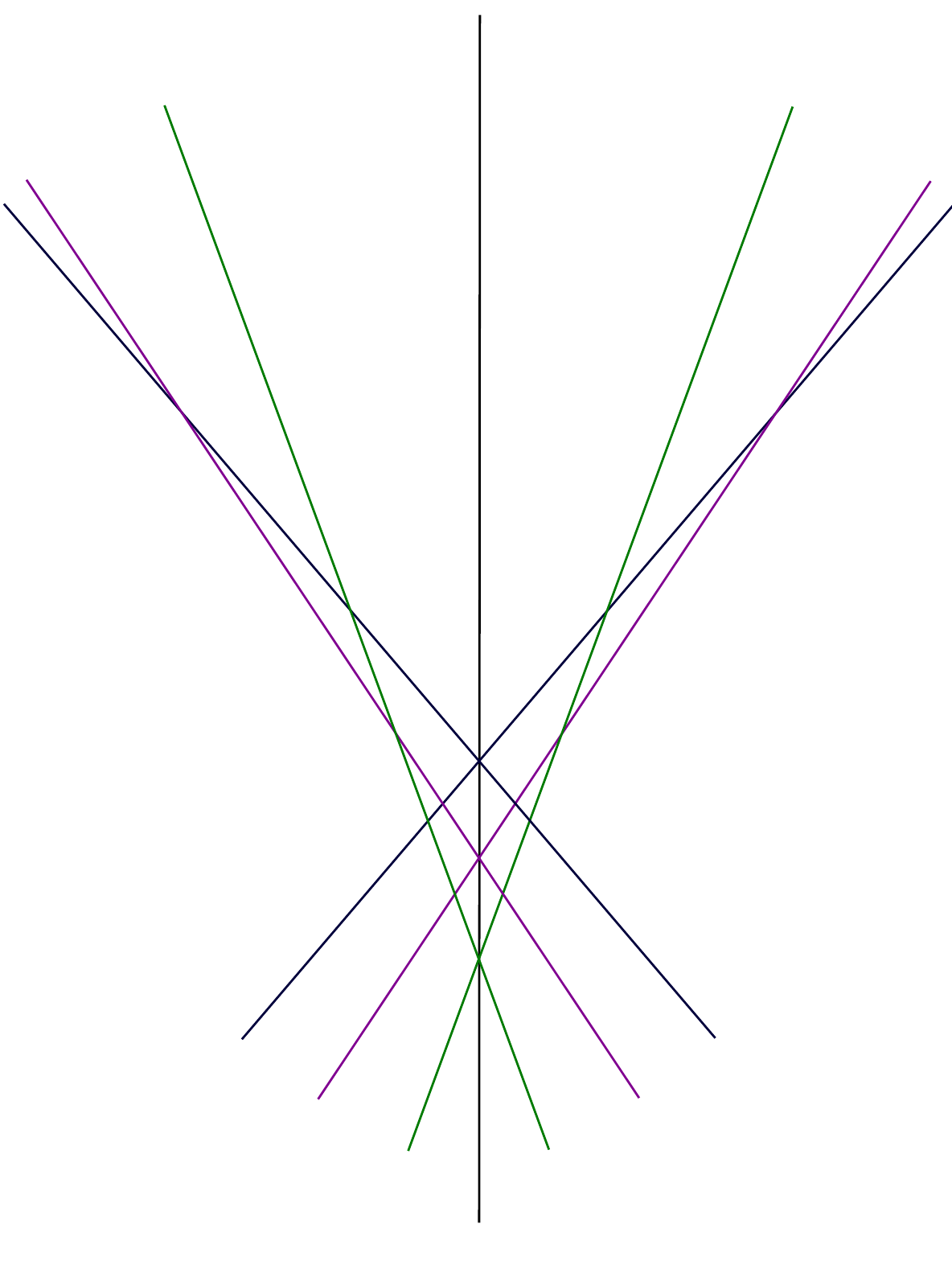}}
    \put(0.6,4.5){\small{\color{green} $\tilde{X}$}}
    \put(2.9,4.5){\small{\color{green} $X$}}
    \put(0.1,4.2){\small{\color{magenta} $\tilde{Y}$}}
    \put(3.4,4.2){\small{\color{magenta} $Y$}}
   \put(-0.15,4.0){\small $\tilde{Z}$}
     \put(3.65,4.0){\small $Z$}
   \put(1.9,4.3){\small $O7$}
  \end{picture}
\caption{\label{fig:weakiiXYZ} Depiction of the three \U1 brane stracks in the presence of the orientifold plane~$O7$.}
\end{figure}

\

Let us now discuss some of the relations among the divisor classes we have just described. Note first that in the Calabi-Yau the locus $s_7\Delta_E^{\rm rem,2}=0$ splits as
\begin{equation}
s_7 \Delta_E^{\rm rem,1}|_{X_3}= s_7^2 s_8^2 +4s_7 s_9 (-s_6 s_8 + s_5 s_9)
=\left(s_7 s_8 - s_9 \left(  \xi + \frac{s_6}{2} \right)\right)\left( s_7 s_8 + s_9 \left(  \xi - \frac{s_6}{2} \right) \right)\,.
\end{equation}
Both components are in the class $4D_{O7}-\frac12(Y+\tilde Y)$, where once again $D_{O7}=[\xi]$ is the class of the O7 plane. From Eqs. \eqref{XXteqs} and \eqref{eq:Y} one also sees that the first factor must be in the class $Z+\tilde X$, while the second must live in $\tilde Z+X$. By using this and the D7-tadpole cancellation condition $8D_{O7}=X_++Y_++Z_+$ (where we have rewritten the divisors in terms of orientifold even and odd combinations, i.e. $D_\pm=D\pm \tilde D$), we obtain 
\beq
Z=4D_{O7}-\frac12(Y+\tilde Y)-\tilde X\,, \quad \tilde Z= \sigma^\ast Z \,.
\eeq

Next lets consider the polynomial $s_7^2\Delta_E^{\rm rem,1}$, whose vanishing produces a divisor in the class $2(X+\tilde X)+(Y+\tilde Y)$. In the double cover Calabi-Yau thereefold, this polynomial factorizes as
\begin{eqnarray}
s_7^2\Delta_E^{\rm rem,1}|_{X_3}\,
    &=&\left(\left(\xi - \frac{s_6}{2}\right) (s_3s_6-s_2s_7) + s_7 (s_3 s_5 - s_1 s_7)\right) \nonumber\\
        &&\times\left(\left(\xi + \frac{s_6}{2}\right) (s_3s_6-s_2s_7) - s_7 (s_3 s_5 - s_1 s_7)\right)\:.
\end{eqnarray}
Once again both factors are in the same homology. The homology class of the first monomial is in the class $2X+Y$, while the second is in $2\tilde X+\tilde Y$, therefore implying 
\beq
Y_-=Y-\tilde Y=-2X_-\,. 
\eeq
Hence we have three unrelated even divisors $D_{O7}$, $X_+$ and $Y_+$ and one odd divisor $X_-$. The even ones are to be related to the base divisors $\pi^*(\bar{K}_B)$, $\pi^*(\cS_7)$ and $\pi^*(\cS_9)$. The first obvious identifications are 
\begin{equation}
\pi^*(\bar{K}_B)=D_{O7}\,,\quad \pi^*(\cS_7)=X_+\,.
\end{equation}
As regard $Y_+$, note that $Y_+=[\Delta_E^{\rm rem,1}]=2(\pi^*(\cS_3)+\pi^*(\cS_5))$. Using Tables~\ref{eq:sis} to write these classes in terms of $\cS_7$ and $\cS_9$ we find 
\begin{equation}
\pi^*(\cS_9)=3D_{O7}-\tfrac12 Y_+\,.
\end{equation}
Similarly as in the previous section, we have the relations $D_{O7}X=D_{O7}\tilde X=X \tilde X$, implying the absence of the conifold singularity on the type IIB side. These relations can be rewritten as 
\beq
2 D_{O7} X_+=X_+^2-X_-^2\,,\quad D_{O7} X_-=0\,.
\eeq

Even though we have a set of three massive $\U1$'s, there are however two linear combinations of the three $\U1$ generators that lead to massless $\U1$ gauge symmetries as expected from the F-theory side. As we have just shown, the relation between the odd part of the D7-brane divisors is $X_-=-\frac12 Y_-=Z_-$. This implies that there is only one combination of the D-brane $\U1$'s that eats the odd axion and become massive. The orthogonal combinations remain massless. The two massless $\U1$ generators are given by
\begin{equation}
\label{Qmassless} Q_1=\tfrac12(Q_X+Q_Y+Q_Z) \qquad\mbox{and}\qquad Q_2=-Q_X+Q_Z\,.
\end{equation}

\subsubsection{Charged matter}

The next step will be to obtain the corresponding matter living at the brane intersections. The schematics of the intersections is given in Fig.\ref{fig:weakiiXYZ}. The corresponding intersections are following, where the sub-indices are the charges $(Q_X,Q_Y,Q_Z)$ under massive \U1's, while the upper indices are the charges $(Q_1,Q_2)$ under the massless \U1's (reported for later use):
\begin{itemize}
\item $X \tilde X\,$: $\cancel{\one_{(2,0,0)}^{(1,-2)}}$. The would-be singlet is localized at the vanishing of the following ideal
\beq
\{\xi,s_7,s_6\}\,.
\eeq
However, this coincides with the $O7$ plane, where no symmetric matter is allowed. Due to that, this field is not part of the spectrum. 
\item $\tilde X Y\,$: $\one_{(1,1,0)}^{(1,-1)}$. This state sits at the vanishing of the ideal 
\beq
\{s_7,\xi+\frac{s_6}{2},s_3\}\,.
\eeq
\item $\tilde X Z\,$: $\one_{(1,0,1)}^{(1,0)}$. The singlet is located at the vanishing of 
\beq
\{s_7,\xi+\frac{s_6}{2},s_5 s_9-s_6s_8 \}\,.
\eeq
\item $X Y\,$: $\one_{(-1,1,0)}^{(0,1)}$. This state sits at the vanishing of the ideal 
\beq
\{s_7,\xi-\frac{s_6}{2},s_3 s_5^2-
s_2 s_5s_6 +s_1 s_6^2\}\,.
\eeq
\item $X Z\,$: $\one_{(-1,0,1)}^{(0,2)}$. This state sits at the vanishing of the ideal 
\beq
\{s_7,\xi-\frac{s_6}{2},s_9\}\,.
\eeq
\item $Y\tilde Y\,$: $\one_{(0,2,0)}^{(1,0)}$. This state sits at the vanishing of the ideal 
\beq
\{s_3s_5-s_1s_7, s_2s_5-s_1s_6, s_3s_6-s_2s_7, {\rm Eq. \eqref{eq:XP5}}\}\,.
\eeq
\item $Y Z\,$: $\one_{(0,1,1)}^{(1,1)}$. This state sits at the vanishing locus of the union of the ideals for $Y$ and $Z$, which occurs to be prime. 
\item $\tilde Y Z\,$: $\one_{(0,-1,1)}^{(0,1)}$. This state sits at the vanishing of the ideal: 
\begin{align}
\begin{split}
\{&2 s_7 s_8-s_6 s_9+2 s_9 \xi, s_6 s_8 -2 s_5 s_9+2 s_8 \xi,\,
2s_3s_6s_8+2\xi s_2s_9-2 s_3 s_5 s_9-s_2 s_6 s_9+2 s_1 s_7 s_9, \\
& s_3 s_8^2-s_2 s_8 s_9+s_1 s_9^2, 2 s_3 s_5 s_8-2 s_2 s_5 s_9+s_1 s_6 s_9+2 \xi s_1 s_9 ,\,
2\xi s_3s_6-s_3s_6^2-2\xi s_2s_7+2 s_3 s_5 s_7+\\
&+s_2 s_6 s_7-2 s_1 s_7^2, 2\xi s_3s_5-s_3 s_5 s_6 -2\xi s_1 s_7+2s_2s_5s_7-s_1 s_6 s_7,\\
& 2\xi  s_2s_5 - 2 s_3 s_5^2 - 2\xi s_1s_6+s_2 s_5 s_6 -s_1 s_6^2 + 2 s_1 s_5 s_7,\, {\rm Eq. \eqref{eq:XP5}}\}\,.
\end{split}
\end{align}
\item $Z \tilde{Z}\,$: $\one_{(0,0,2)}^{(1,2)}$. The ideal associated to this state is given by 
\beq
\{s_8,s_9,{\rm Eq. \eqref{eq:XP5}}\}\,.
\eeq
\end{itemize}

Let us focus on the charges of these fields under the two massless $\U1$ generators \eqref{Qmassless}. As in the SM example of Section \ref{sec:ToricMSSM}, there are fields that have different charges under the three D7-brane massive $\U1$'s, but have the same charges under the two massless $\U1$'s. Correspondingly if one goes away from the weak coupling limit (i.e. take $\epsilon$ finite), the corresponding matter curves join: there is one matter curve for each pair of massless charges. 
Said differently, we can look back at the F-theory Table \ref{tab:poly5_matter} and consider the $\epsilon$ scaling for the matter loci, taking only the leading order in $\epsilon$ for the various ideals. One notices that some curves split into two irreducible loci.  The splitting occurs for the loci associated to the singlets $\one_{(1,0)}$ and $\one_{(0,1)}$. The correspondence for the matter curves works as follows
\beq
\text{
\begin{tabular}{|l||c|c|c|c|c|c|c|c|}
\hline
Type IIB & $\one_{(1,1,0)}$ & $\one_{(1,0,1)}$ & $\one_{(0,2,0)}$ & $\one_{(-1,0,1)}$ & $\one_{(0,1,1)}$ & $\one_{(0,0,2)}$ & $\one_{(-1,1,0)}$ & $\one_{(0,-1,1)}$ \\ \hline
F-theory & $\one_{(1,-1)}$ & \multicolumn{2}{c|}{$\one_{(1,0)}$} & $\one_{(0,2)}$ & $\one_{(1,1)}$ & $\one_{(1,2)}$ & \multicolumn{2}{c|}{$\one_{(0,1)}$}\\
\hline
\end{tabular}
}\,,\nonumber
\eeq
where for the type IIB matter we reported only the massive $\U1$ charges.

To make the phenomenon clearer, let us consider one of the splitting. Take the curve ${\bf 1}_{(1,0)}$ in F-theory. Its locus is given by the following ideal (it is not a complete intersection, as it can be inferred from the more implicit form in Table \ref{tab:poly5_matter}):
\begin{eqnarray}
V(I_{(2)}) &=& \left\{ s_3s_6s_8 - s_2s_7s_8 - s_3s_5s_9 + s_1s_7s_9,\,\,\, s_3s_6s_7 - s_2s_7^2  - s_3^2 s_9, 
\right. \nonumber \\
&&\,\,\,\,\, s_3s_5s_7 - s_1s_7^2- s_3^2 s_8,\,\,\, s_2s_5s_7 - s_1s_6s_7 - s_2s_3s_8 + s_1s_3s_9, \nonumber \\
&&\,\,\,\,\, \left. s_2s_5s_6s_8 - s_1s_6^2 s_8 - s_2^2 s_8^2  - s_2s_5^2 s_9
     + s_1s_5s_6s_9 + 2\,s_1s_2s_8s_9 - s_1^2 s_9^2 \right\} \:.\nonumber
\end{eqnarray}
At leading order in $\epsilon$ (remember that only $s_8$ and $s_9$ scales with $\epsilon$) one has
\begin{eqnarray}
V(I_{(2)})^{\rm w.c.} &=& \left\{ -s_3(s_5s_9  - s_6s_8) + s_7(s_1s_9- s_2s_8),\,\,\, (s_3s_6 - s_2s_7)s_7 , 
\right. \nonumber \\
&&\,\,\,\,\, (s_3s_5 - s_1s_7)s_7,\,\,\, (s_2s_5 - s_1s_6)s_7 , \,\,\, 
\left. -(s_2s_5 - s_1s_6) ( s_5s_9 - s_6s_8)  \right\} \:.\nonumber
\end{eqnarray}
The vanishing locus associated with this ideal is then the union of the loci given by the  ideals\footnote{To make it manifest, one can notice that the first equation of $V(I_{(2)})^{\rm w.c.}$ can be rewritten as $s_8(s_3s_6-s_2s_7)-s_9(s_3s_5-s_1s_7)$.}
\begin{eqnarray}
\left\{ s_7,\,\,\, s_5s_9-s_6s_8 \right\} &\mbox{and}& \left\{  s_3s_5-s_1s_7, s_2s_5-s_1s_6, s_3s_6-s_2s_7  \right\} \:.\nonumber
\end{eqnarray}
The last one further splits on the CY threefold. We recognize the loci of the curves $\one_{(1,0,1)}$ and  $\one_{(0,2,0)}$ in type IIB.

\subsubsection{Fluxes and chiralities}

Finding all admisible gauge fluxes amounts to finding all solutions for the D5-tadpole cancellation condition, that in this case takes the form
\begin{align}
0\,=\,& X_- F^X_+ +Y_-F^Y_++Z_-F^Z_+X_+ +F^X_- +Y_+F^Y_-+Z_-+F^Z_-\\
=\,& X_-(F^X_+-2F^Y_++F^Z_+)+X_+(F^X_--F^Z_-)+Y_+(F^Y_--F^Z_-)\,.
\end{align}
We find three types of independent solutions to this equation:
\begin{itemize}
\item D5-tadpole canceling even fluxes are
\begin{align}\label{eq:evenF5}
( F^X_+,F^Y_+,F^Z_+)_1=&\tfrac12(1,1,1)F_1\,,\\
( F^X_+,F^Y_+,F^Z_+)_2=&(-1,0,1)F_2\,,\\ 
( F^X_+,F^Y_+,F^Z_+)_{\lambda}=&\lambda(D_{O7},0,0)\,. 
\end{align}
The first two are the fluxes along the massless $\U1$ generators, with $F_1,F_2\in H^{(1,1)}_+(X_3)$, and the last one is a flux for a massive $\U1$, with $\lambda$ a suitable rational number in agreement with flux quantization. 
\item As regard the orientifold odd sector we only get the general solution $F^X_- =F^Y_-=F^Z_-$ which corresponds to a shift in the $B$-field. 
\item There is also a mixed flux solution
\begin{align}
( F^Z_+,F^Y_-)_\alpha=\alpha(Y_+,-X_-)\,, 
\end{align}
where again $\alpha$ is a rational number compatible with flux quantization.
\end{itemize}
The matching to the vertical fluxes in F-theory proceeds as follows: The fluxes $\mathcal{F}_1$ and $\mathcal{F}_2$ are related to $F_1$ and $F_2$ in the same way as in Section \ref{sec:WCMSSM}, while the $\Lambda$ flux requires to take a linear combination of $F_1$, $F_2$, $\lambda$ and $\alpha$ fluxes. If we take 
\begin{align}
F_1=0\,,\quad F_2=-\left(2 D_{O7}- \frac12 X_+ - \frac12 Y_+\right)\Lambda\,,\quad  \lambda=-2\Lambda\,,\quad 2\alpha=\Lambda\,,
\end{align} 
the FI terms, D3-tadpoles as well as the chiral indices induced by the $\Lambda$ flux match the result in the type IIB setup. 
Hence, one concludes that the remaining type IIB flux combination cannot be among the harmonic vertical fluxes in F-theory.


%
\section{$\SU3\times\SU2\times\U{1}^2$ model}
\label{sec:F5Top}
A variation of the $F_{5}$ polytope model which allows to incorporate non-Abelian symmetries and matter has been constructed in refs. \cite{Lin:2014qga,Lin:2016vus}, for standard model like theories with one extra $\U1$. Their approach consists in modifying the fibration by toric construction known as a top \cite{Candelas:1996su,Candelas:1997pv}, and elaborates on the classification of all possible tops for fibers cut as hypersurfaces in any of the sixteen 2D toric ambient spaces \cite{Bouchard:2003bu}.  There are only five inequivalent $\SU3\times\SU2$ tops of $dP_2$, out of which we focus on the one that has been studied more widely, denoted as ${\rm I}\times {\rm A}$ in \cite{Lin:2014qga}. 

\subsection{F-theory description}
\subsubsection{Geometric setup}

 \begin{figure}[htb]
\centering
\begin{minipage}{.56\textwidth}
  \centering
  \includegraphics[scale=.4]{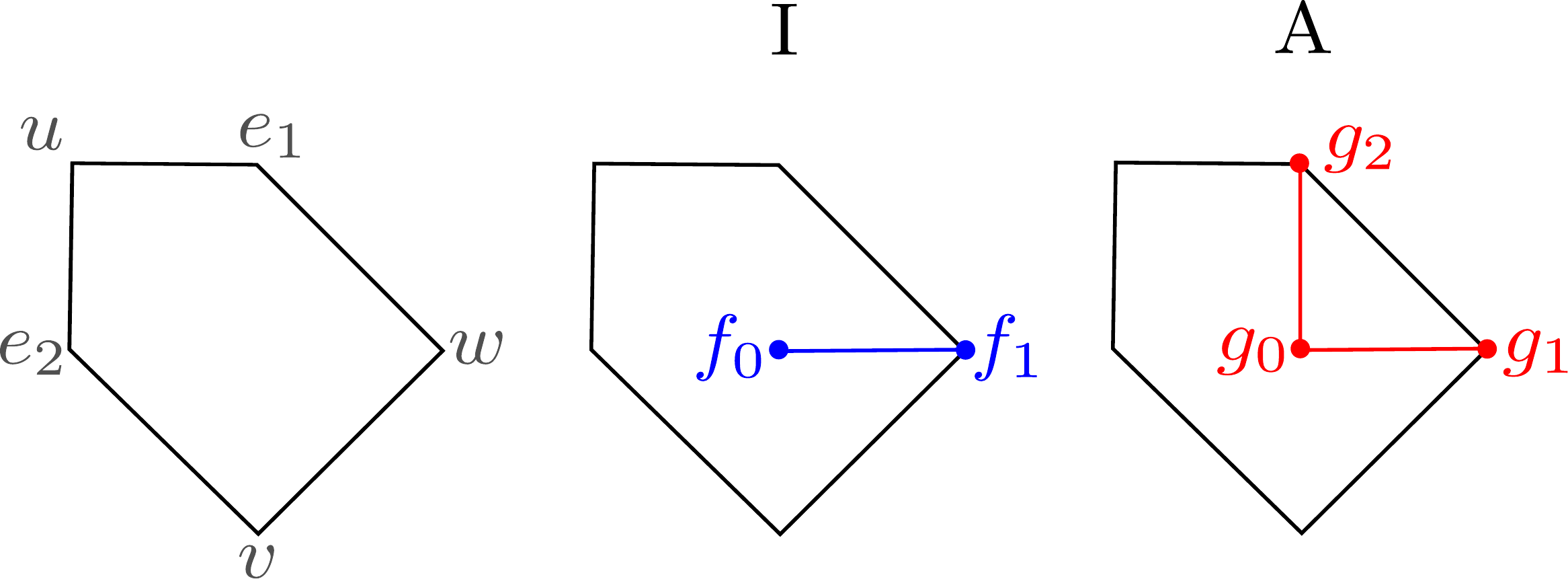}
\end{minipage}%
\begin{minipage}{.44\textwidth}
\centering
{\footnotesize
  \begin{tabular}{|c|c|}\hline
Section & Line Bundle\\ \hline
$u$ & $\mathcal{O}(H-E_1-E_2+\cS_9+K_B)$ \\ \hline
$v$ & $\mathcal{O}(H-E_2+G_1+\cS_9-\cS_7)$\\ \hline
$w$ & $\mathcal{O}(H-E_1-F_1-G_2)$\\ \hline
$e_1$ & $\mathcal{O}(E_1)$\\ \hline
$e_2$ & $\mathcal{O}(E_2)$\\ \hline
\end{tabular}}
\\
\vspace{2mm}
 \renewcommand{\arraystretch}{1.4}
{\footnotesize
  \begin{tabular}{|c|c||c|c|c|}\hline
$f_0$ & $f_1$ & $g_0$ & $g_1$ & $g_2$ \\ \hline
$\cW_2-F_1$ & $F_1$ & $\cW_3-G_1-G_2$ & $G_1$ & $G_2$\\ \hline
\end{tabular}}
\end{minipage}
\caption{\label{fig:poly5_Top} The tops I and A used to engineer the non Abelian gauge symmetry $\SU3\times\SU2$ over $F_5$ \cite{Lin:2014qga,Lin:2016vus}. The table contains the divisor classes of the $\mathbb{P}_{F_{11}}$ ambient space as well as the top coordinates. Note that some divisor classes for the ambient space coordinates get modified in the presence of the top. }
\end{figure}
Inducing non-Abelian gauge enhancements first requires specifying additional base divisors along which the fiber degenerates. In this case we define $\cW_2=\{w_2=0\}$ and $\cW_3=\{w_3=0\}$ as the base divisors where the $\SU2$ and $\SU3$ gauge symmetries live. 

The top construction automatically provides the (toric) resolution divisors that make the fiber smooth over $w_2=0$ and $w_3=0$ as well. In the hypersurface equation, the presence of additional divisors is manifest by the presence of additional blow up coordinates $f_0$, $f_1$ and $g_0$, $g_1$ and $g_2$ for $\SU2$ and $\SU3$, respectively. The idea is to refine the sections $s_i$ of Section \ref{sec:F5} such that they $f,g,\Delta$ have the right vanishing orders for $w_1$ and $w_2$; they will then be of the form $s_i=\tilde s_i f_0^{l_0}  f_1^{l_1}  g_0^{m_0} g_1^{m_1} g_2^{m_2}$, with suitable integers $l_j$, $m_j$.

For the top model ${\rm I}\times {\rm A}$ of \cite{Lin:2014qga,Lin:2016vus}, the fiber equation is given by $p_{F_5}^{{\rm I}\times {\rm A}} =0$, with 
\begin{align}
\label{eq:pF5Top}
\begin{split}
p_{F_5}^{{\rm I}\times {\rm A}} =\, & s_1 e_1^2  e_2^2  f_0 g_0 g_1^2  u^3 +  s_2 e_1  e_2^2  f_0 g_0 g_1 u^2 v + 
 s_3  e_2^2  f_0 g_0 u v^2 + s_5  e_1^2  e_2 g_1 u^2 w \\
 &+ s_6 e_1  e_2  u v w + 
 s_7 e_2 g_0 g_2 v^2 w + s_8 e_1^2  f_1 g_1 g_2  u w^2 + s_9 e_1  f_1 g_2  v w^2 \,,
 \end{split}
\end{align}
where by abuse of notation we have written $s_i$ instead of $\tilde s_i$. As already mentioned, for each of the top coordinates $f_i$, $g_i$ there is a toric divisor corresponding to a $\mathbb{P}^1$ fibered over either $\cW_2$ or $\cW_3$, in such a way that the different fiber $\mathbb{P}^1$'s intersect according to the affine Dynkin diagram of the Lie algebra under consideration. Therefore the divisors classes satisfy
\beq
\hat \pi^{*}(\cW_2)=[f_0]+[f_1]\,,\quad \hat \pi^{*}(\cW_3)=[g_0]+[g_1]+[g_2]\,,
\eeq
where $\hat \pi^{*}(\cW_2)$ and $\hat\pi^{*}(\cW_3)$ are divisors in the Calabi-Yau fourfold $Y_4$ obtained by pulling back the base divisors  $\cW_2$ and $\cW_3$ by the projection map $\hat{\pi}:Y_4\rightarrow B_3$. As done in the previous sections, we will use $\cW_2$ and $\cW_3$ to denote the pull-back to the fourfold as well as base divisors. 
The pictures of the top as well as the divisor classes for fiber ambient space and top coordinates are given in Fig. \ref{eq:pF5Top}. 

In order for $p_{F_5}^{{\rm I}\times {\rm A}}$ to give a Calabi-Yau after introducing the top, some of the divisor classes associated with the $s_i$ have to be modified:
\beq
\hspace{-4mm}\text{
{\footnotesize
\renewcommand{\arraystretch}{1.4}
\begin{tabular}{|c|c|c|c|c|c|c|c|}
\hline
$s_1$ & $s_2$ & $s_3$ &  $s_7$   \\ \hline
$3\bar{K}_B-\cS_7-\cS_9-\cW_2-\cW_3$ & $2\bar{K}_B-\cS_9-\cW_2-\cW_3$ & $\bar{K}_B+\cS_7-\cS_9-\cW_2-\cW_3$ & $\cS_7-\cW_3$  \\
\hline
\end{tabular}}
}\:.\nonumber
\eeq
The inequivalent sections of the elliptic fibration can once again been represented by the divisors $S_0=[e_2]$, $S_1=[e_1]$ and $S_2=[u]$. The exceptional divisors are $D^{\SU2}=F_1$, $D^{\SU3}_{1}=G_1$ and $D^{\SU3}_{2}=G_2$. Due to the presence of these, the Shioda maps have to be modified such that the $\U1$ generators are orthogonal to the Cartan generators of the non-Abelian gauge symmetry:
\begin{align}
\label{eq:ShiodaF51Top}
\sigma_1&= S_1 - S_0 - \bar{K}_B+\frac12 D^{\SU2}+\frac13(2D^{\SU3}_{1}+D^{\SU3}_{2}) \, , \\
\label{eq:ShiodaF52Top}
\sigma_2 &= S_2 - S_0 -\bar{K}_B -\cS _9+\frac13(2D^{\SU3}_{1}+D^{\SU3}_{2})\, .
\end{align}
After mapping the fiber equation to the Weierstrass equation, one can work out the $b_i$'s sections:
\begin{align}
\begin{split}
b_2=&\frac14 s_6^2 - s_5 s_7 w_3\,,\\
\end{split}
 &\\
\begin{split}
b_4=&-\frac{1}{12} w_2 w_3 (-s_3 s_6^2 s_8 + 3 s_3 s_5 s_6 s_9 - s_2 s_6^2 s_9 - 
   2 s_3 s_5 s_7 s_8 w_3 + 3 s_2 s_6 s_7 s_8 w_3\\
 & \phantom{-\frac{1}{12} w_2 w_3 (}- 2 s_2 s_5 s_7 s_9 w_3 + 
   3 s_1 s_6 s_7 s_9 w_3 + 2 s_3^2 s_8^2 w_2 w_3 - 2 s_2 s_3 s_8 s_9 w_2 w_3\\
&\phantom{-\frac{1}{12} w_2 w_3 (} + 
   2 s_2^2 s_9^2 w_2 w_3 - 6 s_1 s_3 s_9^2 w_2 w_3 - 6 s_1 s_7^2 s_8 w_3^2)\,,\\
 \end{split}
 &\\
 \begin{split}
b_6=&-\frac{1}{108} w_2^2 w_3^2 (-3 s_3^2 s_6^2 s_8^2 + 18 s_3^2 s_5 s_6 s_8 s_9 - 
   6 s_2 s_3 s_6^2 s_8 s_9 - 27 s_3^2 s_5^2 s_9^2 + 18 s_2 s_3 s_5 s_6 s_9^2\\
   &\phantom{-\frac{1}{108} w_2^2 w_3^2 (} - 
   3 s_2^2 s_6^2 s_9^2 - 24 s_3^2 s_5 s_7 s_8^2 w_3 + 
   18 s_2 s_3 s_6 s_7 s_8^2 w_3+ 6 s_2 s_3 s_5 s_7 s_8 s_9 w_3 \\
    &\phantom{-\frac{1}{108} w_2^2 w_3^2 (}+ 
   18 s_2^2 s_6 s_7 s_8 s_9 w_3 - 90 s_1 s_3 s_6 s_7 s_8 s_9 w_3 - 
   24 s_2^2 s_5 s_7 s_9^2 w_3 + 54 s_1 s_3 s_5 s_7 s_9^2 w_3\\
   &\phantom{-\frac{1}{108} w_2^2 w_3^2 (} + 
   18 s_1 s_2 s_6 s_7 s_9^2 w_3 + 8 s_3^3 s_8^3 w_2 w_3 - 
   12 s_2 s_3^2 s_8^2 s_9 w_2 w_3 - 12 s_2^2 s_3 s_8 s_9^2 w_2 w_3\\
     &\phantom{-\frac{1}{108} w_2^2 w_3^2 (}  + 
   72 s_1 s_3^2 s_8 s_9^2 w_2 w_3 + 8 s_2^3 s_9^3 w_2 w_3 - 
   36 s_1 s_2 s_3 s_9^3 w_2 w_3 - 27 s_2^2 s_7^2 s_8^2 w_3^2\\
   &\phantom{-\frac{1}{108} w_2^2 w_3^2 (}  + 
   72 s_1 s_3 s_7^2 s_8^2 w_3^2 + 18 s_1 s_2 s_7^2 s_8 s_9 w_3^2 - 
   27 s_1^2 s_7^2 s_9^2 w_3^2)\,,\\
 \end{split}
\end{align}
which can be used to construct $f$, $g$ and the discriminant. After doing so, one finds that at codimension-one there are indeed two singularities, one over the locus $w_2=0$, exhibiting the right vanishing orders to coincide with an $A_1$ singularity. The other lives at $w_3=0$ and corresponds to an $A_2$ singularity. At codimension-two we find several loci corresponding to the location of charged matter. They are summarized in Table~\ref{tab:poly5top_matter}, together with the representations for the matter associated with those singularities. Note that the singlet sector, which was already discussed in the previous sections suffers from slight modifications when the top is introduced. In particular, note that the sections $w_1$ and $w_2$ enter in the definition of the singlet curves that can not be written as complete intersections. 

In comparison to the minimal standard model of Section \ref{sec:ToricMSSM}, the presence of the additional $\U1$ symmetry gives many more matter representations which could be of great use for model building. For example, for the model of Section \ref{sec:ToricMSSM}, there is only one doublet curve, hence the Higgs fields in that model must be vector-like. Here instead, one has three doublets, distinguished by their $\U1$ charges. Hence there is in principle the possibility to put, leptons, up-type and down-type Higgses in different curves.
\begin{table}[H]
\begin{center}
\footnotesize
\renewcommand{\arraystretch}{1.6}
\begin{tabular}{|c|c|}
\hline
Representation &Locus \\ \hline\hline
$(\three,\two)_{(\frac16,-\frac13)}$ &  $\{w_3 = w_2 = 0 \}$ \\ \hline\hline
$(\overline{\three},\one)_{(-\frac23,\frac13)}$ &  $\{w_3 = s_3 = 0 \}$ \\ \hline
$(\overline{\three},\one)_{(\frac13,\frac43)}$ &  $\{w_3 = s_9 = 0 \}$ \\ \hline
$(\overline{\three},\one)_{(\frac13,-\frac23)}$ &  $\{w_3 =-s_6 s_7 + s_3 s_9 w_2= 0 \}$ \\ \hline
$(\overline{\three},\one)_{(-\frac23,-\frac23)}$ &  $\{w_3 = -s_6 s_8 + s_5 s_9 = 0 \}$ \\ \hline
$(\overline{\three},\one)_{(\frac13,\frac13)}$ &  $\{w_3 = s_3 s_5^2 - s_2 s_5 s_6 + s_1 s_6^2 = 0 \}$ \\ \hline\hline
$(\one,\two)_{(\frac12,-1)}$ &  $\{w_2 = s_7 = 0 \}$ \\ \hline
$(\one,\two)_{(\frac12,1)}$ &  $\{w_2 =-s_6 s_8 s_9 + s_5 s_9^2 + s_7 s_8^2 w_3= 0 \}$ \\ \hline
$(\one,\two)_{(\frac12,0)}$ &  $\{w_2 = s_3^2 s_5^2 - s_2 s_3 s_5 s_6 + s_1 s_3 s_6^2 + s_2^2 s_5 s_7 w_3 - 
 2 s_1 s_3 s_5 s_7 w_3 - s_1 s_2 s_6 s_7 w_3 + s_1^2 s_7^2 w_3^2
= 0 \}$ \\ \hline\hline
$\one_{(1,-1)}$ &  $V(I_{(1)}):=\{s_3 = s_7 = 0 \}$ \\ \hline

$\one_{(1,0)}$ & $\begin{array}{c} V(I_{(2)}):=\{
-s_3 s_6 s_7 + s_3^2 s_9 w_2 + s_2 s_7^2 w_3=0\,,\\ s_3 s_5 s_7 - s_3^2 s_8 w_2 - s_1 s_7^2 w_3= 0 \}\backslash V(I_{(1)}) \end{array}$ \\ \hline

$\one_{(-1,-2)}$ & $V(I_{(3)}):=\{s_8 = s_9 = 0 \}$ \\ \hline

$\one_{(-1,-1)}$ & $\begin{array}{c} V(I_{(4)}):=\{ s_2 s_8 s_9 - s_3 s_8^2 - s_9^2 s_1=0 \\ s_5 s_9^2 - s_6 s_8 s_9 + s_8^2 s_7w_3 = 0 \} \backslash(V(I_{(3)}) \end{array}$ \\ \hline

$\one_{(0,2)}$ &  $V(I_{(5)}):=\{ s_9 = s_7 = 0 \}$ \\ \hline

$\one_{(0,1)}$ & $\begin{array}{c} V(I_{(6)}):=\{s_9^2 (-s_6 s_7 + s_3 s_9 w_2) (-s_6 s_7 s_8 + s_5 s_7 s_9 + s_3 s_8 s_9 w_2)\\
 + 
 s_7^2 s_9 (-2 s_6 s_7 s_8^2 + s_5 s_7 s_8 s_9 + 2 s_3 s_8^2 s_9 w_2 + 
    s_1 s_9^3 w_2) w_3 
    + s_7^4 s_8^3 w_3^2=0\,,\\
    s_3 s_9^3 w_2 (-s_6 s_7 + s_3 s_9 w_2)+ 
 s_7^2 s_9 (s_6 s_7 s_8 + s_9 (-s_5 s_7 + s_2 s_9 w_2)) w_3 
 - s_7^4 s_8^2 w_3^2= 0 \} \\ \backslash(V(I_{(1)})\cup V(I_{(3)})\cup V(I_{(5)})) \end{array}$ \\ \hline
\end{tabular}
\caption{\label{tab:poly5top_matter}Charged matter under $\SU3\times\SU2\times\U{1}^2$ with the expressions for their corresponding codimension-two loci.}
\end{center}
\end{table}

\subsubsection{Fluxes and Chiral Matter}
After computing the quartic intersections in the Calabi-Yau fourfold one can determine the inequivalent vertical flux directions. 
The $G_4$ fluxes in this model were originally computed in \cite{Lin:2016vus}. Here we have followed a slightly different notation in order to keep a more uniform structure throughout the paper.

As expected, we obtain the usual gauge flux directions along the $\U1$ generators
\beq
G_4^{\mathcal{F}_1}=\sigma_1\wedge \mathcal{F}_1\,,\quad G_4^{\mathcal{F}_2}=\sigma_1\wedge \mathcal{F}_2\,,
\eeq
as well as a version of the $\Lambda$-flux, 
\beq
G_4^{\Lambda}=\Lambda \left[S_0^2 + [K_{B}^{-1}](-[K_{B}^{-1}] + S_2) + \cS_9 (-\cS_7 + \cS_9) +\frac13  [K_{B}^{-1}](2G_1+G_2) \right]\,.
\eeq
This expression coincides with the $\Lambda$-fux obtained in the previous section (see Eq. \eqref{eq:G4F5}) if we set $G_1$ and $G_2$ to zero.

Additionally to the previous fluxes, we obtain four extra inequivalent flux directions which are due to the tunning of the complex structure in the presence of the top: 
\begin{align}
\Scale[0.9]{G_4^{A1}\,=}&\,\Scale[0.9]{\frac{1}{2}A_1 \left[2 F_1^2 - 2 (G_1 + G_2 - 2 (-[K_{B}^{-1}]+ S_0)) \cW_2 + F_1 (-2 (-[K_{B}^{-1}] - \cS_9 + \cW_2) + \cW_3) \right]\,, }\nonumber\\
\Scale[0.9]{G_4^{A2}\,=}&\,\Scale[0.9]{\frac{1}{6}A_2 \left[2 (2 G_1 + G_2 - 3 ([K_{B}^{-1}]+ S_0)) \cW_2 + 3 F_1 (2 S_1 -\cS_7 + \cW_2 + \cW_3) \right]\,,} \nonumber\\
\Scale[0.9]{G_4^{A3}\,=}&\,\Scale[0.9]{\frac{1}{6}A_3 \left[6 G_1^2 + G_1 (6 [K_{B}^{-1}] - 6 S_7 + 4 \cS_9 - 8 \cW_3) + 2 G_2 (\cS_9 - 2 \cW_3) - 
 3 (F_1 - 4 ([K_{B}^{-1}] + S_0)) \cW_3 \right]\,,}\nonumber\\
\Scale[0.9]{G_4^{A4}\,=}&\,\Scale[0.9]{\frac{1}{6}A_4 \left[ -2 G_2( [K_{B}^{-1}]+ \cS_7 - 2 \cW_3) + 3 (F_1 - 2 ([K_{B}^{-1}] + S_0)) \cW_3 + 
 2 G_1 (3 G_2 - 2[K_{B}^{-1}] + \cS_7 + \cW_3) \right]\,.}\nonumber
 \end{align}
This coincides with the observation of \cite{Lin:2016vus} that in this model there are five extra flux directions in addition to the massless \U1 $G_4$ fluxes. Our flux expressions can be matched to those found in \cite{Lin:2016vus} up to SR-ideal components.
The chiralities can be straightforwardly computed by using the curve representatives provided in \cite{Lin:2016vus}.

\subsection{The weak coupling limit}
One can see that the weak coupling limit remains the same as in the $F_{5}$ model, and that $b_2$, $b_4$ and $b_6$ scale accordingly with $\epsilon$ provided $s_8\rightarrow\epsilon s_8$ and $s_9\rightarrow \epsilon s_9$, while all other base sections remain independen of $\epsilon$. The weak discriminant then reads
\begin{equation}\label{eq:deltaE3}
\Delta_E=-\frac14 w_2^2 \cdot w_3^3\cdot s_7 \cdot \left[ (s_3s_5-s_1s_7 w_3)^2-(s_3s_6-s_2s_7 w_3)(s_2s_5-s_1s_6)
   \right] \cdot 
   \left[s_7 s_8^2 w_3- s_6 s_8 s_9 + s_5 s_9^2\right]\:,
\end{equation}
from which we can read out the $\U3$ and $\SU2$ factors as well as the three $\U1$ factors which change just slightly in comparison with Eq. \eqref{eq:deltaE2}. 
 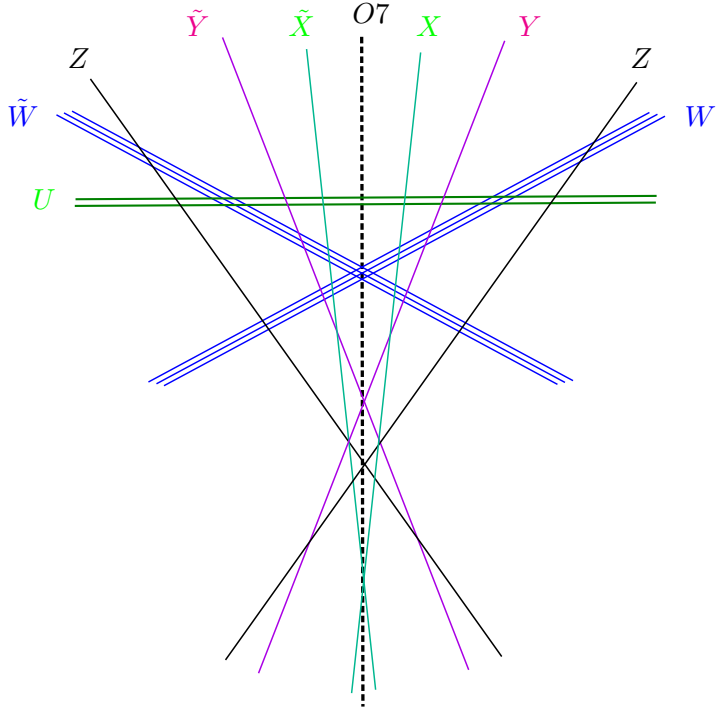
\begin{figure}[htb]
  \centering
  \setlength{\unitlength}{0.1\textwidth}
  \begin{picture}(5.3,6)
    \put(0,0){\includegraphics[scale=.7]{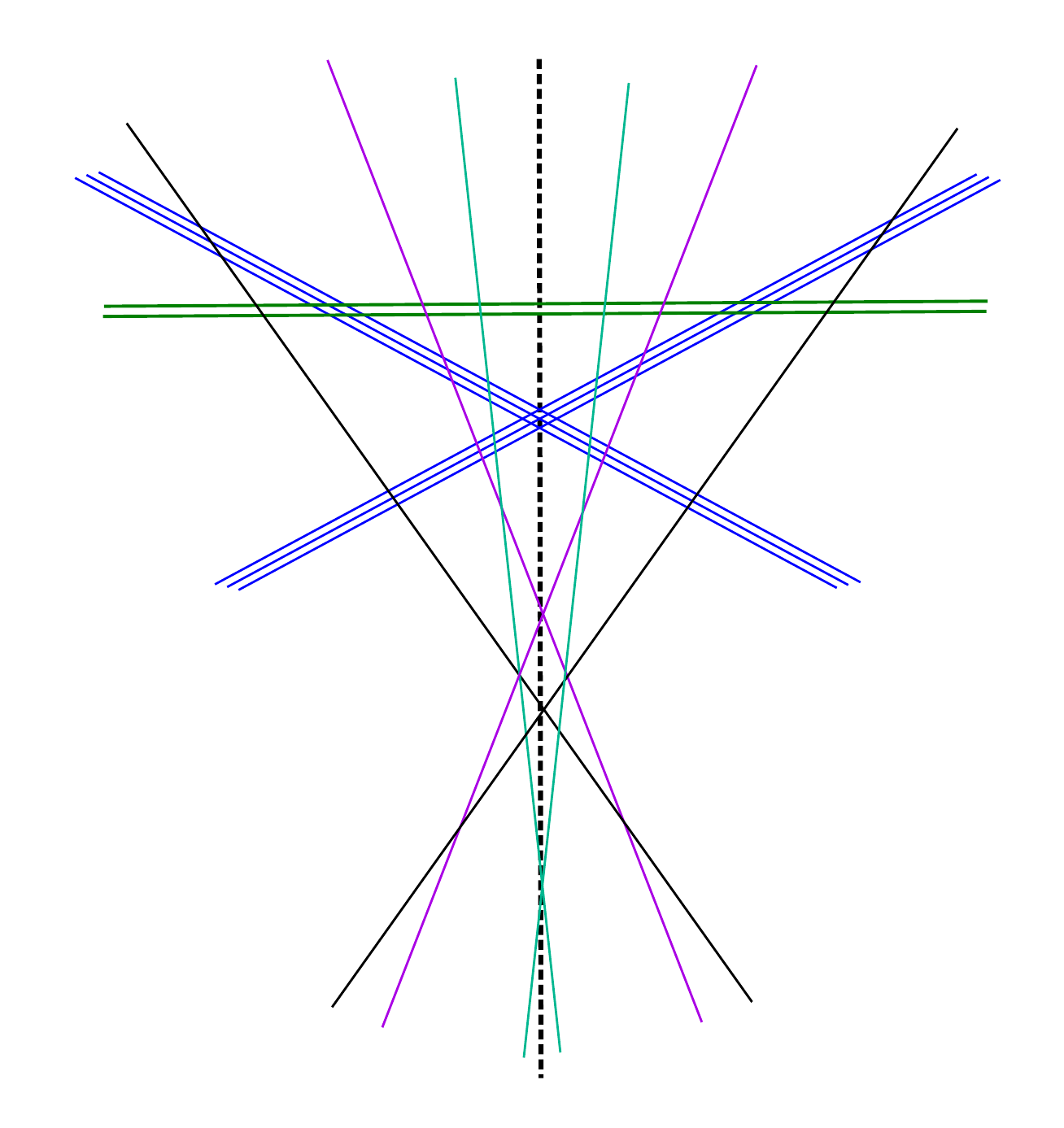}}
    \put(2.2,5.5){\small{\color{green} $\tilde{X}$}}
    \put(3.2,5.5){\small{\color{green} $X$}}
        \put(0.2,4.15){\small{\color{green} $U$}}
        \put(2.7,5.6){\small $O7$}
    \put(1.4,5.5){\small{\color{magenta} $\tilde{Y}$}}
    \put(4.0,5.5){\small{\color{magenta} $Y$}}
   \put(0.0,4.8){\small{\color{blue} $\tilde{W}$}}
     \put(5.3,4.8){\small{\color{blue} $W$}}
      \put(4.8,5.25){\small{ $Z$}}
      \put(0.4,5.25){\small{ $Z$}}
  \end{picture}
\caption{\label{fig:weakii} Schematics of the brane setup for the $\SU3\times \SU2 \times \U1\times \U1$ model. The $\SU2$ brane lies on a symplectic cycle. The two $\U1$ symmetries in F-theory result from two geometrically massless combinations out of three $\U1$ symmetries in type IIB.}
\end{figure}
The Calabi-Yau Equation in this case reads 
\beq\label{eq:XP5top}
\xi^2=\frac{s_6^2}{4}-s_5 s_7 w_3\,.
\eeq
Note that in this case, we must not forbid one but three conifold points, namely: $\xi=s_6=s_5=s_7=0$, $\xi=s_6=s_7=w_3=0$ and $\xi=s_6=s_5=w_3=0$. Moreover the CY threefold has now $h^{1,1}_-=2$ (and $h^{1,1}_+= h^{1,1}(B_3)\geq 2$): the two independent loci $s_5=0$ and $s_7=0$ both split into two divisors mapped to each other by the orientifold involution $\xi\mapsto -\xi$, and in different homology classes.

\subsubsection{D7 brane setup}
After the introduction of $w_2$ and $w_3$, the new brane stacks that one obtains are
\begin{itemize}
 \item  \textbf{$\mathbf{\U3}$ stack:} In the  Calabi-Yau, the locus $\{w_3=0\}$ splits into the following components: 
 \begin{equation}\label{WWtop}
W\,\equiv\,\,\,\, \left\{ w_3=0, \,\,\, \xi-\frac12 s_6 = 0 \right\} \qquad\mbox{and}\qquad
\tilde W \,\equiv\,\,\,\, \left\{ w_3=0, \,\,\, \xi+\frac12 s_6 = 0 \right\} \:.
\end{equation}
The $\U3$ symmetry results from wrapping three D7 branes on each of 
these divisors. Since $W$ and $\tilde W$ are in different homology classes, the $\U1\subset \U3$ symmetry  is geometrically massive. 
\item \textbf{\SU2 stack:} There are two D7-branes wrapping the invariant irreducible divisor $U\equiv \{w_2=0\}$. The two branes are image to each other and support an $Sp(1)\cong SU(2)$ gauge symmetry. Once again, the diagonal $\U1$ is projected out by the orienfold action, but there remains the possibility of having an orientifold-odd gauge flux with along $U$. 
\end{itemize}
In addition we have again the three $\U1$ divisors inherited from $F_5$, which however suffer from slight modifications in the presence of the top.
\begin{itemize}
 \item  \textbf{$\mathbf{\U1_1}$ stack:} Over the Calabi-Yau, the locus $\{s_7=0\}$ splits into two components: 
 \begin{equation}\label{XXtop}
X\,\equiv\,\,\,\, \left\{ s_7=0, \,\,\, \xi-\frac12 s_6 = 0 \right\} \qquad\mbox{and}\qquad
\tilde X \,\equiv\,\,\,\, \left\{ s_7=0, \,\,\, \xi+\frac12 s_6 = 0 \right\} \:.
\end{equation}
\item  \textbf{$\mathbf{\U1_2}$ stack:} Let us now consider the following factor in Eq.  \eqref{eq:deltaE3},
\begin{align}\label{eq:disc2top}
\Delta_E^{\rm rem,1}=(s_3s_5-s_1s_7w_3)^2-(s_3s_6-s_2s_7w_3)(s_2s_5-s_1s_6)
\end{align}
The ideal generated by $\Delta_E^{\rm rem,1}$ together with the Calabi-Yau equation decomposes into two prime, non complete intersection ideals, one of which is given by the following expression
\begin{eqnarray}\label{eq:Ytop}
Y\, &\equiv &\left\{ \left(\xi - \frac{s_6}{2}\right) (s_3s_6-s_2s_7 w_3) + s_7 w_3 (s_3 s_5 - s_1 s_7 w_3)\,, \right.\\
&&\quad \left( \xi+\frac{s_6}{2}\right) (s_2s_5 - s_1s_6)  - s_5 (s_3 s_5 - s_1 s_7 w_3)\,,\\
 &&\quad \quad \left. \left( \xi-\frac{s_6}{2}\right) (s_3s_5 - s_1s_7 w_3) + s_7 w_3 (s_2s_5 - s_1s_6) \,,{\rm Eq. \eqref{eq:XP5top}} \right\}\,,
 \end{eqnarray}
while its image $\tilde Y=\sigma^* Y$ is obtained by changing $\xi\mapsto -\xi$ in Eq. \eqref{eq:Ytop}.

\item  \textbf{$\mathbf{\U1_3}$ stack:} The remaining locus to be analyzed reads
\begin{align}\label{eq:disc3top} 
\Delta_E^{\rm rem,2}=&s_7 w_3 s_8^2 - s_6 s_8 s_9 + s_5 s_9^2 \:,
\end{align}
that in the Calabi-Yau splits into two components:  
\begin{align}
Z\,\equiv\,\,\,\, &\{2 s_7 w_3 s_8 - s_6 s_9 - 2 s_9 \xi, s_6 s_8 - 2 s_5 s_9 - 2 s_8 \xi,{\rm Eq \eqref{eq:XP5top}}\}\,,\\
\tilde Z\,\equiv\,\,\,\, &\{2 s_7 w_3 s_8 - s_6 s_9+ 2 s_9 \xi, s_6 s_8 - 2 s_5 s_9+ 2 s_8 \xi,{\rm Eq \eqref{eq:XP5top}}\}\,.
\end{align}
\end{itemize}
Note that in all cases the discriminant components split into divisors in the Calabi-Yau which are in different homology classes. Therefore they receive a geometrical mass and therefore, any massless \U1 is going to be a linear combination of these. In particular, the massless generators are 
\begin{equation}
\label{QmasslessF5top} Q_1=\tfrac12(\tfrac13 Q_W+Q_X+Q_Y+Q_Z) \qquad\mbox{and}\qquad Q_2=-\tfrac13 Q_W-Q_X+Q_Z\,,
\end{equation}
in terms of the massive \U1 generators $Q_W,Q_X,Q_Y,Q_Z$.

For this case as well, one can check that some homology relations are satisfied. First note that one can multiply Eq. \eqref{eq:disc3top} by $s_7 w_3$ to obtain 
\begin{equation}\label{eq:split1}
s_7 w_3 \Delta_E^{\rm rem,1}|_{X_3}=\left(s_7 w_3 s_8 - s_9 \left(  \xi + \frac{s_6}{2} \right)\right)\left( s_7 w_3 s_8 + s_9 \left(  \xi - \frac{s_6}{2} \right) \right)\,.
\end{equation}
note that both of these factors must be in the class $4D_{O7}-(W+\tilde W)-U-\tfrac12 (Y+\tilde Y)$, and that the first is in the class $Z+\tilde X+ \tilde W$, from which we can deduce the following relation
\beq
Z=4D_{O7}-(W+2\tilde W)-U-\tilde X-\frac12(Y+\tilde Y)\,, \quad \tilde Z= \sigma^\ast Z \,, 
\eeq
in agreement with the D7 tadpole $8D_{O7}=3W_++2U+X_++Y_++Z_+$, where we have written everything in terms of orientifold even and odd divisors. 

In a similar fashion we can multiply $s_7^2 w_3^2$ with $\Delta_E^{\rm rem,1}$ in order to deduce
\begin{eqnarray}
s_7^2 w_3^2\Delta_E^{\rm rem,1}|_{X_3}\,
    &=&\left(\left(\xi - \frac{s_6}{2}\right) (s_3s_6-s_2s_7 w_3) + s_7 w_3 (s_3 s_5 - s_1 s_7 w_3)\right) \nonumber\\
        &&\times\left(\left(\xi + \frac{s_6}{2}\right) (s_3s_6-s_2s_7 w_3) - s_7 w_3 (s_3 s_5 - s_1 s_7 w_3)\right)\:.
\end{eqnarray}
Since both factors are equal in homolgy and the first is in the class $Y+2X+2W$, while the second is its orientifold image, we can derive the following relations among the odd divisors:
\beq
Z_-=-\frac{Y_-}{2}=X_-+W_-\,.
\eeq
Hence we can choose $W_-$ and $X_-$ as generators of $H^{1,1}_-(X_3)$. Couplings between the $\U1$ gauge fields to the corresponding two odd axions will give a mass to two out of the four $\U1$ directions in type IIB. 

The absence of the conifold points imposes certain restrictions on the brane  intersections: 
\beq
X \tilde X=D_{O7} X= D_{O7} \tilde{X}\,,\quad W \tilde W=D_{O7} W= D_{O7} \tilde{W}\,,\quad X\tilde W=\tilde{X} W=0\,,
\eeq
which written in terms of orientifold odd and even classes, translate into
\begin{align}
\begin{split}
X_+^2-X_-^2=D_{O7} X_+\,,\quad W_+^2-W_-^2&=D_{O7} W_+\,,\quad  D_{O7} X_-=D_{O7} W_-=0\,,\\
\quad X_+ W_+=X_- W_-\,,&\quad X_-W_+=X_+W_-\,.
\end{split}
\end{align}
Finally it is convenient to remark some relations between divisors in the base and divisors in the Calabi-Yau threefold:
\begin{align}
\begin{split}
\pi^*(\bar{K}_B)=D_{O7}\,,\quad \pi^*(\cS_7)=X_++W_+\,,&\quad \pi^*(\cS_9)=3D_{O7}-U-W_+-\tfrac12 Y_+\,,\\
\quad  \pi^*(\cW_3)=W_+\,,&\quad \pi^*(\cW_2)=U\,.
\end{split}
\end{align}

\subsubsection{Charged matter}
There is a rich spectrum of matter fields living at the intersections of the divisors introduced above. All matter fields in the type IIB theory are going to be characterized by four $\U1$ charges in addition to their corresponding representation under the non-Abelian gauge symmetries. 
In the non-Abelian matter spectrum one finds (where once again the upper indices give the charges under the massless $\U1$ generators):\\ \\
\textbf{One bifundamental:}
\begin{itemize}
\item $W U\,$: $(\three,\two)_{(1,0,0,0)}^{(\frac16,-\frac13)}$. This state sits at the vanishing of the ideal 
\beq
\{w_3,\xi-\frac{s_6}{2},w_2\}\,.
\eeq
\end{itemize}
\textbf{Six triplets:}
\begin{itemize}
\item $ \tilde W X\,$: $\cancel{(\overline{\three},\one)_{(-1,-1,0,0)}}^{(-\frac23,\frac43)}$. The would-be triplet is localized at the vanishing of the following ideal
\beq
\{\xi,s_7,w_3,s_6\}\,.
\eeq
which is one of the forbidden conifold points. Due to that, this field is absent from the matter spectrum.
\item $\tilde W \tilde X\,$: $(\overline{\three},\one)_{(-1,1,0,0)}^{(\frac13,-\frac23)}$. The state is located at the vanishing of 
\beq
\{w_3,s_7,\xi+\frac{s_6}{2}\}\,.
\eeq
\item $\tilde W Y\,$: $(\overline{\three},\one)_{(-1,0,1,0)}^{(\frac13,\frac13)}$. Taking the union of the ideals for $\tilde W$ and $Y$ (i.e. intersection of the corresponding vanishing loci) one finds that it decomposes into two prime ideals, the first being the conifold point $\{w_3,\xi,s_5,s_6\}$ and the other being 
\beq
\{w_3,\xi+\frac{s_6}{2},s_3\}\,.
\eeq
\item $\tilde W \tilde Y\,$: $(\overline{\three},\one)_{(-1,0,-1,0)}^{(-\frac23,\frac13)}$. This state sits at the vanishing of the ideal 
\beq
\{w_3,\xi+\frac{s_6}{2},s_3 s_5^2-s_2 s_5 s_6+s_1 s_6^2\}\,.
\eeq
\item $\tilde W Z\,$: $(\overline{\three},\one)_{(-1,0,0,1)}^{(\frac13,\frac43)}$. This state sits at the vanishing of the ideal 
\beq
\{w_3,\xi+\frac{s_6}{2},-s_6 s_8+s_5 s_9\}\,.
\eeq
\item $\tilde W \tilde Z\,$: $(\overline{\three},\one)_{(-1,0,0,-1)}^{(-\frac23,-\frac23)}$. From the union of the ideals for $\tilde W$ and $\tilde Z$, one obtains to prime ideals, one corresponding again to the conifold point $\{w_3,\xi,s_5,s_6\}$ and the other being
\beq
\{w_3,\xi+\frac{s_6}{2},s_9\}\,.
\eeq
\item $W \tilde W\,$: $(\overline{\three},\one)_{(2,0,0,0)}^{(\frac13,-\frac23)}$. This state is the antisymmetric representation of $\U3$ and sits on top of the orientifold plane: 
\beq
\{w_3, s_6,\xi\}\,.
\eeq
\end{itemize}
\textbf{Three doublets:}
\begin{itemize}
\item $U \tilde X\,$: $(\one,\two)_{(0,1,0,0)}^{(-\frac12,1)}$. This field is localized at the vanishing of the following ideal
\beq
\{s_7,w_2,\xi+\frac{s_6}{2}\}\,.
\eeq
\item $U Y\,$: $(\one,\two)_{(0,0,1,0)}^{(\frac12,0)}$. This state lives at the union of the corresponding ideals for $U$ and $\tilde{Y}$ which is prime.
\item $U Z\,$: $(\one,\two)_{(0,0,0,1)}^{(\frac12,1)}$. This field lives at the union of the generating ideals for  $U$ and $\tilde{Z}$ and it is also prime.
\end{itemize}
\textbf{Eight singlets:}
\begin{itemize}
\item $X \tilde X\,$: $\cancel{\one_{(0,2,0,0)}^{(1,-2)}}$. The would-be singlet is localized at the vanishing of the following ideal
\beq
\{\xi,s_7,s_6\}\,.
\eeq
However, this coincides with the $O7$ plane, where no symmetric matter is allowed. Due to that, this field is not part of the spectrum. 
\item $\tilde X Y\,$: $\one_{(0,1,1,0)}^{(1,-1)}$. This state sits at the vanishing of the ideal 
\beq
\{s_7,\xi+\frac{s_6}{2},s_3\}\,.
\eeq
\item $\tilde X Z\,$: $\one_{(0,1,0,1)}^{(1,0)}$. The singlet is located at the vanishing of 
\beq
\{s_7,\xi+\frac{s_6}{2},s_5 s_9-s_6s_8 \}\,.
\eeq
\item $X Y\,$: $\one_{(0,-1,1,0)}^{(0,1)}$. This state sits at the vanishing of the ideal 
\beq
\{s_7,\xi-\frac{s_6}{2},s_3 s_5^2-
s_2 s_5s_6 +s_1 s_6^2\}\,.
\eeq
\item $X Z\,$: $\one_{(0,-1,0,1)}^{(0,2)}$. This state sits at the vanishing of the ideal 
\beq
\{s_7,\xi-\frac{s_6}{2},s_9\}\,.
\eeq
\item $Y\tilde Y\,$: $\one_{(0,0,2,0)}^{(1,0)}$. This state sits at the vanishing of the ideal 
\beq
\{s_3s_5-s_1s_7, s_2s_5-s_1s_6, s_3s_6-s_2s_7, {\rm Eq. \eqref{eq:XP5top}}\}\,.
\eeq
\item $Y Z\,$: $\one_{(0,0,1,1)}^{(1,1)}$. This state sits at the vanishing locus of the union of the ideals for $Y$ and $Z$, which occurs to be prime. 
\item $\tilde Y Z\,$: $\one_{(0,0,-1,1)}^{(0,1)}$. This state sits at the vanishing of the ideal: 
\begin{align}
\begin{split}
\{&2 s_7 s_8-s_6 s_9+2 s_9 \xi, s_6 s_8 -2 s_5 s_9+2 s_8 \xi,\,
2s_3s_6s_8+2\xi s_2s_9-2 s_3 s_5 s_9-s_2 s_6 s_9+2 s_1 s_7 s_9, \\
& s_3 s_8^2-s_2 s_8 s_9+s_1 s_9^2, 2 s_3 s_5 s_8-2 s_2 s_5 s_9+s_1 s_6 s_9+2 \xi s_1 s_9 ,\,
2\xi s_3s_6-s_3s_6^2-2\xi s_2s_7+2 s_3 s_5 s_7+\\
&+s_2 s_6 s_7-2 s_1 s_7^2, 2\xi s_3s_5-s_3 s_5 s_6 -2\xi s_1 s_7+2s_2s_5s_7-s_1 s_6 s_7,\\
& 2\xi  s_2s_5 - 2 s_3 s_5^2 - 2\xi s_1s_6+s_2 s_5 s_6 -s_1 s_6^2 + 2 s_1 s_5 s_7,\, {\rm Eq. \eqref{eq:XP5top}}\}\,.
\end{split}
\end{align}
\item $Z \tilde{Z}\,$: $\one_{(0,0,0,2)}^{(1,2)}$. The ideal associated to this state is given by 
\beq
\{s_8,s_9,{\rm Eq. \eqref{eq:XP5top}}\}\,.
\eeq
\end{itemize}

Note that, additionally to the recombination of singlets, the matter curves associated with the states $(\overline{\three},\one)_{(2,0,0,0)}^{(\frac13,-\frac23)}$ and $(\overline{\three},\one)_{(-1,1,0,0)}^{(\frac13,-\frac23)}$ recombine to the matter curve  $(\overline{\three},\one)_{(\frac13,-\frac23)}$ in F-theory. In fact the locus of the last one is $\{w_3,-s_6 s_7+s_3 s_9 w_2\}$: the second equation becomes simply $s_6 s_7$ in the weak coupling limit.

\subsubsection{Fluxes and chiralities}

Next we have to find all gauge flux directions consistent with the D5-tadpole which reads
\begin{align}
0\,=\,&W_-(3F^W_+-2F_+^Y+F^Z_+)+X_-(F^X_+-2F_+^Y+F^Z_+)\\
&+3W_+(F^W_--F^Z_-)+U(F^U_--F^Z_-)+X_+(F^X_--F^Z_-)+Y_+(F^Y_--F^Z_-)\,. 
\end{align}
Once again we distinguish between three different types of flux solutions: 
\begin{itemize}
\item Even fluxes are
\begin{align}\label{eq:evenF5top1}
(F^W_+,F^X_+,F^Y_+,F^Z_+ )_1\,=\,&\tfrac12(\tfrac13,1,1,1)F_1\,,\\ \label{eq:evenF5top2}
(F^W_+,F^X_+,F^Y_+,F^Z_+)_2\,=\,&(-\tfrac13,-1,0,1)F_2\,,\\ 
( F^W_+,F^X_+,F^Y_+,F^Z_+)_{\lambda_1}=\,&\lambda_1(D_{O7},0,0,0)\,,\\
( F^W_+,F^X_+,F^Y_+,F^Z_+)_{\lambda_2}=\,&\lambda_2(0,D_{O7},0,0)\,. 
\end{align}
where the first two are identified with the massless $\U1$ directions. The two forms $F_1$ and $F_2$ belong to $H^{(1,1)}_+(X_3)$. The coefficients $\lambda_1$ and $\lambda_2$ are rational parameters in agreement with flux quantization. 
\item Further we have some mixed directions
\begin{align}\label{eq:mixedF5top}
\begin{split}
(F^W_+,F^U_-)_{\alpha_1}=\alpha_1 (\tfrac13 U,-\tfrac12 	W_-)\,,\quad&\quad (F^W_+,F^X_-)_{\alpha_2}=\alpha_2 (\tfrac13 X_+,- 	W_-)\,,\\
(F^W_+,F^Y_-)_{\alpha_3}=\alpha_3 (\tfrac13Y_+,-W_-)\,,\quad&\quad (F^X_+,F^U_-)_{\alpha_4}=\alpha_4 (U,-\tfrac12 X_-)\,,\\
(F^X_+,F^Y_-)_{\alpha_5}=&\alpha_5 (Y_+,- X_-)\,,
\end{split}
\end{align}
where again $\alpha_i$, $i=1,\ldots,5$ are rational numbers compatible with flux quantization. There are three additional flux directions 
\begin{align}
(F^W_+,F^W_-)\sim (W_+,-W_-)\,,\quad (F^X_+,F^X_-)\sim ( X_+,- 	X_-)\,,\quad (F^X_+,F^X_-)\sim (W_+,-W_-)\,,
\end{align}
which are consistent with the D5-tadpole. In analogy with the observations made in previous sections, we can see that the first two are equivalent to the $\lambda_i$ fluxes, while the third is trivial owed to the fact that $X\tilde W=0$. 
\item For the orientifold odd sector we get the usual uniform distribution of odd fluxes over all brane divisors $F^W_-=F^U_-=F^X_- =F^Y_-=F^Z_-$ which is irrelevant for chirality, D3-tadpoles and FI computations. In addition to that there is a component which could be of relevance, namely
\beq\label{eq:oddF5top}
(F^W_-,F^X_-)\sim(\tfrac13 X_-,-W_-)\,.
\eeq
However this purely odd flux is equivalent to the $\alpha_2$ flux in Eq.~\eqref{eq:mixedF5top}. 
\end{itemize}
Therefore, we obtain nine flux directions, controlled by the two-forms $F_1,F_2$ and the seven parameters $\lambda_1,\lambda_2,\alpha_1,...,\alpha_5$.

The chiralities for the type IIB matter spectrum have been summarized in Appendix~\ref{app:ap1}.

\subsubsection*{Type IIB fluxes vs F-theory $G_4$-flux}

We work out now the match between F-theory and type IIB fluxes.
The fluxes $G_4^{\mathcal{F}_1}$ and $G_4^{\mathcal{F}_2}$ match straightforwardly with the $F_1$ and $F_2$ fluxes defined in Eq. \eqref{eq:evenF5top1} and \eqref{eq:evenF5top2}, when we set $F_1=\pi^{*}(\mathcal{F}_1)$ and $F_2=\pi^{*}(\mathcal{F}_2)$. In that case the F-theoretic as well as the type IIB contributions to chiralities, D3-tadpoles and FI-terms coincide.  

For the remaining F-theory fluxes one has to find a combination of type IIB flux directions which reproduce the effects of the $\Lambda$-flux as well as the four $A_i$-fluxes. We obtain the following match 
\begin{itemize}
\item $G_4^\Lambda$:
\begin{align}
\begin{split}
 F_1=0\,,\quad F_2=-\left(2 D_{O7}- U- \frac32 W_+  - \frac12 X_+ - \frac12 Y_+\right)\Lambda\,, \\ 
\lambda_1=\alpha_2=0\,, \qquad\lambda_2=-2\Lambda\,,\quad \alpha_1=\alpha_4 =2\alpha_3=2\alpha_5=\Lambda\,,
\end{split}
\end{align}
\item $G_4^{A1}$: 
\begin{align}
F_1=0\,,\quad F_2=- U A_1\,,\quad \alpha_1=-A_1\,,\quad \lambda_1=\lambda_2=\alpha_2=\alpha_3=\alpha_4=\alpha_5=0\,,
\end{align}
\item $G_4^{A2}$: 
\begin{align}
F_1=\frac12 U A_2\,,\quad F_2= 0\,,\quad \alpha_4=-A_2\,,\quad \lambda_1=\lambda_2=\alpha_1=\alpha_2=\alpha_3=\alpha_5=0\,,
\end{align}
\item $G_4^{A3}$: 
\begin{align}
F_1=F_2=- \frac12 W_+ A_3\,,\quad \alpha_1=2\alpha_3=A_3\,,\quad \lambda_1=\lambda_2=\alpha_2=\alpha_4=\alpha_5=0\,,
\end{align}
\item $G_4^{A4}$: 
\begin{align}
F_1=\frac12 W_+ A_4\,,\quad F_2=0 \,,\quad \lambda_1=-\frac23 A_4\,,\quad \lambda_2=\alpha_1=\alpha_2=\alpha_3=\alpha_4=\alpha_5=0\,.
\end{align}
\end{itemize}
Again we see that there are two flux directions in the type IIB model which can not be found in the vertical cohomology of the Calabi-Yau fourfold: the first is the $\alpha_2$-flux, which as discussed around Eq. \eqref{eq:oddF5top} can be reinterpreted as a fully orientifold-odd flux direction; the second is a combination of $\lambda_2$- and $\alpha_5$-fluxes with $\alpha_5=4\lambda_2$.

\newpage

\section{Charge 3 states, discrete symmetries and massive \U1's} 
\label{sec:F3F1}
In this section we would like to discuss some additional models which exhibit some interesting properties from the F-theory point of view. The first one is a toric hypersurface fibration based on the toric ambient space $\mathbb{P}_{F_3}$. The resulting model 
has a $\U{1}$ gauge symmetry with three singlets, one of which has charge $q=3$ under the $\U1$ \cite{Klevers:2014bqa}. This model is appealing for various reasons: Charges higher than two might seem exotic from the perturbative point of view. Additionally, this model can be obtained as the Higgsed version of an \SU{2} theory in which the charge three state originates from a decomposition of a three index symmetric representation of $\SU{2}$ \cite{Klevers:2016jsz}. The latter is a truly exotic matter representation, only conceivable from the point of view of intersecting $[p,q]$ 7-branes with multi-pronged strings. 

The second model corresponds to a Higgsed version of the $F_3$ fibration, and has only a $\mathbb{Z}_3$ discrete symmetry. This model is also a toric hypersurface fibration based on the toric space $\mathbb{P}_{F_1}$ \cite{Klevers:2014bqa,Cvetic:2015moa}. 

\subsection{A \U{1} model with charge three singlet}
\begin{figure}[H]
\centering
\begin{minipage}{.56\textwidth}
  \centering
  \includegraphics[scale=.4]{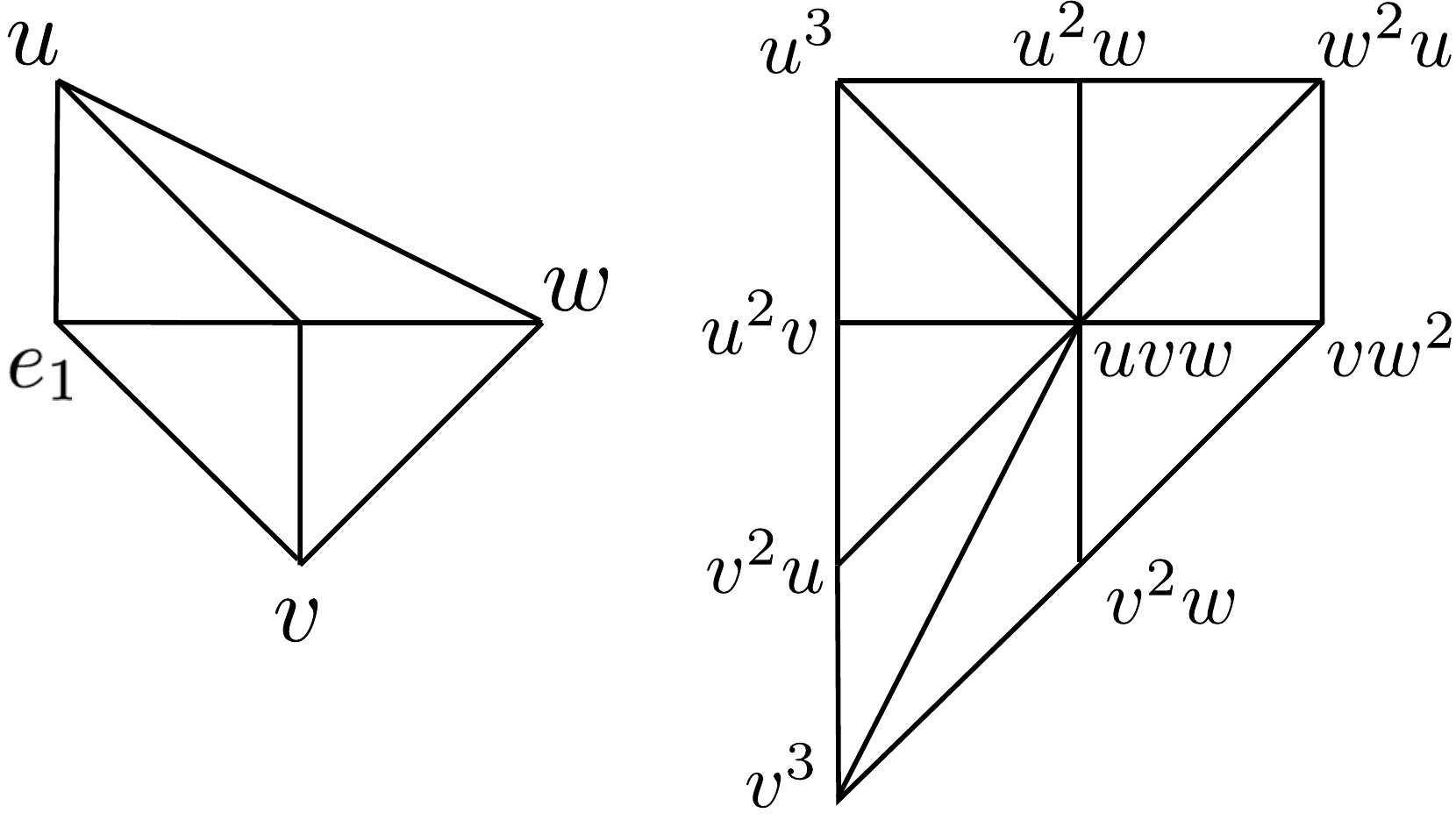}
\end{minipage}%
\begin{minipage}{.44\textwidth}
{\footnotesize
  \begin{tabular}{|c|c|}\hline
Section & Line Bundle\\ \hline
$u$ & $\mathcal{O}(H-E_1+\cS_9+K_B)$ \\ \hline
$v$ & $\mathcal{O}(H-E_1+\cS_9-\cS_7)$\\ \hline
$w$ & $\mathcal{O}(H)$\\ \hline
$e_1$ & $\mathcal{O}(E_1)$\\ \hline
\end{tabular}}
\end{minipage}
\caption{\label{fig:F3_toric} The polytope $F_{3}$ and its dual. The table on the right provides the line bundle classes for the coordinates in $\mathbb{P}_{F_{3}}$.}
\end{figure}

The toric ambient space $\mathbb{P}_{F_3}=dP_1$ is shown in Fig. \ref{fig:F3_toric}.
In addition to the sections $s_i$ that we have in the $\mathbb{P}_{F_5}$ fibration of Section~\ref{sec:F5}, here we have to introduce the additional section $s_4$ of the line bundle $\mathcal{O}(2\cS_7-\cS_9)$. The fiber is cut by the following cubic polynomial
\beq \label{eq:PF3}
	p_{F_3}=s_1 u^3e_1^2 + s_2 u^2 ve_1^2  + s_3 u v^2 e_1^2  +  s_4 v^3e_1^2 +
  s_5 u^2 w e_1
+  s_6 u v w e_1 +  s_7 v^2 w e_1 + s_8 u w^2 + s_9 v w^2 \, .
\eeq
After mapping $p_{F_3}$ to the Weierstrass form we obtain $f$, $g$ and $\Delta$, which we take from ref. \cite{Klevers:2014bqa} and summarize in Appendix \ref{app:ap2} for completeness. 

In Eq. \eqref{eq:PF3} one immediately recognizes a rational section at $e_1=0$, which in the (birationally equivalent) Weierstrass model becomes the zero section $S_0$ of the elliptic fibration. 
Additionally, as discussed in \cite{Klevers:2014bqa} there is an extra non toric section, whose coordinates $S_1\,:\, [\x_1:y_1:z_1]$ in the Weierstrass form can be found in Appendix \ref{app:ap2}. This produces a massless \U1 gauge symmetry in the four dimensional effective theory. Its corresponding generator is given by the Shioda map
\beq
\sigma=S_1-S_0-3\bar{K}_B+\cS_7-2 \cS_9\,.
\eeq

There are no codimension-one singularities. At codimension-two one finds three $I_2$ fibers corresponding to singlets charged under the $\U{1}$ symmetry. The loci for the corresponding singlets are given in Table~\ref{tab:F3_matter}. 
\begin{table}[ht!]
\begin{center}
\footnotesize
\renewcommand{\arraystretch}{1.4}
\begin{tabular}{|c|c|}
\hline
 Representation&  Locus  \\ \hline
$\one_{3}$ &  $V(I_{(3)}):=\{s_8 = s_9 = 0\}$   \\ \hline

$\one_{2}$ &  $
 V(I_{(2)}):=
 \{s_4 s_8^3- s_3 s_8^2 s_9 + s_2 s_8 s_9^2 -s_1 s_9^3
 =s_7 s_8^2  + s_5 s_9^2\!-\! s_6 s_8 s_9 =0\}\backslash\ V(I_{(3)})
 $
 \\ \hline
$\one_{1}$ & $\begin{array}{c}
V(I_{(1)}):=\{y_{1}=f z_1^4+3 x_1^2=0\}\backslash\ (V(I_{(2)})\cup V(I_{(3)}))
\end{array}$ \\ \hline
\end{tabular}
\caption{\label{tab:F3_matter} The loci for the charged matter representations under the  $\U1$ symmetry. The charges are written as subscripts. The locus for the singlet $\one_{1}$ is given in terms of the sections $x_1$, $y_1$ and $z_1$ given in Appendix~\ref{app:ap2}. }
\end{center}
\end{table}

\

Let us now discuss the type IIB limit of this model, which we can reach upon setting the following $\epsilon$ scalings for the sections $s_i$:
\beq
 s_1 \rightarrow \epsilon^1 s_1, \quad s_5 \rightarrow \epsilon^1 s_5, \quad s_8 \rightarrow \epsilon^1 s_8, \quad s_i \rightarrow \epsilon^0 s_i \quad {\rm (}i\neq 1,5,8{\rm )}\:.
\eeq
The location of the D7-branes can be read out from the irreducible components of $\Delta_E=0$ with 
\begin{align}
\begin{split}
\Delta_E=-\frac14 s_9&\, (-s_3 s_6 s_7 + s_2 s_7^2 + s_3^2 s_9 + 
   s_4 (s_6^2-4s_2 s_9))\\
  &\times (s_2^2 s_8^2 + 
   s_2 (-s_5 s_6 s_8 + s_5^2 s_9 - 2 s_1 s_8 s_9) + 
   s_1 (s_6^2 s_8 - s_5 s_6 s_9 + s_1 s_9^2))\,.
\end{split}
\end{align}
We 
encounter three factors, two of which split in the Calabi-Yau threefold 
$$\xi^2=\frac{s_6^2}{4}-s_2 s_9\:,$$ 
while the one in the middle gives and orientifold invariant D7-brane. The brane configuration looks schematically as depicted on the right hand side of Figure \ref{fig:F1F3}. 

In the Calabi-Yau treefold the locus $s_9=0$ splits into the following components:
\begin{equation}\label{XXF3}
X\,\equiv\,\,\,\, \left\{ s_9=0, \,\,\, \xi-\frac12 s_6 = 0 \right\} \qquad\mbox{and}\qquad
\tilde X \,\equiv\,\,\,\, \left\{ s_9=0, \,\,\, \xi+\frac12 s_6 = 0 \right\} \:.
\end{equation}
The invariant brane is given by:
\beq
{\rm W}= \{-s_3 s_6 s_7 + s_2 s_7^2 + s_3^2 s_9 + 
   s_4 (s_6^2-4s_2 s_9)=0 \}\cap X_3\,.
\eeq
Finally we have the remaining locus. Here we do not write explicitly the ideals of the two components. We instead compute their homology classes. We start 
by noticing that
\begin{align}\label{eq:discF3}
\begin{split}
s_9^2\Delta^{\rm rem}_E=&\left(s_8 \left(\frac{s_6}{2}-\xi\right)-s_9\left(s_1 s_9+s_5\left(\frac{s_6}{2}-\xi\right)^2\right)\right)\\
&\times\left(s_8 \left(\frac{s_6}{2}+\xi\right)-s_9\left(s_1 s_9+s_5\left(\frac{s_6}{2}+\xi\right)^2\right)\right)\,.
\end{split}
\end{align}
The class of each factor is $\tfrac12(8D_{O7}-{\rm W}-(X+\tilde{X}))$, while the class of the product of the two must be $Y+\tilde Y+2(X+\tilde X)$, where $Y$ and $\tilde Y$ correspond to the split divisors for $\Delta^{\rm rem}_E$ in the Calabi-Yau threefold. Identifying the first factor with $Y+2X$ we obtain the following homology relation: 
\beq\label{YF1F3eq}
Y=\frac12 \left(8D_{O7}-{\rm W}-(3X-\tilde X)\right)\,.
\eeq
We see that $Y_-=-2X_-$. 
Both the splitting loci are made up of a brane and its image in different homology classes. Hence the corresponding gauge bosons are massive. The massless combination has the following generator:
\beq
Q=2Q_X+Q_Y\,.
\eeq
The matter representations living at the brane intersections are summarized in table \ref{tab:F3IIBmatter}. We also include the corresponding curves in F-theory. We notice once more that curves that have the same charges under the massless \U1 
merge away from the weak coupling limit ($\epsilon\neq 0$). Moreover we see that the charge 3 state arises also in perturbative type IIB theory.
\begin{table}[H]
\begin{center}
\footnotesize
\renewcommand{\arraystretch}{1.4}
\begin{tabular}{|c|c|c|c|c|c|c|}
\hline
\multirow{ 2}{*}{Type IIB} & $X \tilde X$ & $X\tilde Y$ & ${\rm W}X$ & $\tilde Y Y$ & $X  Y$ & $Y{\rm W}$ \\ \cline{2-7}
& $\cancel{\one_{(2,0)}}$ & $\one_{(1,1)}$ & $\one_{(-1,0)}$ & $\one_{(0,-2)}$ & $\one_{(1,-1)}$ & $\one_{(0,1)}$  \\ \hline\hline
F-theory & - & $\one_{3}$ & \multicolumn{2}{c|}{$\one_{-2}$} & \multicolumn{2}{c|}{$\one_{1}$}\\
\hline
\end{tabular}
\caption{\label{tab:F3IIBmatter} The matter arising at the different brane intersections and their recombination pattern in the F-theory model. Note that the singlet $\one_{(2,0)}$ is not present because the divisors  $X$ and $\tilde X$ only intersect on top of the orientifold plane.}
\end{center}
\end{table}

Let us finish with an observation. By tuning the complex structure moduli of the F-theory model under consideration, one can un-Higgs the massless \U1 symmetry to an \SU2 with exotic three-index symmetric representation matter \cite{Klevers:2016jsz}. One could wonder what happens in the weak coupling limit. In fact one realizes that there is no $\epsilon$ scaling that does not destroy the spectrum.\footnote{The complex structure must tune such that $s_5 =s_8 \sigma_5$, $s_6 =s_8 \sigma_7 +s_9 \sigma_5$ , $s_7 =s_9 \sigma_7$ \cite{Klevers:2016jsz}. To have the proper weak coupling scaling, one imposes $s_8\rightarrow \epsilon s_8$. But this would also require in particular $s_6\rightarrow s_9\sigma_5$ that makes the CY threefold have an $A_1$ singularity over a curve. Moreover the invariant $W$ brane would factorize giving an extra brane wrapping the $s_9=0$ locus.} This is actually expected, as in perturbative type IIB there are no states in the three-index symmetric representations.
\subsubsection*{Fluxes}
In F-theory, the generic vertical $G_4$ flux is
\beq
G_4=\sigma\wedge \mathcal{F}+\Lambda (S_0^2+\bar{K}_B(S_0+\cS_9)-\cS_7 \cS_9+\cS_9^2)\,, 
\eeq
with $\mathcal{F}$ a vertical divisor.

On the type IIB side, the D5-tadpole cancellation conditions involving the gauge fluxes on the D7-branes~is
\beq
X_-( F^X_+-2F^Y_+)+X_+( F^X_--F^Y_-)+{\rm W}(F^{\rm W}_--F^Y_-)=0\,,
\eeq
from which we obtain the following flux directions
\beq
( F^X_+,F^Y_+)=(2,1)F\,,\quad ( F^X_+,F^Y_+)_\lambda=\lambda(D_{O7},0)\,,\quad ( F^X_+,F^{\rm W}_-)_\beta=\beta({\rm W},-X_-)\,,
\eeq
in addition to the odd component $ F^X_-=F^Y_-=F^{\rm W}_-$ corresponding to a choice of $B$-field.

The matching of the flux components in type IIB with the ones in F-theory proceeds as follows. The massless $\U1$ gauge fluxes match under the condition $F=\pi^*(\mathcal{F})$. As for the $\Lambda$-flux we see once more that it is matched by a linear combination of $F$-, $\lambda$- and $\beta$-fluxes: 
\beq
F=\frac12 \Lambda X_+\,,\quad \lambda=\Lambda\,, \quad \beta=-\frac12 \Lambda\,.
\eeq
We note again that in the type IIB limit we have one additional flux direction in comparison to the F-theory uplift.

\subsection{A model with $\mathbb{Z}_3$ discrete gauge symmetry}

Let us start from the F-theory \U1 model just described and let us give a vacuum expectation value (VEV) to the charge 3 state. We expect to break the massless \U1 symmetry to the $\mathbb{Z}_3$ subgroup that preserve this VEV. Under this discrete group, the other two states have the same charge, since $-2=1$ mod $3$.

Geometrically, this VEV can be viewed as 
a complex structure deformation of the Weierstrass model. The same space can be described by 
cutting the fiber out of the toric ambient space $\mathbb{P}_{F_1}=\mathbb{P}^2$ instead of $dP_1$, so that in comparison with Fig. \ref{fig:F3_toric} we do not have the divisor $E_1$ (the zero section). The deformation can be read in the extra monomial in the fiber equation (i.e. an extra node in the dual polytope) which we denote by $s_{10}$ and which is a section of $\mathcal{O}(2\cS_9-\cS_7)$. The corresponding fiber equation reads
\beq \label{eq:pF1}
p_{F_{1}} = s_1 u^3 + s_2 u^2 v+ s_3 u v^2 + s_4 v^3 + s_5 u^2 w + s_6 u v w + s_7 v^2 w + s_8 u w^2+s_9 v w^2 +s_{10} w^3\,.
\eeq
This equation does not define an elliptic, but a genus-one fibration \cite{Braun:2014oya,Morrison:2014era}
given that it does not have sections but merely three-sections, out of which we can choose $S_{(3)}=\{u=0\}\cap Y_4$ as a representative.

In order to get the Weierstrass form for this model one take its Jacobian\footnote{That is the group of degree zero line bundles over $[p_{F_1}]$.} $J([p_{F_1}])$, which does exhibit a zero section corresponding to the trivial line bundle. The final expressions for $f$ and $g$ for this model have been presented in Appendix~\ref{app:ap2}. 
One can show that at codimension-one there are no singularities in the fourfold. Moreover there are no extra sections in the Weierstrass model. Hence, as we expected, there are neither non-Abelian nor Abelian continuous gauge symmetries in this model. Instead, the presence of the three-section together with the fact that the model can be understood in terms of a Higgs mechanism in which a charge three singlet picks a VEV
are supporting evidence for the presence of a discrete $\mathbb{Z}_3$ gauge symmetry\footnote{More formally, the presence of the discrete symmetry can infered explicitly after finding the Tate-Shafarevich goup of the Jacobian fibration $J([p_{F_1}])$ \cite{Cvetic:2015moa}.} \cite{Cvetic:2015moa}.

As explained in \cite{Klevers:2014bqa}, looking back at Eq. \eqref{eq:pF1} one finds that an 
$I_2$ 
fiber develops whenever $p_{F_{1}}$ factorizes as 
\beq \label{eq:cod2}
p_{F_{1}} = s_1 (u+\alpha_1 v+\alpha_2 w)( u^2+ \beta_1 v^2 + \beta_2 w^2+ \beta_3 uv+ \beta_4 vw+\beta_5 uw)\,,
\eeq
with $\alpha_i$, $\beta_i$ suitable polynomials in the $s_i$. From the na\"ive counting of parameters we can deduce that this type of factorization occurs at codimension-two.\footnote{In \cite{Klevers:2014bqa} it is shown that once one has constructed the ideal $I_{(\alpha_i,\beta_i,s_i)}$ after comparing monomial in \eqref{eq:pF1} and \eqref{eq:cod2}, and constructing the corresponding elimination ideal $I_{(s_i)}=I_{(s_i,\alpha,\beta)}\cap K[s_i]$ in the polynomial ring $K[s_i]$ generated by the sections $s_i$ only, the ideal $I_{(s_i)}$ is generated by 50 polynomials and that its codimension in $K[s_i]$ is indeed two.} The corresponding singlet 
carries charge $1\equiv -2\,{\rm mod}\,3$ under the discrete symmetry. We see that after the transition the two curves that have the same charges under the surviving symmetries join together into one curve.

\

\begin{figure}[h]
\centering
\begin{minipage}{.5\textwidth}
 \centering
  \setlength{\unitlength}{0.2\textwidth}
  \begin{picture}(4,5)
    \put(0,0){\includegraphics[scale=.5]{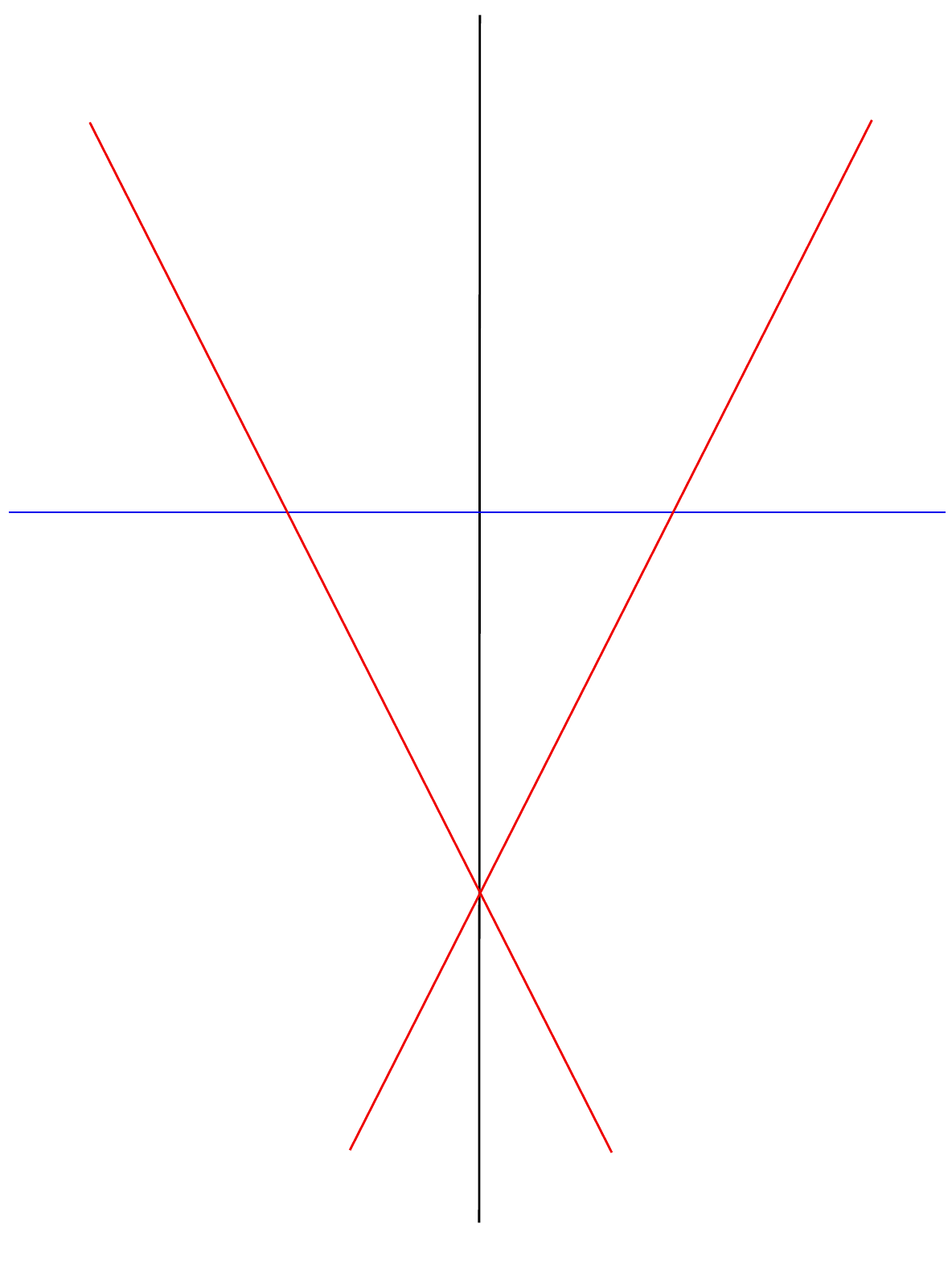}}
    \put(-0.3,2.7){\small{\color{blue} ${\rm W}$}}
    \put(0.1,4.2){\small{\color{red} $\tilde{Z}$}}
    \put(3.3,4.2){\small{\color{red} $Z$}}
   \put(1.9,4.3){\small $O7$}
  \end{picture}
\end{minipage}%
\begin{minipage}{.5\textwidth}
 \centering
  \setlength{\unitlength}{0.2\textwidth}
  \begin{picture}(4,5)
    \put(0,0){\includegraphics[scale=.5]{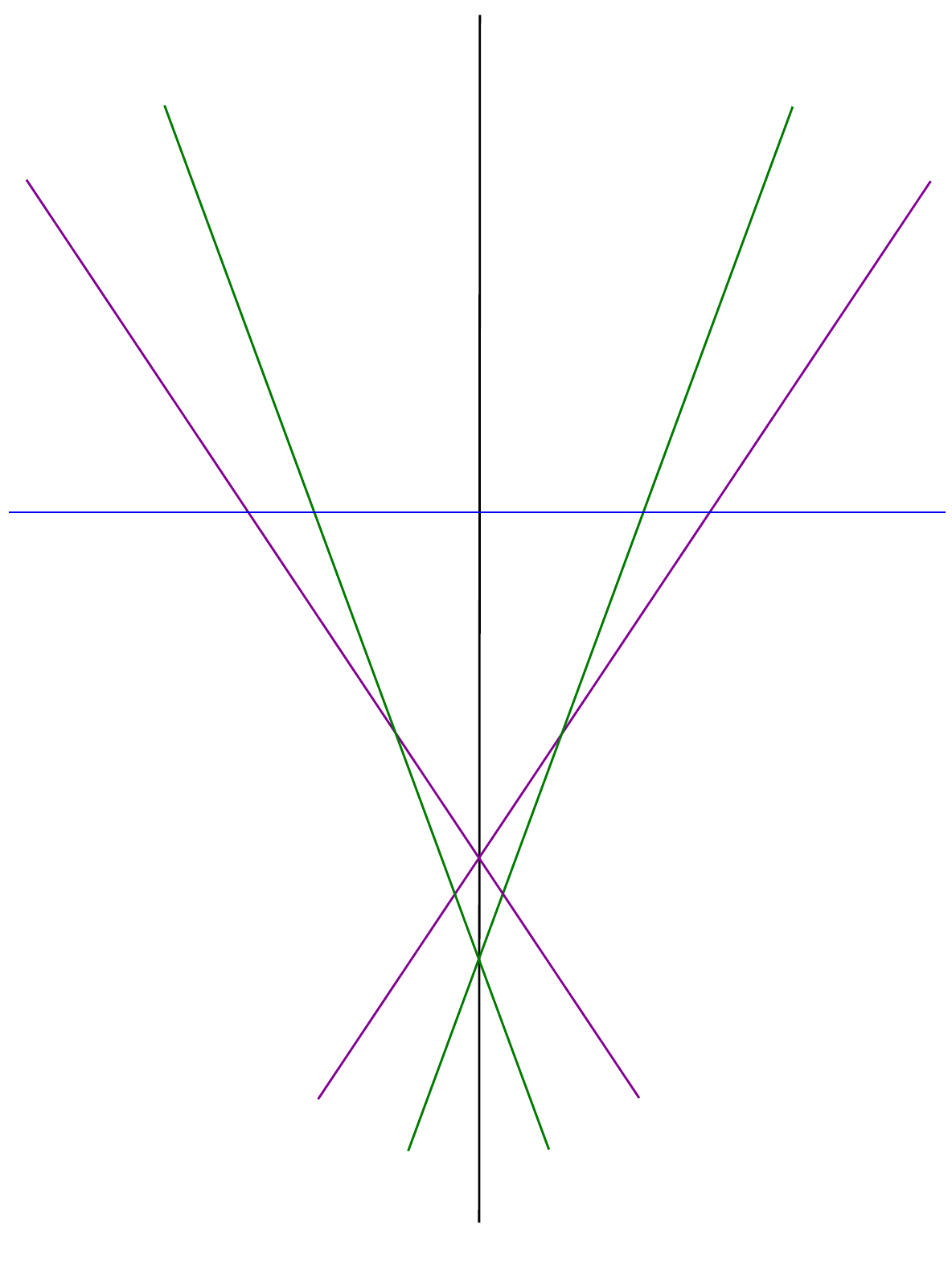}}
    \put(-0.3,2.7){\small{\color{blue} ${\rm W}$}}
    \put(0.6,4.5){\small{\color{green} $\tilde{X}$}}
    \put(2.9,4.5){\small{\color{green} $X$}}
    \put(0.1,4.2){\small{\color{magenta} $\tilde{Y}$}}
    \put(3.4,4.2){\small{\color{magenta} $Y$}}
   \put(1.9,4.3){\small $O7$}
  \end{picture}
\end{minipage}
\caption{\label{fig:F1F3} The intersecting brane configurations for the fibrations based on $P_{F_1}$ (left) and $P_{F_3}$ (right). In the limit $s_{10}=0$, the brane $Z$ splits into two.}
\end{figure}
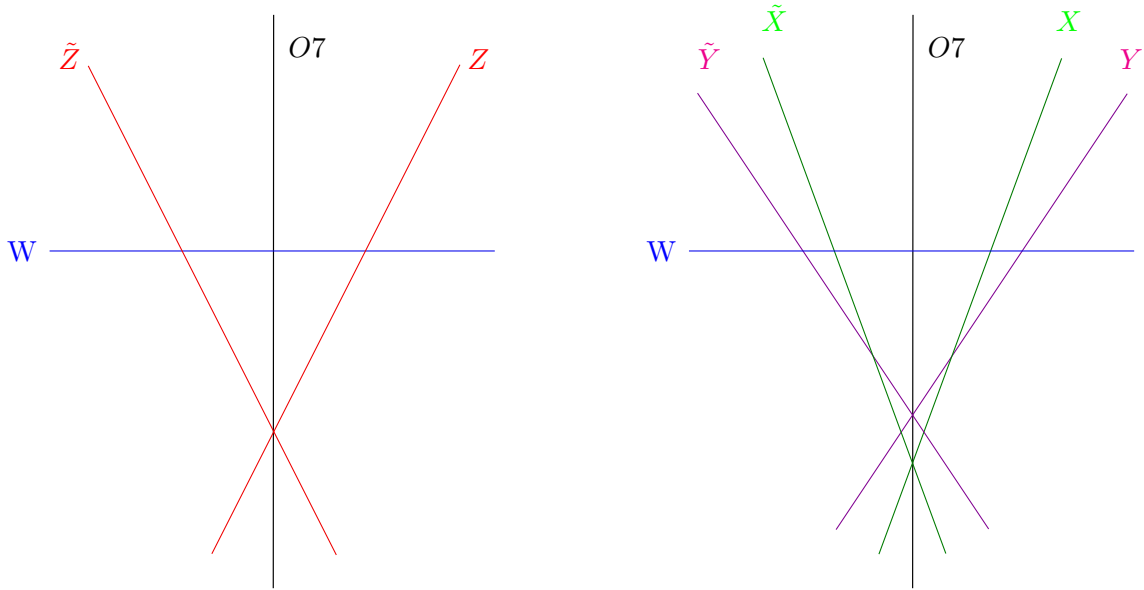

The type IIB limit of this model is obtained after setting the following $\epsilon$-dependence on the sections $s_i$:
\beq
 s_1 \rightarrow \epsilon^1 s_1, \quad s_5 \rightarrow \epsilon^1 s_5, \quad s_8 \rightarrow \epsilon^1 s_8, \quad s_{10} \rightarrow \epsilon^1 s_{10}, \quad s_i \rightarrow \epsilon^0 s_i \quad {\rm (}i\neq 1,5,8,10{\rm )}\,.
\eeq
In this limit the locus of the D7-branes is given by
\begin{align}\label{Z3DeltaE}
\begin{split}
\Delta_E=-\frac14 &\, (-s_3 s_6 s_7 + s_2 s_7^2 + s_3^2 s_9 + 
   s_4 (s_6^2-4s_2 s_9))\\
    &\times [ -s_{10}^2 s_2^3 + 
 s_{10} (s_1 s_6^3 - s_2 s_6 (s_5 s_6 + 3 s_1 s_9) + s_2^2 (s_6 s_8 + 2 s_5 s_9))\\
  &\,\,\,+s_9(s_2^2 s_8^2 + 
   s_2 (-s_5 s_6 s_8 + s_5^2 s_9 - 2 s_1 s_8 s_9) + 
   s_1 (s_6^2 s_8 - s_5 s_6 s_9 + s_1 s_9^2))]\,.
\end{split}
\end{align}
In the Calabi-Yau threefold once again we have an invariant D7-brane ${\rm W}$, in addition to the remaining part of the discriminant which splits into two divisors $Z$ and $\tilde Z$ (see the left-hand side of Fig.~\ref{fig:F1F3}). One can easily prove the following homology relations 
\beq
Z_+=Z+\tilde Z=8 D_{O7}-{\rm W}\,, \quad Z_-=Z-\tilde Z=-3X_-\,,
\eeq
where we have used the divisor $X_-=X-\tilde X$, with $X$ and $\tilde X$ as defined in Eq. \eqref{XXF3}. Recall that in this model the absence of the conifold point is given in terms of the divisor $X_+$, $D_{O7}X_+(2D_{O7}-X_+)=0$\,, even though there is no physical brane configuration along this divisor. 

At weak coupling, the spectrum of this model is identical as in the Morrison-Park model of Section~\ref{Sec:MPexample}, with a singlet with charge $2$ (under the \U1 living on the $Z$-brane) living at the intersection of the banes $Z$ and $\tilde Z$ away from the orientifold plane, and a singlet with charge $1$ living at the intersection of $Z$ with the invariant brane ${\rm W}$. However, in contrast to the Morrison-Park model, the divisors $Z$ and $\tilde Z$ are in different homology classes. For this reason, the $\U1$ gauge symmetry in our case becomes massive. Hence, at the perturbative level we only have a global $\U{1}$ symmetry under which which the two states are distinguished. The matching of states between the F-theory and the type IIB model is summarized in Table \ref{tab:matchF1}.
\begin{table}[ht!]
\begin{center}
\beq
\text{
\renewcommand{\arraystretch}{1.4}
\begin{tabular}{|c|c|c|}
\hline
\multicolumn{2}{|c|}{Type IIB} & F-theory  \\ \hline
$\tilde Z Z$  & $\one_{-2}$ &\multirow{ 2}{*}{$\one$} \\
 $Z {\rm W}$ & $\one_{1}$ & \\
\hline
\end{tabular}
}
\eeq
\caption{\label{tab:matchF1}On the left column, the states in type IIB with the charges under the massive \U1. On the right column the F-theory state (charged under the discrete $\mathbb{Z}_3$ symmetry).}
\end{center}\end{table}

If we set $s_{10}\equiv 0$, the setup just outlined becomes the one described in the \U1 model above, i.e. \eqref{Z3DeltaE}  coincides with Eq. \eqref{eq:discF3}. This is exactly the same Higgs mechanism described on the F-theory side. Let us see how it works in type IIB:
The field that takes non-zero VEV is $\one_{1,1}$ sitting at the intersection $X\tilde Y$ (see Table~\ref{tab:F3IIBmatter}). 
The elements of $\U1_X\times\U1_Y$ are $e^{i\alpha Q_X+\beta Q_Y}$.  The state $\one_{1,1}$ transforms by the phase $e^{i(\alpha+\beta)}$. The non-zero VEV is then left invariant by the $\U1_Z$ subgroup with elements $e^{i\beta(-Q_X+Q_Y)}e^{2\pi i k Q_X}$. Since all the states in the model have integer charges, the last factor can be neglected and one gets a \U1 symmetry with generator $Q_Z=-Q_X+Q_Y$.
Accordingly, the D7-branes wrapping the  $X$ ($\tilde{X}$) and the $\tilde Y$ ($Y$) divisors recombine to give the brane wrapping the $\tilde{Z}$ ($Z$) divisor. 
The other states, $\one_{(-1,0)}$ and $\one_{(0,1)}$ have the same charge $Q_Z$ and the corresponding curves join together in the matter curve for $\one_1$. The same happens for $\one_{(0,-2)}$ and $\one_{(1,-1)}$ joining together in $\one_{-2}$.

The massive $\U1_Z$ symmetry is broken at non-perturbative level by instanton effects. In the present model, there is for instance a D1-instanton wrapping the curve $\Sigma=\mathcal{C}-\tilde{\mathcal{C}}$, where $\mathcal{C}$ is a holomorphic curve intersecting the divisor $X$ at one point.\footnote{This curve will exist generically. If the minimal intersection number is large than one, the surviving discrete group may be bigger.} Its charge under the D7-brane \U1 living on the $Z$ divisor is equal to\cite{Martucci:2015dxa}
$$
q_{D1} = \Sigma \cdot Z = \frac12 \Sigma \cdot Z_-= -  \frac32 \Sigma \cdot X_- = -3 \Sigma\cdot X = -3 \:.
$$
In the present model we have $h^{1,1}_-(X_3)=1$ and hence $\Sigma$ is the minimal (odd) curve that can be wrapped by a D1-brane. Hence the D1-instantons break the massive \U1 to its $\mathbb{Z}_3$ subgroup whose elements are $e^{i\beta Q_Z}$ where $\beta=\frac{k}{3}$ with $k=0,1,2$. One should also check the D3-instantons, i.e. D3-branes wrapping invariant four-cycles in $X_3$ and possibly with flux, but the argument above works in the same way, giving the same discrete symmetry.

We then see that in type IIB the discrete symmetry arises at the non-perturbative level and is a subgroup of the massive \U1. The two states have the same $\mathbb{Z}_3$ symmetry. When we go away from the perturbative weak coupling limit, the matter curves supporting such states join together into one curve, supporting a state with again the same $\mathbb{Z}_3$ charge.\footnote{Notice that the elements $e^{\frac{2\pi i k}{3} (-Q_x+Q_Y) }$ of the discrete symmetry in type IIB can be identified with those $e^{\frac{2\pi i k'}{3} (2Q_x+Q_Y) }$ of the  $\mathbb{Z}_3$ subgroup of the massless \U1 (that is indeed the discrete symmetry identified in F-theory): the difference is by the phase $e^{2\pi i Q_X}$, that is always equal to one in this mode, due to the fact that the states have integer $Q_X$ charges.}
This agrees with what claimed in Section~\ref{sec:ToricMSSM}: when two type IIB matter curves associated with states distinguished by massive \U1 charges (but with the same massless \U1 charges) join together, then one can conclude that the actual symmetry is a discrete subgroup (possibly trivial) under which the joining states have the same charges.

Let us comment on the possible couplings. 
In the F-theory model studied in this section, we expect a perturbative cubic coupling $\one^3$ at the triple self-intersection of the $I_1$ locus (or at the self-intersection of the corresponding matter curve). 
Hence, the respective type IIB coupling should also be of order one. 
In fact, in type IIB we have two states corresponding to the F-theory one. 
Most of  the triple couplings involving $\one_{-2}$ and $\one_1$ are allowed only by the unbroken discrete $\mathbb{Z}_3$ symmetry, but are forbidden by the massive \U1. If all of them would be of this type, we would have a discrepancy with what is predicted by F-theory. However, this does not happen, because there exists one triple coupling allowed by the massive \U1, i.e.
$
  \one_{-2}\one_{1}^2
$. 
This is very similar to what happens with the down Yukawa coupling of Section~\ref{sec:ToricMSSM}, where only one of the possible coupling was allowed by the massive \U1 and that was actually the one corresponding to the F-theory Yukawa.

In  \cite{Martucci:2015dxa,Martucci:2015oaa} it was stressed that the coupling terms allowed by the massless \U1 symmetries can be divided into two categories: the `perturbative' and the `non-perturbative' couplings. The first ones are of order one and are typically associated with the points of enhanced symmetry. The second ones are exponentially suppressed as they are mediated by membrane instantons with finite size (also after the F-theory limit). To distinguish among them in F-theory, one needs to find the homological relations between the fiber components wrapped by the matter M2-branes involved in the coupling. In the first case, the homological relation is inside the fiber homology, while in the second it is satisfied only in the homology of the CY fourfold.
In future investigations, it would be interesting to analyze the fiber structure in the present simple model and see what is the fate of the type IIB instantonic couplings.

\subsubsection*{Fluxes and chiralities}

There is a single flux direction on the F-theory side, 
\beq
G_4=\frac{\Lambda}{9}\left(3 S_{(3)}(3 S_{(3)}+ 3 \bar{K}_B -2\cS_7-2\cS_9)- 2 \cS_7^2 + 5 \cS_7 \cS_9 - 2 \cS_9^2\right)\,.
\eeq
This is in agreement with the vanishing of the following Chern-Simons coefficients \cite{Lin:2015qsa}
\beq
\int_{Y_4} G_4 \wedge S_{(3)} \wedge D_\alpha = \int_{Y_4} G_4 \wedge D_\alpha \wedge D_\beta=0\,,
\eeq
with $D_\alpha$, $D_\beta$ being any of the vertical divisors in the compactification. The contribution of the $G_4$ flux to the D3-tadpole reads
\beq\label{ref:tadpoleF1}
\frac12 \int G_4\wedge G_4=\frac{\Lambda^2}{18} (\cS_7 - 2 \cS_9) (2 \cS_7 - \cS_9) (-3 \bar{K}_B +\cS_7 + \cS_9)\,. 
\eeq

On the type IIB side we need to impose the D5-tadpole cancellation condition
\beq
Z_- F^Z_++Z_+ F^Z_-+{\rm W} F^{\rm W}_-=0\,.
\eeq
The allowed fluxes are
\beq
(F^Z_+)_\lambda=\lambda D_{O7}\,,\quad (F^Z_+,F^{\rm W}_-)_\beta=\beta ({\rm W},-Z_-)\,.
\eeq

In order to match the D3-tadpole contribution of the $G_4$-flux (Eq. \eqref{ref:tadpoleF1}) with the type IIB one, we need to impose 
\beq
\lambda=-\frac{\Lambda}{3}\,,\quad \beta=\frac{\Lambda}{6}\,,
\eeq
up to terms proportional to $X_+D_{O7}(2D_{O7}-X_+)$ (that on the smooth CY threefold are zero). This is the type IIB flux that corresponds to the $\Lambda$ F-theory flux.  
Again, there is a massive type IIB flux that is not described in F-theory by a harmonic vertical four-form flux.

Finally, one could use the flux match to compute chiralities in F-theory, where the matter locus is very complicated to deal with.
In fact, in type IIB we easily compute 
\begin{align}
\chi(\one_{-2})=&\frac14 (48 D_{O7}^2 + {\rm W}^2 - 9 X_+^2 + 2 D_{O7} (-7 {\rm W} + 9 X_+)) (D_{O7} \lambda + 
   {\rm W} \beta)\,\\
\chi(\one_{1})=&\frac12 {\rm W}(D_{O7} (-8 D_{O7} +{\rm W}  ) \lambda + 
   (-8 D_{O7} {\rm W} + {\rm W}^2 + 18 D_{O7} X_+ - 9 X_+^2) \beta)\,.
\end{align}
Adding these two chiralities we arrive at a weak coupling expression for the chirality of the $\mathbb{Z}_3$ singlet under the $\Lambda$-flux:
\begin{align}\label{eq:singWC}
\chi(\one)_{\rm w.c.}=\frac{\Lambda}{8} (-(8 D_{O7} - {\rm W}) (-2 D_{O7} + {\rm W})^2 + 6 D_{O7} (-2 D_{O7} + 3 {\rm W}) X_+ + 
   3 (2 D_{O7} - 3 {\rm W}) X_+^2) \,.
\end{align}
This can be written in terms of the base divisors, recalling that 
\beq
D_{O7}=\pi^*(\bar{K}_B)\,,\quad X_+=\pi^*(\cS_9)\,,\quad  {\rm W}=\pi^*(2\bar{K}_B+2\cS_7-\cS_9). 
\eeq
The resulting chirality at weak coupling is expected to match the F-theory result, i.e.  
\beq 
\chi(\one)_F=\chi(\one)_{\rm w.c.}\,.
\eeq
This is true when the type IIB CY three-fold has no conifold singularity. If we do not require this, then the intersection $\cS_9 \bar{K}_B(2\bar{K}_B-\cS_9)$ is generically non-zero, and we can only claim that
\beq \label{chiFwc}
\chi(\one)_F=\chi(\one)_{\rm w.c.}+a_1 \cS_9 \bar{K}_B(2\bar{K}_B-\cS_9)\,.
\eeq

Note now that this model has a very special feature: it is symmetric upon exchange of the $\mathbb{P}^2$ coordinates $v$ and $w$. In the dual polytope this corresponds to a pairwise exchange of the sections $s_i$ which essentially amounts to exchange between the divisors $\cS_7$ and $\cS_9$. Due to this symmetry, the $F_1$ model exhibits a second weak coupling limit, which is based on the following $\epsilon$-scalings for the sections $s_i$
\beq
 s_1 \rightarrow \epsilon^1 s_1, \quad s_2 \rightarrow \epsilon^1 s_2, \quad s_3 \rightarrow \epsilon^1 s_3, \quad s_{4} \rightarrow \epsilon^1 s_{4}, \quad s_i \rightarrow \epsilon^0 s_i \quad {\rm (}i\neq 1,2,3,4{\rm )}\,.
\eeq
In this limit we have the following relations to the base divisors
\beq
D_{O7}=\pi^*(\bar{K}_B)\,,\quad X_+=\pi^*(\cS_7)\,,\quad  {\rm W}=\pi^*(2\bar{K}_B+2\cS_9-\cS_7) 
\eeq
such that when we plug these expressions in Eq. \eqref{eq:singWC}, we obtain a different weak coupling limit of the F-theory chirality $\chi(\one)^{\prime}_{\rm w.c.}$ satisfying
\beq \label{chiFwcprime}
\chi(\one)_F=\chi(\one)^{\prime}_{\rm w.c.}+a_2 \cS_7 \bar{K}_B(2\bar{K}_B-\cS_7)\,.
\eeq
By comparing the two expressions \eqref{chiFwc} and \eqref{chiFwcprime} we are able to obtain the general F-theory expression for the chirality of the singlet $\one$ in F-theory:
\begin{align}
\chi(\one)_F=\Lambda(\cS_7 - 2 \cS_9) (2 \cS_7 - \cS_9) (-3 \bar{K}_B + \cS_7 + \cS_9)\,. 
\end{align}
It is remarkable that we could have access to this quantity by looking at the weak coupling limit of the F-theory compactification. Note that the locus of the $\mathbb{Z}_3$ charged singlet is generated by fifty non-transversally intersecting polynomials and therefore, the direct computation of the chiral index turns cumbersome.

\section{Conclusions}
\label{sec:conclusion}

In this work we  considered F-theory compactifications with interesting phenomenological features, like an MSSM spectrum, a set of massles \U1 symmetries, charge three states or discrete symmetries.
For each model, we showed that a weak coupling limit exists: we
worked out the $\epsilon$ scaling of the sections defining the corresponding elliptic fibration, such that the resulting perturbative type IIB configuration presents the same spectrum as the F-theory one. This is not always possible, as it happens for example when  the spectrum includes exotic matter states.
As a first result, this shows once more that perturbative type IIB is a powerful setup for model building, where several of the  features of the F-theory models can be realized.

We were able to match  the gauge group and  the matter content with the corresponding type IIB model. In the F-theory models, we worked out all the harmonic vertical four-form fluxes.
We saw that  these $G_4$ fluxes can describe three types of D7-brane fluxes: 1) even fluxes along massless \U1 generators; 2) even and odd fluxes along massive \U1 generators, provided that they cancel the D5-tadpole; 3) odd diagonal fluxes along an \Sp1 stack (the diagonal \U1 gauge boson is projected out by the orientifold projection, but the corresponding flux survives if along an odd form). However, not all the D5-tadpole canceling massive \U1 fluxes are described by  harmonic vertical $G_4$. This answers a question raised in \cite{Krause:2012yh}, as anticipated in the introduction. These extra fluxes may be described by non-harmonic \cite{Grimm:2009yu} or non-vertical four-forms \cite{Braun:2014xka}. In type IIB they provide more freedom in model building: in particular, they may allow to adjust the fluxes to obtain the right number of generations. It would be extremely interesting to study further this point, trying to identify the proper four-form in F-theory.

We finally found that some F-theory matter curves splits at $\epsilon\rightarrow 0$. This is related to the fact that at zero string coupling one recovers a continuous \U1 symmetry \cite{Braun:2014nva} that at finite coupling is broken by non-perturbative effects (like D1-instantons) \cite{Martucci:2015dxa,Martucci:2015oaa}. 
In fact, in type IIB, geometrically massive \U1 symmetries are preserved as global symmetries at the perturbative level, and are generically broken by non-perturbative effects to some discrete subgroup. Correspondingly, at weak coupling there are distinct curves for states that have different massive \U1 charges.
In F-theory this distinction is not present: the elliptic fibration is only sensitive to the true (unbroken) symmetries. Hence states that have the same surviving discrete symmetry charges but different broken massive \U1 charges live on the same curve.
Said differently, the splitting of the matter curves at $\epsilon\rightarrow 0$ is a manifestation of the fact that the full massive \U1 symmetry  is restored at zero coupling (in fact, in \cite{Braun:2014nva} it was shown that at $\epsilon\rightarrow 0$ a new closed two-form arises that corresponds to this \U1 becoming a massless unbroken symmetry).
We aim to come back to this point, by applying the approach of \cite{Martucci:2015dxa}. In the explicit model of Section~\ref{sec:QdP7}, we moreover showed how to use this splitting to infer which flux localizes the zero mode wave functions away from the down Yukawa point, in such a way to suppress this coupling with respect to the order one top Yukawa coupling.

We believe that our constructions may be useful for future investigations, especially when one needs to test some F-theory ideas in the most known perturative regime.


%
\subsection*{Acknowledgments}

We would like to thank Andreas Braun, Denis Klevers, Ling Lin, Fernando Quevedo, Timo Weigand for useful and stimulating discussions. We would like to dedicate this work to Anne Gatti for her lifetime dedication to the ICTP, and for the great care and and frendship she has unconditionally offered to physicists and scientists from all around the globle, among which the authors of this paper are to be counted. 
The work of R.V. is supported by the Programme ``Rita Levi Montalcini for young researchers'' of the Italian Ministry of Research.

%
%
\appendix

\section{Chiral matter in the $\SU{3}\times\SU{2}\times\U{1}^2$ model}
\label{app:ap1}
\begin{table}[H]
\begin{center}
\footnotesize
\renewcommand{\arraystretch}{1.5}
\begin{tabular}{|c|c|}
\hline
Representation & Chirality \\ \hline\hline
$(\three,\two)_{(1,0,0,0)}^{(\frac16,-\frac13)}$ &  $\frac{1}{12} U W_+ (F_1 - 2 F_2 + 6 D_{O7} \lambda_1 + \alpha_1 (-6 D_{O7} + 2 U + 3 W_+) + 
   2 \alpha_2 X_+ + 3 \alpha_4 X_+ + 2 \alpha_3 Y_+)$ \\ \hline\hline
   
$(\overline{\three},\one)_{(-\frac23,\frac13)}$ &  $\begin{array}{c} \frac{1}{12} W_+ X_+ (2 F_1 - 4 F_2 - 3 D_{O7} \lambda_1 + 3 D_{O7} \lambda_2 - 2 (\alpha_1-3\alpha_4) U\\ - 2 \alpha_2 (6 D_{O7} - 3 W_++ X_+) - 2( \alpha_3- 3 \alpha_5)Y_+)\end{array}$ \\ \hline

$(\overline{\three},\one)_{(-1,0,1,0)}^{(\frac13,\frac13)}$ &   $\begin{array}{c} \tfrac{1}{12}W_+[(F_1 + F_2 - 3 D_{O7} \lambda_1 - \alpha_1 U - \alpha_2 X_+) (-4 D_{O7} + 2 (W_+ + X_+) + Y_+) \\ +3 \alpha_5 X_+ (2 (W_+ + X_+) + Y_+) - 
 \alpha_3 (2 D_{O7} (6 W_+ + Y_+) - (3 W_+ - Y_+) (2 (W_+ + X_+) + Y_+))]\end{array}$ \\ \hline
 
 $(\overline{\three},\one)_{(-1,0,-1,0)}^{(-\frac23,\frac13)}$ &   $\begin{array}{c} \tfrac{1}{12}W_+[(2 F_1 - F_2 - 3 D_{O7} \lambda_1 + \alpha_1 U + \alpha_2 X_+) (-4 D_{O7} + 2 (W_+ + X_+) - 
     Y_+) - 6 D_{O7} \lambda_1 Y_+) \\ +3 \alpha_5 X_+ (2 (W_+ + X_+) + Y_+) - 
 \alpha_3 (2 D_{O7} (6 W_+ + Y_+) - (3 W_+ - Y_+) (2 (W_+ + X_+) + Y_+))]\end{array}$ \\ \hline

$(\overline{\three},\one)_{(-1,0,0,1)}^{(\frac13,\frac43)}$ &  $\frac{1}{12} W_+ (-10 D_{O7} + 2 (U + 2 W_+ + X_+) + Y_+) (-F_1 - 4 F_2 + 
   3 D_{O7} \lambda_1 + \alpha_1 U + \alpha_2 X_+ + \alpha_3 Y_+)$ \\ \hline
   
$(\overline{\three},\one)_{(\frac13,-\frac23)}$ &  $\tfrac{1}{12} W_+ (-6 D_{O7} + 2 (U + W_+) + Y_+) (2 F_1 + 2 F_2 + 3 D_{O7} \lambda_1 + 
   \alpha_1 U + \alpha_2 Xp + \alpha_3 Y_+)$ \\ \hline

$(\overline{\three},\one)_{(2,0,0,0)}^{(\frac13,-\frac23)}$ &  $\tfrac16 D_{O7} W_+ (F_1 + 2 (-F_2 + 3 D_{O7} \lambda_1 + \alpha_1 U + \alpha_2 X_+ + \alpha_3 Y_+))$ \\ \hline\hline

$(\one,\two)_{(0,1,0,0)}^{(-\frac12,1)}$ &  $-\frac14 U X_+ (F_1 - 2 F_2 + 2 D_{O7} \lambda_2 +  \alpha_4 (2D_{O7} + 2U-X_+) -( \alpha_1 -2 \alpha_2) W_+ + 
2 \alpha_5 Y_+)$ \\ \hline

$(\one,\two)_{(0,0,1,0)}^{(\frac12,0)}$ &  $\frac14 U( F_1 Y_++2 (2 D_{O7} - W_+- X_+) (\alpha_1 W_+- 2 \alpha_3 W_+ + (\alpha_4 - 2 \alpha_5) X_+) )$ \\ \hline

$(\one,\two)_{(0,0,0,1)}^{(\frac12,1)}$ &  $-\frac14 U[ (F_1 + 2 F_2) (-8 D_{O7} + 2 U + 3 W_+ + X_+ + Y_+)+(2 D_{O7} - W_+ - X_+) (\alpha_1 W_+ + \alpha_4 X_+)]$ \\ \hline\hline

$\one_{(0,1,1,0)}^{(1,-1)}$ &   $\begin{array}{c} \tfrac14 X_+[(F_1 - F_2+\alpha_4 U - \alpha_3 W_+) (-4 D_{O7} + 2 (W_+ + X_+)+Y_+)+ \lambda_2 D_{O7} (-4 D_{O7} + 2 X_+ + Y_+) \\ -\alpha_5 (2 D_{O7} (-2 X_+ + Y_+) + (X_+ - Y_+) (2 (W_+ + X_+) + Y_+))]\end{array}$ \\ \hline

 $\one_{(0,1,0,1)}^{(1,0)}$ &   $ -\tfrac14 X_+ (-10 D_{O7} + 2 (U + 2 W_+ + X_+) + Y_+) (F_1 + D_{O7} \lambda_2 + 
   \alpha_4 U + \alpha_5 Y_+)-\lambda_2 D_{O7}W_+]$ \\ \hline

$\one_{(0,-1,1,0)}^{(0,1)}$ &   $\begin{array}{c} \tfrac14 X_+[(F_2 -\alpha_4 U - \alpha_3 W_+) (4 D_{O7} - 2 (W_+ + X_+) + Y_+)-\lambda_2 D_{O7}  (4 D_{O7} - 2 X_+ + Y_+)\\ +\alpha_5((2 (W_+ + X_+) - Y_+) (X_+ + Y_+) - 2 D_{O7} (2 X_+ + Y_+))]\end{array}$ \\ \hline

$\one_{(0,-1,0,1)}^{(0,2)}$ & $ -\tfrac14 X_+ [(6 D_{O7} - 2 (U + W_+) - Y_+) (-2 F_2 + 
      D_{O7} \lambda_2 + \alpha_4 U + \alpha_5 Y_+)-D_{O7} \lambda_2 W_+]$ \\ \hline

$\one_{(0,0,2,0)}^{(1,0)}$ &  $\frac18 F_1 (8 D_{O7} (W_+ + X_+) - 4 (W_+ + X_+)^2 - 2 D_{O7} Y_+ + Y_+^2)$ \\ \hline

$\one_{(0,0,1,1)}^{(1,1)}$ & $\begin{array}{c} -\tfrac14 [(F_1 + 
   F_2) (-2 (W_+ + X_+) (-2 D_{O7} + W_+ + X_+) + (-8 D_{O7} + 2 U + 3 W_+ +X_+) Y_+ + Y_+^2)\\ +(\alpha_3 W_+ + \alpha_5 X_+) (32 D_{O7}^2 + (W_+ + X_+) (4 U + 6 W_+ + 2 X_+ + Y_+) \\
   - 2 D_{O7} (4 U + 12 W_+ + 8 X_+ + Y_+))-4 D_{O7}(\alpha_3 W_+^2 + \alpha_5 X_+^2)] \end{array}$ \\ \hline

 $\one_{(0,0,-1,1)}^{(0,1)}$ & $\begin{array}{c} \frac14 [ F_2 (-2 (W_+ + X_+)^2 - (2 U + 3 W_+ + X_+) Y_+ - Y_+^2 + 
   4 D_{O7} (W_+ + X_+ + 2 Y_+))\\
 +(\alpha_3 W_+ + \alpha_5 X_+) (32 D_{O7}^2 + (W_+ + X_+) (4 U + 6 W_+ + 2 X_+ + 3 Y_+)\\ - 
   2 D_{O7} (4 U + 12 W_+ + 8 X_+ + 3 Y_+)) -4 D_{O7}(\alpha_3 W_+^2 + \alpha_5 X_+^2)]\end{array}$ \\ \hline
   
$\one_{(0,0,0,2)}^{(1,2)}$ & $\tfrac18 (F_1 + 2 F_2) (6 D_{O7} - 2 (U + W_+) - Y_+) (8 D_{O7} - 2 (U + 2 W_+ + X_+) - Y_+)$ \\ \hline
\end{tabular}
\caption{\label{tab:poly5topIIB}Chiral Indices for the matter in type IIB limit of the $\SU3\times\SU2\times\U{1}^2$ model}
\end{center}
\end{table}
\section{Additional data for hypersurfaces based on $F_3$ and $F_1$}
\label{app:ap2}
In this appendix we merely reproduce a part of appendix B of \cite{Klevers:2014bqa}, which provides $f$ and $g$ for the genus-one fibration based on $F_1$ as well as the coordinates of the non-troic section in $F_3$

\subsection{$f$ and $g$ for the Wierstrass model based on $\mathbb{P}_{F_1}$}
\small
\begin{align}
\begin{split}
\label{eq:fF1}
f&=\frac{1}{48} (-(s_6^2 - 4 (s_5 s_7 + s_3 s_8 + s_2 s_9))^2 + 
    24 (-s_6 (s_{10} s_2 s_3 - 9 s_1 s_{10} s_4 + s_4 s_5 s_8 \\
&\phantom{=}+ s_2 s_7 s_8 + s_3 s_5 s_9 +
          s_1 s_7 s_9) + 
      2 (s_{10} s_3^2 s_5 + s_1 s_7^2 s_8 + s_2 s_3 s_8 s_9 + s_1 s_3 s_9^2 \\
&\phantom{=}+  s_7 (s_{10} s_2^2 - 3 s_1 s_{10} s_3 + s_3 s_5 s_8 + s_2 s_5 s_9) + 
         s_4 (-3 s_{10} s_2 s_5 + s_2 s_8^2 + (s_5^2 - 3 s_1 s_8) s_9)))) 
\end{split}
\end{align}
\begin{align}
\label{eq:gF1}
\begin{split}
g&=\frac{1}{864} ((s_6^2 - 4 (s_5 s_7 + s_3 s_8 + s_2 s_9))^3 - 
   36 (s_6^2 - 4 (s_5 s_7 + s_3 s_8 + s_2 s_9)) \\
&\phantom{=}\times (-s_6 (s_{10} s_2 s_3 - 9 s_1 s_{10} s_4 + 
         s_4 s_5 s_8 + s_2 s_7 s_8 + s_3 s_5 s_9 + s_1 s_7 s_9) \\
&\phantom{=}+  2 (s_{10} s_3^2 s_5 + s_1 s_7^2 s_8 + s_2 s_3 s_8 s_9 + s_1 s_3 s_9^2 + 
         s_7 (s_{10} s_2^2 - 3 s_1 s_{10} s_3 + s_3 s_5 s_8 + s_2 s_5 s_9) \\
&\phantom{=}+  s_4 (-3 s_{10} s_2 s_5 + s_2 s_8^2 + (s_5^2 - 3 s_1 s_8) s_9))) + 
   216 ((s_{10} s_2 s_3 - 9 s_1 s_{10} s_4 + s_4 s_5 s_8 \\
&\phantom{=}+ s_2 s_7 s_8 + s_3 s_5 s_9 + 
        s_1 s_7 s_9)^2 + 4 (-s_1 s_{10}^2 s_3^3 - s_1^2 s_{10} s_7^3 - 
         s_4^2 (27 s_1^2 s_{10}^2 + s_{10} s_5^3 \\
&\phantom{=}+ s_1 (-9 s_{10} s_5 s_8 + s_8^3)) + 
         s_{10} s_3^2 (-s_2 s_5 + s_1 s_6) s_9 - s_1 s_3^2 s_8 s_9^2 \\
&\phantom{=}-  s_7^2 (s_{10} (s_2^2 s_5 - 2 s_1 s_3 s_5 - s_1 s_2 s_6) + 
            s_1 s_8 (s_3 s_8 + s_2 s_9)) \\
&\phantom{=}-  s_3 s_7 (s_{10} (-s_2 s_5 s_6 + s_1 s_6^2 + s_2^2 s_8 + 
               s_3 (s_5^2 - 2 s_1 s_8) + s_1 s_2 s_9) \\
&\phantom{=}+  s_9 (s_2 s_5 s_8 - s_1 s_6 s_8 + s_1 s_5 s_9)) + 
         s_4 (-s_{10}^2 (s_2^3 - 9 s_1 s_2 s_3) \\
&\phantom{=}+ s_{10} (s_6 (-s_2 s_5 s_6 + s_1 s_6^2 + s_2^2 s_8) + 
               s_3 (s_5^2 s_6 - s_2 s_5 s_8 - 3 s_1 s_6 s_8)) \\
&\phantom{=}+ (s_{10} (2 s_2^2 s_5 + 3 s_1 s_3 s_5 - 
                  3 s_1 s_2 s_6) + 
               s_8 (-s_3 s_5^2 + s_2 s_5 s_6 - s_1 s_6^2 - s_2^2 s_8 + 
                  2 s_1 s_3 s_8)) s_9 \\
&\phantom{=}+ (-s_2 s_5^2 + s_1 s_5 s_6 + 
               2 s_1 s_2 s_8) s_9^2 - s_1^2 s_9^3 + 
            s_7 (s_{10} (2 s_2 s_5^2 - 3 s_1 s_5 s_6 + 3 s_1 s_2 s_8 + 
                  9 s_1^2 s_9) \\
&\phantom{=}- s_8 (s_2 s_5 s_8 - s_1 s_6 s_8 + s_1 s_5 s_9))))))
\end{split}
\end{align}
\normalsize

These equations also provide the right expressions $f$ and $g$ in the case of $F_{11}$, $F_{5}$ and $F_{3}$ after setting some suitable sections to zero. For example, in order to obtain $f$ and $g$ for $F_3$ one sets $s_{10}=0$.

\subsection{The Weierstrass coordinates of the non-toric section $S_1$ in $F_3$}
\small
\begin{align}
\begin{split}
y_1 &=\frac{1}{2} (2 s_1^3 s_9^9+s_1 (2 s_2 (s_5^2-3 s_1 s_8)-3 s_1 s_5 s_6) s_9^8+((s_3 s_5^2-s_2 s_6 s_5+s_1 (s_6^2-s_5 s_7)) s_5^2 \\
  &\phantom{=} +6 s_1 (s_2^2  + s_1 s_3) s_8^2+ (-2 s_2^2 s_5^2+2 s_1 s_2 s_6 s_5+s_1 (3 s_1 (s_6^2+2 s_5 s_7)-4 s_3 s_5^2)) s_8) s_9^7  \\
  &\phantom{=}  -s_8 (2 (s_2^3+6 s_1 s_3 s_2+ 3 s_1^2 s_4) s_8^2- (s_5 s_6 s_2^2+(6 s_3 s_5^2-4 s_1 (s_6^2+2 s_5 s_7)) s_2 \\
  &\phantom{=}+s_1 (6 s_4 s_5^2+2 s_3 s_6 s_5-9 s_1 s_6 s_7)) s_8+s_5 (3 s_4 s_5^3 + 2 s_3 s_6 s_5^2-3 s_2 s_7 s_5^2-2 s_2 s_6^2 s_5+s_1 s_6 s_7 s_5 \\
  &\phantom{=}+2 s_1 s_6^3)) s_9^6+s_8^2 (s_1 s_6^4-s_2 s_5 s_6^3+s_3 s_5^2 s_6^2+7 s_1 s_5 s_7 s_6^2 + 9 s_4 s_5^3 s_6-8 s_2 s_5^2 s_7 s_6+s_1 s_5^2 s_7^2\\
  &\phantom{=}+6 (s_3 (s_2^2+s_1 s_3)+2 s_1 s_2 s_4) s_8^2-s_3 s_5^3 s_7+(-4 s_3^2 s_5^2-8 s_2 s_4 s_5^2  - 6 s_1 s_4 s_6 s_5+s_2^2 s_6^2+6 s_1^2 s_7^2 \\
  &\phantom{=} +2 s_2 (s_2 s_5+7 s_1 s_6) s_7+s_3 (2 s_1 (s_6^2+2 s_5 s_7)-6 s_2 s_5 s_6)) s_8) s_9^5- s_8^3 (s_8 (6 s_2 s_8-5 s_5 s_6) s_3^2 \\
  &\phantom{=}-5 s_6 s_7 (s_5^2-2 s_1 s_8) s_3+5 s_7 (s_6 s_8 s_2^2-s_5 (s_6^2+s_5 s_7) s_2+2 s_1 s_7 s_8 s_2 + s_1 s_6 (s_6^2+2 s_5 s_7))\\
  &\phantom{=} +s_4 (5 (2 s_6^2+s_5 s_7) s_5^2-10 (s_3 s_5+s_2 s_6) s_8 s_5+6 (s_2^2+2 s_1 s_3) s_8^2)) s_9^4 + s_8^4 (2 (s_3^3+6 s_2 s_4 s_3+3 s_1 s_4^2) s_8^2 \\
   &\phantom{=} -(6 s_4^2 s_5^2+s_3^2 s_6^2-4 (s_2^2+2 s_1 s_3) s_7^2+2 s_3 (s_3 s_5-3 s_2 s_6) s_7+ 2 s_4 (s_2 s_6^2+7 s_3 s_5 s_6-3 s_1 s_7 s_6 \\
  &\phantom{=} +2 s_2 s_5 s_7)) s_8+5 (s_4 s_5 s_6 (s_6^2+2 s_5 s_7)+s_7 (s_7 (2 s_1 s_6^2-s_2 s_5 s_6+s_1 s_5 s_7)- s_3 s_5 (s_6^2+s_5 s_7)))) s_9^3 \\
  &\phantom{=} -s_8^5 (3 s_8 (2 s_2 s_8-3 s_5 s_6) s_4^2+(s_6^4+(7 s_5 s_7-4 s_3 s_8) s_6^2+2 s_2 s_7 s_8 s_6+s_5^2 s_7^2-8 s_3 s_5 s_7 s_8 \\
  &\phantom{=}+ 6 s_8 (s_8 s_3^2+s_1 s_7^2)) s_4+s_7 (s_6 s_8 s_3^2-(s_6^3+8 s_5 s_7 s_6-6 s_2 s_7 s_8) s_3+s_7 (9 s_1 s_6 s_7+s_2 (s_6^2-s_5 s_7)))) s_9^2  \\
  &\phantom{=}+ s_8^6 (3 s_8 (-s_6^2-2 s_5 s_7+2 s_3 s_8) s_4^2+s_7 (2 s_6^3+s_5 s_7 s_6-2 s_3 s_8 s_6+4 s_2 s_7 s_8) s_4+s_7^2 (2 s_8 s_3^2-2 s_6^2 s_3\\
  &\phantom{=} - 3 s_5 s_7 s_3+3 s_1 s_7^2+2 s_2 s_6 s_7)) s_9+s_8^7 (-2 s_8^2 s_4^3+3 s_6 s_7 s_8 s_4^2+s_7^2 (-s_6^2+s_5 s_7-2 s_3 s_8) s_4\\
    &\phantom{=}+s_7^3 (s_3 s_6-s_2 s_7)))\,, \\
  \end{split}
&\\
\begin{split}
x_1 &= \frac{1}{12} (12 s_1^2 s_9^6+4 (2 s_2 (s_5^2-3 s_1 s_8)-3 s_1 s_5 s_6) s_9^5+((s_6^2-4 s_5
   s_7) s_5^2+12 (s_2^2+2 s_1 s_3) s_8^2 \\
  &\phantom{=} -4 (4 s_3 s_5^2+ s_2 s_6 s_5-3 s_1 (s_6^2+2 s_5 s_7)) s_8) s_9^4-2 s_8 (-4 (s_6 s_7+3 s_4 s_8) s_5^2 \\
  &\phantom{=}+(s_6^3-10 s_3 s_8 s_6+4 s_2 s_7
   s_8) s_5 + 2 s_8 (9 s_1 s_6 s_7+6 s_1 s_4 s_8+s_2 (s_6^2+6 s_3 s_8))) s_9^3 \\
  &\phantom{=} +s_8^2 (s_6^4-2 s_5 s_7 s_6^2-8 s_5^2 s_7^2+12 (s_3^2+2 s_2 s_4) s_8^2- 4 (9 s_4 s_5 s_6-s_7 (5 s_2 s_6+6 s_1 s_7)\\
  &\phantom{=}+s_3 (s_6^2+2 s_5 s_7)) s_8) s_9^2-2 s_8^3 (12 s_3 s_4 s_8^2+2 (s_7 (s_3 s_6+4 s_2 s_7)  - 3 s_4 (s_6^2+2 s_5 s_7)) s_8\\
  &\phantom{=}  
  +s_6 s_7 (s_6^2-4 s_5 s_7)) s_9+s_8^4 ((s_6^2-4 s_5 s_7) s_7^2+4 (2 s_3 s_7-3 s_4 s_6) s_8 s_7+12 s_4^2 s_8^2))\,, 
  \end{split}
  &\\
  \begin{split}
  z_1 &= s_7 s_8^2+s_9 (s_5 s_9-s_6 s_8)\, .
\end{split}
\end{align}

\bibliographystyle{utphys}	

\begin{thebibliography}{100}

\bibitem{Vafa:1996xn}
C.~Vafa, ``{Evidence for F theory},''
  \href{http://dx.doi.org/10.1016/0550-3213(96)00172-1}{{\em Nucl.Phys.}
  {\bfseries B469} (1996) 403--418},
\href{http://arxiv.org/abs/hep-th/9602022}{{\ttfamily arXiv:hep-th/9602022
  [hep-th]}}.

\bibitem{Morrison:1996na}
D.~R. Morrison and C.~Vafa, ``{Compactifications of F theory on Calabi-Yau
  threefolds. 1},'' \href{http://dx.doi.org/10.1016/0550-3213(96)00242-8}{{\em
  Nucl.Phys.} {\bfseries B473} (1996) 74--92},
\href{http://arxiv.org/abs/hep-th/9602114}{{\ttfamily arXiv:hep-th/9602114
  [hep-th]}}.

\bibitem{Morrison:1996pp}
D.~R. Morrison and C.~Vafa, ``{Compactifications of F theory on Calabi-Yau
  threefolds. 2.},'' \href{http://dx.doi.org/10.1016/0550-3213(96)00369-0}{{\em
  Nucl.Phys.} {\bfseries B476} (1996) 437--469},
\href{http://arxiv.org/abs/hep-th/9603161}{{\ttfamily arXiv:hep-th/9603161
  [hep-th]}}.

\bibitem{Braun:2014oya}
V.~Braun and D.~R. Morrison, ``{F-theory on Genus-One Fibrations},''
\href{http://arxiv.org/abs/1401.7844}{{\ttfamily arXiv:1401.7844 [hep-th]}}.

\bibitem{Morrison:2014era}
D.~R. Morrison and W.~Taylor, ``{Sections, multisections, and U(1) fields in
  F-theory},''
\href{http://arxiv.org/abs/1404.1527}{{\ttfamily arXiv:1404.1527 [hep-th]}}.

\bibitem{Donagi:2008ca}
R.~Donagi and M.~Wijnholt, ``{Model Building with F-Theory},''
  \href{http://arxiv.org/abs/0802.2969}{{\ttfamily arXiv:0802.2969 [hep-th]}}.

\bibitem{Beasley:2008dc}
C.~Beasley, J.~J. Heckman, and C.~Vafa, ``{GUTs and Exceptional Branes in
  F-theory - I},'' \href{http://dx.doi.org/10.1088/1126-6708/2009/01/058}{{\em
  JHEP} {\bfseries 01} (2009) 058},
\href{http://arxiv.org/abs/0802.3391}{{\ttfamily arXiv:0802.3391 [hep-th]}}.

\bibitem{Beasley:2008kw}
C.~Beasley, J.~J. Heckman, and C.~Vafa, ``{GUTs and Exceptional Branes in
  F-theory - II: Experimental Predictions},''
  \href{http://dx.doi.org/10.1088/1126-6708/2009/01/059}{{\em JHEP} {\bfseries
  01} (2009) 059},
\href{http://arxiv.org/abs/0806.0102}{{\ttfamily arXiv:0806.0102 [hep-th]}}.

\bibitem{Donagi:2008kj}
R.~Donagi and M.~Wijnholt, ``{Breaking GUT Groups in F-Theory},'' {\em
  Adv.Theor.Math.Phys.} {\bfseries 15} (2011) 1523--1604,
\href{http://arxiv.org/abs/0808.2223}{{\ttfamily arXiv:0808.2223 [hep-th]}}.

\bibitem{Marsano:2009ym}
J.~Marsano, N.~Saulina, and S.~Schafer-Nameki, ``{F-theory Compactifications
  for Supersymmetric GUTs},''
  \href{http://dx.doi.org/10.1088/1126-6708/2009/08/030}{{\em JHEP} {\bfseries
  08} (2009) 030},
\href{http://arxiv.org/abs/0904.3932}{{\ttfamily arXiv:0904.3932 [hep-th]}}.

\bibitem{Blumenhagen:2009yv}
R.~Blumenhagen, T.~W. Grimm, B.~Jurke, and T.~Weigand, ``{Global F-theory
  GUTs},'' \href{http://dx.doi.org/10.1016/j.nuclphysb.2009.12.013}{{\em
  Nucl.Phys.} {\bfseries B829} (2010) 325--369},
\href{http://arxiv.org/abs/0908.1784}{{\ttfamily arXiv:0908.1784 [hep-th]}}.

\bibitem{Grimm:2009yu}
T.~W. Grimm, S.~Krause, and T.~Weigand, ``{F-Theory GUT Vacua on Compact
  Calabi-Yau Fourfolds},''
  \href{http://dx.doi.org/10.1007/JHEP07(2010)037}{{\em JHEP} {\bfseries 1007}
  (2010) 037},
\href{http://arxiv.org/abs/0912.3524}{{\ttfamily arXiv:0912.3524 [hep-th]}}.

\bibitem{Marsano:2009gv}
J.~Marsano, N.~Saulina, and S.~Schafer-Nameki, ``{Monodromies, Fluxes, and
  Compact Three-Generation F-theory GUTs},''
  \href{http://dx.doi.org/10.1088/1126-6708/2009/08/046}{{\em JHEP} {\bfseries
  0908} (2009) 046},
\href{http://arxiv.org/abs/0906.4672}{{\ttfamily arXiv:0906.4672 [hep-th]}}.

\bibitem{Marsano:2009wr}
J.~Marsano, N.~Saulina, and S.~Schafer-Nameki, ``{Compact F-theory GUTs with
  U(1) (PQ)},'' \href{http://dx.doi.org/10.1007/JHEP04(2010)095}{{\em JHEP}
  {\bfseries 1004} (2010) 095},
\href{http://arxiv.org/abs/0912.0272}{{\ttfamily arXiv:0912.0272 [hep-th]}}.

\bibitem{Krause:2011xj}
S.~Krause, C.~Mayrhofer, and T.~Weigand, ``{$G_4$ flux, chiral matter and
  singularity resolution in F-theory compactifications},''
  \href{http://dx.doi.org/10.1016/j.nuclphysb.2011.12.013}{{\em Nucl.Phys.}
  {\bfseries B858} (2012) 1--47},
\href{http://arxiv.org/abs/1109.3454}{{\ttfamily arXiv:1109.3454 [hep-th]}}.

\bibitem{Mayrhofer:2012zy}
C.~Mayrhofer, E.~Palti, and T.~Weigand, ``{U(1) symmetries in F-theory GUTs
  with multiple sections},''
\href{http://arxiv.org/abs/1211.6742}{{\ttfamily arXiv:1211.6742 [hep-th]}}.

\bibitem{Borchmann:2013hta}
J.~Borchmann, C.~Mayrhofer, E.~Palti, and T.~Weigand, ``{SU(5) Tops with
  Multiple U(1)s in F-theory},''
\href{http://arxiv.org/abs/1307.2902}{{\ttfamily arXiv:1307.2902 [hep-th]}}.

\bibitem{Braun:2013nqa}
V.~Braun, T.~W. Grimm, and J.~Keitel, ``{Geometric Engineering in Toric
  F-Theory and GUTs with U(1) Gauge Factors},''
\href{http://arxiv.org/abs/1306.0577}{{\ttfamily arXiv:1306.0577 [hep-th]}}.

\bibitem{Braun:2013yti}
V.~Braun, T.~W. Grimm, and J.~Keitel, ``{New Global F-theory GUTs with U(1)
  symmetries},''
\href{http://arxiv.org/abs/1302.1854}{{\ttfamily arXiv:1302.1854 [hep-th]}}.

\bibitem{Cvetic:2013uta}
M.~Cveti{\v c}, A.~Grassi, D.~Klevers, and H.~Piragua, ``{Chiral
  Four-Dimensional F-Theory Compactifications With SU(5) and Multiple
  U(1)-Factors},''
\href{http://arxiv.org/abs/1306.3987}{{\ttfamily arXiv:1306.3987 [hep-th]}}.

\bibitem{Garcia-Etxebarria:2014qua}
I.~García-Etxebarria, T.~W. Grimm, and J.~Keitel, ``{Yukawas and discrete
  symmetries in F-theory compactifications without section},''
  \href{http://dx.doi.org/10.1007/JHEP11(2014)125}{{\em JHEP} {\bfseries 11}
  (2014) 125},
\href{http://arxiv.org/abs/1408.6448}{{\ttfamily arXiv:1408.6448 [hep-th]}}.

\bibitem{Krippendorf:2014xba}
S.~Krippendorf, D.~K. Mayorga~Pena, P.-K. Oehlmann, and F.~Ruehle, ``{Rational
  F-Theory GUTs without exotics},''
  \href{http://dx.doi.org/10.1007/JHEP07(2014)013}{{\em JHEP} {\bfseries 1407}
  (2014) 013},
\href{http://arxiv.org/abs/1401.5084}{{\ttfamily arXiv:1401.5084 [hep-th]}}.

\bibitem{Krippendorf:2015kta}
S.~Krippendorf, S.~Schafer-Nameki, and J.-M. Wong, ``{Froggatt-Nielsen meets
  Mordell-Weil: A Phenomenological Survey of Global F-theory GUTs with
  U(1)s},'' \href{http://dx.doi.org/10.1007/JHEP11(2015)008}{{\em JHEP}
  {\bfseries 11} (2015) 008},
\href{http://arxiv.org/abs/1507.05961}{{\ttfamily arXiv:1507.05961 [hep-th]}}.

\bibitem{Lawrie:2015hia}
C.~Lawrie, S.~Schafer-Nameki, and J.-M. Wong, ``{F-theory and All Things
  Rational: Surveying U(1) Symmetries with Rational Sections},''
  \href{http://dx.doi.org/10.1007/JHEP09(2015)144}{{\em JHEP} {\bfseries 09}
  (2015) 144},
\href{http://arxiv.org/abs/1504.05593}{{\ttfamily arXiv:1504.05593 [hep-th]}}.

\bibitem{Sen:1997gv}
A.~Sen, ``{Orientifold limit of F theory vacua},''
  \href{http://dx.doi.org/10.1103/PhysRevD.55.R7345}{{\em Phys. Rev.}
  {\bfseries D55} (1997) R7345--R7349},
\href{http://arxiv.org/abs/hep-th/9702165}{{\ttfamily arXiv:hep-th/9702165
  [hep-th]}}.

\bibitem{Clingher:2012rg}
A.~Clingher, R.~Donagi, and M.~Wijnholt, ``{The Sen Limit},''
  \href{http://dx.doi.org/10.4310/ATMP.2014.v18.n3.a2}{{\em Adv. Theor. Math.
  Phys.} {\bfseries 18} no.~3, (2014) 613--658},
\href{http://arxiv.org/abs/1212.4505}{{\ttfamily arXiv:1212.4505 [hep-th]}}.

\bibitem{Collinucci:2016hgh}
A.~Collinucci and I.~García-Etxebarria, ``{E$_{6}$ Yukawa couplings in
  F-theory as D-brane instanton effects},''
  \href{http://dx.doi.org/10.1007/JHEP03(2017)155}{{\em JHEP} {\bfseries 03}
  (2017) 155},
\href{http://arxiv.org/abs/1612.06874}{{\ttfamily arXiv:1612.06874 [hep-th]}}.

\bibitem{Collinucci:2008pf}
A.~Collinucci, F.~Denef, and M.~Esole, ``{D-brane Deconstructions in IIB
  Orientifolds},'' \href{http://dx.doi.org/10.1088/1126-6708/2009/02/005}{{\em
  JHEP} {\bfseries 02} (2009) 005},
\href{http://arxiv.org/abs/0805.1573}{{\ttfamily arXiv:0805.1573 [hep-th]}}.

\bibitem{Collinucci:2008zs}
A.~Collinucci, ``{New F-theory lifts},''
  \href{http://dx.doi.org/10.1088/1126-6708/2009/08/076}{{\em JHEP} {\bfseries
  08} (2009) 076},
\href{http://arxiv.org/abs/0812.0175}{{\ttfamily arXiv:0812.0175 [hep-th]}}.

\bibitem{Collinucci:2009uh}
A.~Collinucci, ``{New F-theory lifts. II. Permutation orientifolds and enhanced
  singularities},'' \href{http://dx.doi.org/10.1007/JHEP04(2010)076}{{\em JHEP}
  {\bfseries 04} (2010) 076},
\href{http://arxiv.org/abs/0906.0003}{{\ttfamily arXiv:0906.0003 [hep-th]}}.

\bibitem{Blumenhagen:2009up}
R.~Blumenhagen, T.~W. Grimm, B.~Jurke, and T.~Weigand, ``{F-theory uplifts and
  GUTs},'' \href{http://dx.doi.org/10.1088/1126-6708/2009/09/053}{{\em JHEP}
  {\bfseries 09} (2009) 053},
\href{http://arxiv.org/abs/0906.0013}{{\ttfamily arXiv:0906.0013 [hep-th]}}.

\bibitem{Collinucci:2010gz}
A.~Collinucci and R.~Savelli, ``{On Flux Quantization in F-Theory},''
  \href{http://dx.doi.org/10.1007/JHEP02(2012)015}{{\em JHEP} {\bfseries 02}
  (2012) 015},
\href{http://arxiv.org/abs/1011.6388}{{\ttfamily arXiv:1011.6388 [hep-th]}}.

\bibitem{Braun:2011zm}
A.~P. Braun, A.~Collinucci, and R.~Valandro, ``{G-flux in F-theory and
  algebraic cycles},''
  \href{http://dx.doi.org/10.1016/j.nuclphysb.2011.10.034}{{\em Nucl.Phys.}
  {\bfseries B856} (2012) 129--179},
\href{http://arxiv.org/abs/1107.5337}{{\ttfamily arXiv:1107.5337 [hep-th]}}.

\bibitem{Krause:2012yh}
S.~Krause, C.~Mayrhofer, and T.~Weigand, ``{Gauge Fluxes in F-theory and Type
  IIB Orientifolds},'' \href{http://dx.doi.org/10.1007/JHEP08(2012)119}{{\em
  JHEP} {\bfseries 1208} (2012) 119},
\href{http://arxiv.org/abs/1202.3138}{{\ttfamily arXiv:1202.3138 [hep-th]}}.

\bibitem{Collinucci:2012as}
A.~Collinucci and R.~Savelli, ``{On Flux Quantization in F-Theory II: Unitary
  and Symplectic Gauge Groups},''
  \href{http://dx.doi.org/10.1007/JHEP08(2012)094}{{\em JHEP} {\bfseries 08}
  (2012) 094},
\href{http://arxiv.org/abs/1203.4542}{{\ttfamily arXiv:1203.4542 [hep-th]}}.

\bibitem{Esole:2012tf}
M.~Esole and R.~Savelli, ``{Tate Form and Weak Coupling Limits in F-theory},''
  \href{http://dx.doi.org/10.1007/JHEP06(2013)027}{{\em JHEP} {\bfseries 06}
  (2013) 027},
\href{http://arxiv.org/abs/1209.1633}{{\ttfamily arXiv:1209.1633 [hep-th]}}.

\bibitem{Braun:2014nva}
A.~P. Braun, A.~Collinucci, and R.~Valandro, ``{The fate of U(1)'s at strong
  coupling in F-theory},''
  \href{http://dx.doi.org/10.1007/JHEP07(2014)028}{{\em JHEP} {\bfseries 1407}
  (2014) 028},
\href{http://arxiv.org/abs/1402.4054}{{\ttfamily arXiv:1402.4054 [hep-th]}}.

\bibitem{Braun:2014pva}
A.~P. Braun, A.~Collinucci, and R.~Valandro, ``{Hypercharge flux in F-theory
  and the stable Sen limit},''
  \href{http://dx.doi.org/10.1007/JHEP07(2014)121}{{\em JHEP} {\bfseries 07}
  (2014) 121},
\href{http://arxiv.org/abs/1402.4096}{{\ttfamily arXiv:1402.4096 [hep-th]}}.

\bibitem{Collinucci:2014taa}
A.~Collinucci and R.~Savelli, ``{F-theory on singular spaces},''
  \href{http://dx.doi.org/10.1007/JHEP09(2015)100}{{\em JHEP} {\bfseries 09}
  (2015) 100},
\href{http://arxiv.org/abs/1410.4867}{{\ttfamily arXiv:1410.4867 [hep-th]}}.

\bibitem{Martucci:2015dxa}
L.~Martucci and T.~Weigand, ``{Non-perturbative selection rules in F-theory},''
  \href{http://dx.doi.org/10.1007/JHEP09(2015)198}{{\em JHEP} {\bfseries 09}
  (2015) 198},
\href{http://arxiv.org/abs/1506.06764}{{\ttfamily arXiv:1506.06764 [hep-th]}}.

\bibitem{Martucci:2015oaa}
L.~Martucci and T.~Weigand, ``{Hidden Selection Rules, M5-instantons and Fluxes
  in F-theory},'' \href{http://dx.doi.org/10.1007/JHEP10(2015)131}{{\em JHEP}
  {\bfseries 10} (2015) 131},
\href{http://arxiv.org/abs/1507.06999}{{\ttfamily arXiv:1507.06999 [hep-th]}}.

\bibitem{Greiner:2017ery}
S.~Greiner and T.~W. Grimm, ``{Three-form periods on Calabi-Yau fourfolds:
  Toric hypersurfaces and F-theory applications},''
  \href{http://dx.doi.org/10.1007/JHEP05(2017)151}{{\em JHEP} {\bfseries 05}
  (2017) 151},
\href{http://arxiv.org/abs/1702.03217}{{\ttfamily arXiv:1702.03217 [hep-th]}}.

\bibitem{Choi:2010su}
K.-S. Choi and T.~Kobayashi, ``{Towards the MSSM from F-theory},''
  \href{http://dx.doi.org/10.1016/j.physletb.2010.08.042}{{\em Phys. Lett.}
  {\bfseries B693} (2010) 330--333},
\href{http://arxiv.org/abs/1003.2126}{{\ttfamily arXiv:1003.2126 [hep-th]}}.

\bibitem{Choi:2010nf}
K.-S. Choi, ``{SU(3) x SU(2) x U(1) Vacua in F-Theory},''
  \href{http://dx.doi.org/10.1016/j.nuclphysb.2010.08.012}{{\em Nucl. Phys.}
  {\bfseries B842} (2011) 1--32},
\href{http://arxiv.org/abs/1007.3843}{{\ttfamily arXiv:1007.3843 [hep-th]}}.

\bibitem{Choi:2013hua}
K.-S. Choi, ``{On the Standard Model Group in F-theory},''
  \href{http://dx.doi.org/10.1140/epjc/s10052-014-2939-7}{{\em Eur. Phys. J.}
  {\bfseries C74} (2014) 2939},
\href{http://arxiv.org/abs/1309.7297}{{\ttfamily arXiv:1309.7297 [hep-th]}}.

\bibitem{Grassi:2014zxa}
A.~Grassi, J.~Halverson, J.~Shaneson, and W.~Taylor, ``{Non-Higgsable QCD and
  the Standard Model Spectrum in F-theory},''
  \href{http://dx.doi.org/10.1007/JHEP01(2015)086}{{\em JHEP} {\bfseries 01}
  (2015) 086},
\href{http://arxiv.org/abs/1409.8295}{{\ttfamily arXiv:1409.8295 [hep-th]}}.

\bibitem{Cvetic:2015txa}
M.~Cvetic, D.~Klevers, D.~K. Mayorga~Peña, P.-K. Oehlmann, and J.~Reuter,
  ``{Three-Family Particle Physics Models from Global F-theory
  Compactifications},'' \href{http://dx.doi.org/10.1007/JHEP08(2015)087}{{\em
  JHEP} {\bfseries 08} (2015) 087},
\href{http://arxiv.org/abs/1503.02068}{{\ttfamily arXiv:1503.02068 [hep-th]}}.

\bibitem{Halverson:2015jua}
J.~Halverson and W.~Taylor, ``{$ {\mathrm{\mathbb{P}}}^1 $-bundle bases and the
  prevalence of non-Higgsable structure in 4D F-theory models},''
  \href{http://dx.doi.org/10.1007/JHEP09(2015)086}{{\em JHEP} {\bfseries 09}
  (2015) 086},
\href{http://arxiv.org/abs/1506.03204}{{\ttfamily arXiv:1506.03204 [hep-th]}}.

\bibitem{Lin:2014qga}
L.~Lin and T.~Weigand, ``{Towards the Standard Model in F-theory},''
\href{http://arxiv.org/abs/1406.6071}{{\ttfamily arXiv:1406.6071 [hep-th]}}.

\bibitem{Lin:2016vus}
L.~Lin and T.~Weigand, ``{G 4 -flux and standard model vacua in F-theory},''
  \href{http://dx.doi.org/10.1016/j.nuclphysb.2016.09.008}{{\em Nucl. Phys.}
  {\bfseries B913} (2016) 209--247},
\href{http://arxiv.org/abs/1604.04292}{{\ttfamily arXiv:1604.04292 [hep-th]}}.

\bibitem{Klevers:2014bqa}
D.~Klevers, D.~K. Mayorga~Peña, P.-K. Oehlmann, H.~Piragua, and J.~Reuter,
  ``{F-Theory on all Toric Hypersurface Fibrations and its Higgs Branches},''
  \href{http://dx.doi.org/10.1007/JHEP01(2015)142}{{\em JHEP} {\bfseries 01}
  (2015) 142},
\href{http://arxiv.org/abs/1408.4808}{{\ttfamily arXiv:1408.4808 [hep-th]}}.

\bibitem{Chen:2010ts}
C.-M. Chen, J.~Knapp, M.~Kreuzer, and C.~Mayrhofer, ``{Global SO(10) F-theory
  GUTs},'' \href{http://dx.doi.org/10.1007/JHEP10(2010)057}{{\em JHEP}
  {\bfseries 1010} (2010) 057},
\href{http://arxiv.org/abs/1005.5735}{{\ttfamily arXiv:1005.5735 [hep-th]}}.

\bibitem{Antoniadis:2012yk}
I.~Antoniadis and G.~K. Leontaris, ``{Building SO(10) models from F-theory},''
  \href{http://dx.doi.org/10.1007/JHEP08(2012)001}{{\em JHEP} {\bfseries 08}
  (2012) 001},
\href{http://arxiv.org/abs/1205.6930}{{\ttfamily arXiv:1205.6930 [hep-th]}}.

\bibitem{Tatar:2012tm}
R.~Tatar and W.~Walters, ``{GUT theories from Calabi-Yau 4-folds with SO(10)
  Singularities},'' \href{http://dx.doi.org/10.1007/JHEP12(2012)092}{{\em JHEP}
  {\bfseries 12} (2012) 092},
\href{http://arxiv.org/abs/1206.5090}{{\ttfamily arXiv:1206.5090 [hep-th]}}.

\bibitem{Blumenhagen:2008zz}
R.~Blumenhagen, V.~Braun, T.~W. Grimm, and T.~Weigand, ``{GUTs in Type IIB
  Orientifold Compactifications},''
  \href{http://dx.doi.org/10.1016/j.nuclphysb.2009.02.011}{{\em Nucl. Phys.}
  {\bfseries B815} (2009) 1--94},
\href{http://arxiv.org/abs/0811.2936}{{\ttfamily arXiv:0811.2936 [hep-th]}}.

\bibitem{Grimm:2011tb}
T.~W. Grimm, M.~Kerstan, E.~Palti, and T.~Weigand, ``{Massive Abelian Gauge
  Symmetries and Fluxes in F-theory},''
  \href{http://dx.doi.org/10.1007/JHEP12(2011)004}{{\em JHEP} {\bfseries 1112}
  (2011) 004},
\href{http://arxiv.org/abs/1107.3842}{{\ttfamily arXiv:1107.3842 [hep-th]}}.

\bibitem{Borchmann:2013jwa}
J.~Borchmann, C.~Mayrhofer, E.~Palti, and T.~Weigand, ``{Elliptic fibrations
  for SU(5) x U(1) x U(1) F-theory vacua},''
\href{http://arxiv.org/abs/1303.5054}{{\ttfamily arXiv:1303.5054 [hep-th]}}.

\bibitem{Cvetic:2013nia}
M.~Cveti{\v c}, D.~Klevers, and H.~Piragua, ``{F-Theory Compactifications with
  Multiple U(1)-Factors: Constructing Elliptic Fibrations with Rational
  Sections},'' \href{http://dx.doi.org/10.1007/JHEP06(2013)067}{{\em JHEP}
  {\bfseries 1306} (2013) 067},
\href{http://arxiv.org/abs/1303.6970}{{\ttfamily arXiv:1303.6970 [hep-th]}}.

\bibitem{Cvetic:2013jta}
M.~Cveti{\v c}, D.~Klevers, and H.~Piragua, ``{F-Theory Compactifications with
  Multiple U(1)-Factors: Addendum},''
  \href{http://dx.doi.org/10.1007/JHEP12(2013)056}{{\em JHEP} {\bfseries 1312}
  (2013) 056},
\href{http://arxiv.org/abs/1307.6425}{{\ttfamily arXiv:1307.6425 [hep-th]}}.

\bibitem{Klevers:2016jsz}
D.~Klevers and W.~Taylor, ``{Three-Index Symmetric Matter Representations of
  SU(2) in F-Theory from Non-Tate Form Weierstrass Models},''
  \href{http://dx.doi.org/10.1007/JHEP06(2016)171}{{\em JHEP} {\bfseries 06}
  (2016) 171},
\href{http://arxiv.org/abs/1604.01030}{{\ttfamily arXiv:1604.01030 [hep-th]}}.

\bibitem{Cvetic:2015moa}
M.~Cvetič, R.~Donagi, D.~Klevers, H.~Piragua, and M.~Poretschkin, ``{F-theory
  vacua with $\mathbb Z_3$ gauge symmetry},''
  \href{http://dx.doi.org/10.1016/j.nuclphysb.2015.07.011}{{\em Nucl. Phys.}
  {\bfseries B898} (2015) 736--750},
\href{http://arxiv.org/abs/1502.06953}{{\ttfamily arXiv:1502.06953 [hep-th]}}.

\bibitem{Blumenhagen:2006xt}
R.~Blumenhagen, M.~Cvetic, and T.~Weigand, ``{Spacetime instanton corrections
  in 4D string vacua: The Seesaw mechanism for D-Brane models},''
  \href{http://dx.doi.org/10.1016/j.nuclphysb.2007.02.016}{{\em Nucl. Phys.}
  {\bfseries B771} (2007) 113--142},
\href{http://arxiv.org/abs/hep-th/0609191}{{\ttfamily arXiv:hep-th/0609191
  [hep-th]}}.

\bibitem{Ibanez:2006da}
L.~E. Ibanez and A.~M. Uranga, ``{Neutrino Majorana Masses from String Theory
  Instanton Effects},''
  \href{http://dx.doi.org/10.1088/1126-6708/2007/03/052}{{\em JHEP} {\bfseries
  03} (2007) 052},
\href{http://arxiv.org/abs/hep-th/0609213}{{\ttfamily arXiv:hep-th/0609213
  [hep-th]}}.

\bibitem{Florea:2006si}
B.~Florea, S.~Kachru, J.~McGreevy, and N.~Saulina, ``{Stringy Instantons and
  Quiver Gauge Theories},''
  \href{http://dx.doi.org/10.1088/1126-6708/2007/05/024}{{\em JHEP} {\bfseries
  05} (2007) 024},
\href{http://arxiv.org/abs/hep-th/0610003}{{\ttfamily arXiv:hep-th/0610003
  [hep-th]}}.

\bibitem{Blumenhagen:2009qh}
R.~Blumenhagen, M.~Cvetic, S.~Kachru, and T.~Weigand, ``{D-Brane Instantons in
  Type II Orientifolds},''
  \href{http://dx.doi.org/10.1146/annurev.nucl.010909.083113}{{\em Ann. Rev.
  Nucl. Part. Sci.} {\bfseries 59} (2009) 269--296},
\href{http://arxiv.org/abs/0902.3251}{{\ttfamily arXiv:0902.3251 [hep-th]}}.

\bibitem{BerasaluceGonzalez:2011wy}
M.~Berasaluce-Gonzalez, L.~E. Ibanez, P.~Soler, and A.~M. Uranga, ``{Discrete
  gauge symmetries in D-brane models},''
  \href{http://dx.doi.org/10.1007/JHEP12(2011)113}{{\em JHEP} {\bfseries 12}
  (2011) 113},
\href{http://arxiv.org/abs/1106.4169}{{\ttfamily arXiv:1106.4169 [hep-th]}}.

\bibitem{BerasaluceGonzalez:2012zn}
M.~Berasaluce-Gonzalez, P.~G. Camara, F.~Marchesano, and A.~M. Uranga, ``{Zp
  charged branes in flux compactifications},''
  \href{http://dx.doi.org/10.1007/JHEP04(2013)138}{{\em JHEP} {\bfseries 04}
  (2013) 138},
\href{http://arxiv.org/abs/1211.5317}{{\ttfamily arXiv:1211.5317 [hep-th]}}.

\bibitem{Donagi:2012ts}
R.~Donagi, S.~Katz, and M.~Wijnholt, ``{Weak Coupling, Degeneration and Log
  Calabi-Yau Spaces},''
\href{http://arxiv.org/abs/1212.0553}{{\ttfamily arXiv:1212.0553 [hep-th]}}.

\bibitem{Morrison:2012ei}
D.~R. Morrison and D.~S. Park, ``{F-Theory and the Mordell-Weil Group of
  Elliptically-Fibered Calabi-Yau Threefolds},''
  \href{http://dx.doi.org/10.1007/JHEP10(2012)128}{{\em JHEP} {\bfseries 1210}
  (2012) 128},
\href{http://arxiv.org/abs/1208.2695}{{\ttfamily arXiv:1208.2695 [hep-th]}}.

\bibitem{Bershadsky:1996nh}
M.~Bershadsky, K.~A. Intriligator, S.~Kachru, D.~R. Morrison, V.~Sadov, {\em
  et~al.}, ``{Geometric singularities and enhanced gauge symmetries},''
  \href{http://dx.doi.org/10.1016/S0550-3213(96)90131-5}{{\em Nucl.Phys.}
  {\bfseries B481} (1996) 215--252},
\href{http://arxiv.org/abs/hep-th/9605200}{{\ttfamily arXiv:hep-th/9605200
  [hep-th]}}.

\bibitem{Greene:1993vm}
B.~R. Greene, D.~R. Morrison, and M.~R. Plesser, ``{Mirror manifolds in higher
  dimension},'' \href{http://dx.doi.org/10.1007/BF02101657}{{\em
  Commun.Math.Phys.} {\bfseries 173} (1995) 559--598},
\href{http://arxiv.org/abs/hep-th/9402119}{{\ttfamily arXiv:hep-th/9402119
  [hep-th]}}.

\bibitem{Bies:2014sra}
M.~Bies, C.~Mayrhofer, C.~Pehle, and T.~Weigand, ``{Chow groups, Deligne
  cohomology and massless matter in F-theory},''
\href{http://arxiv.org/abs/1402.5144}{{\ttfamily arXiv:1402.5144 [hep-th]}}.

\bibitem{Bies:2017fam}
M.~Bies, C.~Mayrhofer, and T.~Weigand, ``{Gauge Backgrounds and Zero-Mode
  Counting in F-Theory},''
\href{http://arxiv.org/abs/1706.04616}{{\ttfamily arXiv:1706.04616 [hep-th]}}.

\bibitem{Grimm:2010ez}
T.~W. Grimm and T.~Weigand, ``{On Abelian Gauge Symmetries and Proton Decay in
  Global F-theory GUTs},''
  \href{http://dx.doi.org/10.1103/PhysRevD.82.086009}{{\em Phys.Rev.}
  {\bfseries D82} (2010) 086009},
  \href{http://arxiv.org/abs/1006.0226}{{\ttfamily arXiv:1006.0226 [hep-th]}}.

\bibitem{Marsano:2011hv}
J.~Marsano and S.~Schafer-Nameki, ``{Yukawas, G-flux, and Spectral Covers from
  Resolved Calabi-Yau's},''
  \href{http://dx.doi.org/10.1007/JHEP11(2011)098}{{\em JHEP} {\bfseries 1111}
  (2011) 098},
\href{http://arxiv.org/abs/1108.1794}{{\ttfamily arXiv:1108.1794 [hep-th]}}.

\bibitem{Lin:2015qsa}
L.~Lin, C.~Mayrhofer, O.~Till, and T.~Weigand, ``{Fluxes in F-theory
  Compactifications on Genus-One Fibrations},''
  \href{http://dx.doi.org/10.1007/JHEP01(2016)098}{{\em JHEP} {\bfseries 01}
  (2016) 098},
\href{http://arxiv.org/abs/1508.00162}{{\ttfamily arXiv:1508.00162 [hep-th]}}.

\bibitem{Dasgupta:1999ss}
K.~Dasgupta, G.~Rajesh, and S.~Sethi, ``{M theory, orientifolds and G -
  flux},'' \href{http://dx.doi.org/10.1088/1126-6708/1999/08/023}{{\em JHEP}
  {\bfseries 08} (1999) 023},
\href{http://arxiv.org/abs/hep-th/9908088}{{\ttfamily arXiv:hep-th/9908088
  [hep-th]}}.

\bibitem{Grimm:2011fx}
T.~W. Grimm and H.~Hayashi, ``{F-theory fluxes, Chirality and Chern-Simons
  theories},'' \href{http://dx.doi.org/10.1007/JHEP03(2012)027}{{\em JHEP}
  {\bfseries 1203} (2012) 027},
\href{http://arxiv.org/abs/1111.1232}{{\ttfamily arXiv:1111.1232 [hep-th]}}.

\bibitem{Grimm:2011sk}
T.~W. Grimm and R.~Savelli, ``{Gravitational Instantons and Fluxes from
  M/F-theory on Calabi-Yau fourfolds},''
  \href{http://dx.doi.org/10.1103/PhysRevD.85.026003}{{\em Phys.Rev.}
  {\bfseries D85} (2012) 026003},
\href{http://arxiv.org/abs/1109.3191}{{\ttfamily arXiv:1109.3191 [hep-th]}}.

\bibitem{Cvetic:2012xn}
M.~Cveti{\v c}, T.~W. Grimm, and D.~Klevers, ``{Anomaly Cancellation And
  Abelian Gauge Symmetries In F-theory},''
  \href{http://dx.doi.org/10.1007/JHEP02(2013)101}{{\em JHEP} {\bfseries 1302}
  (2013) 101},
\href{http://arxiv.org/abs/1210.6034}{{\ttfamily arXiv:1210.6034 [hep-th]}}.

\bibitem{Witten:1996md}
E.~Witten, ``{On flux quantization in M theory and the effective action},''
  \href{http://dx.doi.org/10.1016/S0393-0440(96)00042-3}{{\em J.Geom.Phys.}
  {\bfseries 22} (1997) 1--13},
\href{http://arxiv.org/abs/hep-th/9609122}{{\ttfamily arXiv:hep-th/9609122
  [hep-th]}}.

\bibitem{Hayashi:2008ba}
H.~Hayashi, R.~Tatar, Y.~Toda, T.~Watari, and M.~Yamazaki, ``{New Aspects of
  Heterotic--F Theory Duality},''
  \href{http://dx.doi.org/10.1016/j.nuclphysb.2008.07.031}{{\em Nucl.Phys.}
  {\bfseries B806} (2009) 224--299},
\href{http://arxiv.org/abs/0805.1057}{{\ttfamily arXiv:0805.1057 [hep-th]}}.

\bibitem{Ibanez:1998qp}
L.~E. Ibanez, R.~Rabadan, and A.~M. Uranga, ``{Anomalous U(1)'s in type I and
  type IIB D = 4, N=1 string vacua},''
  \href{http://dx.doi.org/10.1016/S0550-3213(98)00791-3}{{\em Nucl. Phys.}
  {\bfseries B542} (1999) 112--138},
\href{http://arxiv.org/abs/hep-th/9808139}{{\ttfamily arXiv:hep-th/9808139
  [hep-th]}}.

\bibitem{Poppitz:1998dj}
E.~Poppitz, ``{On the one loop Fayet-Iliopoulos term in chiral four-dimensional
  type I orbifolds},''
  \href{http://dx.doi.org/10.1016/S0550-3213(98)00793-7}{{\em Nucl. Phys.}
  {\bfseries B542} (1999) 31--44},
\href{http://arxiv.org/abs/hep-th/9810010}{{\ttfamily arXiv:hep-th/9810010
  [hep-th]}}.

\bibitem{Aldazabal:2000dg}
G.~Aldazabal, S.~Franco, L.~E. Ibanez, R.~Rabadan, and A.~M. Uranga, ``{D = 4
  chiral string compactifications from intersecting branes},''
  \href{http://dx.doi.org/10.1063/1.1376157}{{\em J. Math. Phys.} {\bfseries
  42} (2001) 3103--3126},
\href{http://arxiv.org/abs/hep-th/0011073}{{\ttfamily arXiv:hep-th/0011073
  [hep-th]}}.

\bibitem{Donagi:2009ra}
R.~Donagi and M.~Wijnholt, ``{Higgs Bundles and UV Completion in F-Theory},''
\href{http://arxiv.org/abs/0904.1218}{{\ttfamily arXiv:0904.1218 [hep-th]}}.

\bibitem{Minasian:1997mm}
R.~Minasian and G.~W. Moore, ``{K theory and Ramond-Ramond charge},''
  \href{http://dx.doi.org/10.1088/1126-6708/1997/11/002}{{\em JHEP} {\bfseries
  11} (1997) 002},
\href{http://arxiv.org/abs/hep-th/9710230}{{\ttfamily arXiv:hep-th/9710230
  [hep-th]}}.

\bibitem{Freed:1999vc}
D.~S. Freed and E.~Witten, ``{Anomalies in string theory with D-branes},'' {\em
  Asian J. Math.} {\bfseries 3} (1999) 819,
\href{http://arxiv.org/abs/hep-th/9907189}{{\ttfamily arXiv:hep-th/9907189
  [hep-th]}}.

\bibitem{Font:2008id}
A.~Font and L.~E. Ibanez, ``{Yukawa Structure from U(1) Fluxes in F-theory
  Grand Unification},''
  \href{http://dx.doi.org/10.1088/1126-6708/2009/02/016}{{\em JHEP} {\bfseries
  02} (2009) 016},
\href{http://arxiv.org/abs/0811.2157}{{\ttfamily arXiv:0811.2157 [hep-th]}}.

\bibitem{Heckman:2008qa}
J.~J. Heckman and C.~Vafa, ``{Flavor Hierarchy From F-theory},''
  \href{http://dx.doi.org/10.1016/j.nuclphysb.2010.05.009}{{\em Nucl. Phys.}
  {\bfseries B837} (2010) 137--151},
\href{http://arxiv.org/abs/0811.2417}{{\ttfamily arXiv:0811.2417 [hep-th]}}.

\bibitem{Cecotti:2009zf}
S.~Cecotti, M.~C.~N. Cheng, J.~J. Heckman, and C.~Vafa, ``{Yukawa Couplings in
  F-theory and Non-Commutative Geometry},''
\href{http://arxiv.org/abs/0910.0477}{{\ttfamily arXiv:0910.0477 [hep-th]}}.

\bibitem{Marchesano:2009rz}
F.~Marchesano and L.~Martucci, ``{Non-perturbative effects on seven-brane
  Yukawa couplings},''
  \href{http://dx.doi.org/10.1103/PhysRevLett.104.231601}{{\em Phys. Rev.
  Lett.} {\bfseries 104} (2010) 231601},
\href{http://arxiv.org/abs/0910.5496}{{\ttfamily arXiv:0910.5496 [hep-th]}}.

\bibitem{Leontaris:2010zd}
G.~K. Leontaris and G.~G. Ross, ``{Yukawa couplings and fermion mass structure
  in F-theory GUTs},'' \href{http://dx.doi.org/10.1007/JHEP02(2011)108}{{\em
  JHEP} {\bfseries 02} (2011) 108},
\href{http://arxiv.org/abs/1009.6000}{{\ttfamily arXiv:1009.6000 [hep-th]}}.

\bibitem{Aparicio:2011jx}
L.~Aparicio, A.~Font, L.~E. Ibanez, and F.~Marchesano, ``{Flux and Instanton
  Effects in Local F-theory Models and Hierarchical Fermion Masses},''
  \href{http://dx.doi.org/10.1007/JHEP08(2011)152}{{\em JHEP} {\bfseries 08}
  (2011) 152},
\href{http://arxiv.org/abs/1104.2609}{{\ttfamily arXiv:1104.2609 [hep-th]}}.

\bibitem{Palti:2012aa}
E.~Palti, ``{Wavefunctions and the Point of $E_8$ in F-theory},''
  \href{http://dx.doi.org/10.1007/JHEP07(2012)065}{{\em JHEP} {\bfseries 07}
  (2012) 065},
\href{http://arxiv.org/abs/1203.4490}{{\ttfamily arXiv:1203.4490 [hep-th]}}.

\bibitem{Font:2013ida}
A.~Font, F.~Marchesano, D.~Regalado, and G.~Zoccarato, ``{Up-type quark masses
  in SU(5) F-theory models},''
  \href{http://dx.doi.org/10.1007/JHEP11(2013)125}{{\em JHEP} {\bfseries 11}
  (2013) 125},
\href{http://arxiv.org/abs/1307.8089}{{\ttfamily arXiv:1307.8089 [hep-th]}}.

\bibitem{Marchesano:2015dfa}
F.~Marchesano, D.~Regalado, and G.~Zoccarato, ``{Yukawa hierarchies at the
  point of E$_{8}$ in F-theory},''
  \href{http://dx.doi.org/10.1007/JHEP04(2015)179}{{\em JHEP} {\bfseries 04}
  (2015) 179},
\href{http://arxiv.org/abs/1503.02683}{{\ttfamily arXiv:1503.02683 [hep-th]}}.

\bibitem{Font:2015slq}
A.~Font, ``{Yukawa couplings in string theory: the case for F-theory GUT's},''
\href{http://dx.doi.org/10.1088/1742-6596/651/1/012009}{{\em J. Phys. Conf.
  Ser.} {\bfseries 651} no.~1, (2015) 012009}.

\bibitem{Carta:2015eoh}
F.~Carta, F.~Marchesano, and G.~Zoccarato, ``{Fitting fermion masses and
  mixings in F-theory GUTs},''
  \href{http://dx.doi.org/10.1007/JHEP03(2016)126}{{\em JHEP} {\bfseries 03}
  (2016) 126},
\href{http://arxiv.org/abs/1512.04846}{{\ttfamily arXiv:1512.04846 [hep-th]}}.

\bibitem{Candelas:1996su}
P.~Candelas and A.~Font, ``{Duality between the webs of heterotic and type II
  vacua},'' \href{http://dx.doi.org/10.1016/S0550-3213(96)00410-5}{{\em
  Nucl.Phys.} {\bfseries B511} (1998) 295--325},
\href{http://arxiv.org/abs/hep-th/9603170}{{\ttfamily arXiv:hep-th/9603170
  [hep-th]}}.

\bibitem{Candelas:1997pv}
P.~Candelas, E.~Perevalov, and G.~Rajesh, ``{Comments on A, B, C chains of
  heterotic and type II vacua},''
  \href{http://dx.doi.org/10.1016/S0550-3213(97)00374-X}{{\em Nucl. Phys.}
  {\bfseries B502} (1997) 594--612},
\href{http://arxiv.org/abs/hep-th/9703148}{{\ttfamily arXiv:hep-th/9703148
  [hep-th]}}.

\bibitem{Bouchard:2003bu}
V.~Bouchard and H.~Skarke, ``{Affine Kac-Moody algebras, CHL strings and the
  classification of tops},''
  \href{http://dx.doi.org/10.4310/ATMP.2003.v7.n2.a1}{{\em Adv. Theor. Math.
  Phys.} {\bfseries 7} no.~2, (2003) 205--232},
\href{http://arxiv.org/abs/hep-th/0303218}{{\ttfamily arXiv:hep-th/0303218
  [hep-th]}}.

\bibitem{Braun:2014xka}
A.~P. Braun and T.~Watari, ``{The Vertical, the Horizontal and the Rest:
  anatomy of the middle cohomology of Calabi-Yau fourfolds and F-theory
  applications},'' \href{http://dx.doi.org/10.1007/JHEP01(2015)047}{{\em JHEP}
  {\bfseries 01} (2015) 047},
\href{http://arxiv.org/abs/1408.6167}{{\ttfamily arXiv:1408.6167 [hep-th]}}.

\end{thebibliography}
\providecommand{\href}[2]{#2}\begingroup\raggedright\endgroup

\end{document}